\documentclass[ twocolumn,twocolappendix]{aastex7}

\usepackage{hyperref}
\usepackage[htt]{hyphenat} 
\usepackage{xspace} 
\usepackage{subcaption}
\usepackage{glossaries}  

\graphicspath{{figures}}

\providecommand{\secref}[1]{\hyperref[#1]{\S\ref{#1}}}
\providecommand{\appref}[1]{\hyperref[#1]{Appendix \ref{#1}}}
\providecommand{\tabref}[1]{\hyperref[#1]{\autoref{#1}}}
\providecommand{\figref}[1]{\hyperref[#1]{\autoref{#1}}}
\providecommand{\eqnref}[1]{\hyperref[#1]{Eq.~\ref{#1}}}
\providecommand{\recref}[1]{\hyperref[#1]{REC-\ref{#1}}}

\newacronym{2D} {2D} {Two-dimensional}
\newacronym{ADC} {ADC} {Analogue-to-Digital Converter}
\newacronym{ADQL} {ADQL} {Astronomical Data Query Language (IVOA standard)}
\newacronym{ADU} {ADU} {Analogue-to-Digital Unit}
\newacronym{AOS} {AOS} {Active Optics System}
\newacronym{API} {API} {Application Programming Interface}
\newacronym{ASPIC} {ASPIC} {Analog Signal Processing Integrated Circuit}
\newacronym{AST} {AST} {NSF Division of Astronomical Sciences}
\newacronym{ATLAS} {ATLAS} {Asteroid Terrestrial-impact Last Alert System}
\newacronym{AURA} {AURA} {Association of Universities for Research in Astronomy}
\newacronym{Adam} {Adam} {Adaptive Moment Estimation}
\newglossaryentry{Alert} {name={Alert}, description={A packet of information for each source detected with signal-to-noise ratio > 5 in a difference image by Alert Production, containing measurement and characterization parameters based on the past 12 months of LSST observations plus small cutouts of the single-visit, template, and difference images, distributed via the internet}}
\newglossaryentry{Alert Production} {name={Alert Production}, description={Executing on the Prompt Processing system, the Alert Production payload processes and calibrates incoming images, performs Difference Image Analysis to identify DIASources and DIAObjects, and then packages the resulting alerts for distribution.}}
\newglossaryentry{Alternate Standard Visit} {name={Alternate Standard Visit}, description={A single observation of an LSST field comprised of one 30 second exposure}}
\newglossaryentry{Archive} {name={Archive}, description={The repository for documents required by the NSF to be kept. These include documents related to design and development, construction, integration, test, and operations of the LSST observatory system. The archive is maintained using the enterprise content management system DocuShare, which is accessible through a link on the project website www.project.lsst.org}}
\newglossaryentry{Archive Center} {name={Archive Center}, description={Part of the LSST Data Management System, the LSST archive center is a data center at NCSA that hosts the LSST Archive, which includes released science data and metadata, observatory and engineering data, and supporting software such as the LSST Software Stack}}
\newglossaryentry{Association Pipeline} {name={Association Pipeline}, description={An application that matches detected Sources or DIASources or generated Objects to an existing catalog of Objects, producing a (possibly many-to-many) set of associations and a list of unassociated inputs. Association Pipelines are used in Alert Production after DIASource generation and in the final stages of Data Release processing to ensure continuity of Object identifiers}}
\newglossaryentry{Association of Universities for Research in Astronomy} {name={Association of Universities for Research in Astronomy}, description={ consortium of US institutions and international affiliates that operates world-class astronomical observatories, AURA is the legal entity responsible for managing what it calls independent operating Centers, including LSST, under respective cooperative agreements with the National Science Foundation. AURA assumes fiducial responsibility for the funds provided through those cooperative agreements. AURA also is the legal owner of the AURA Observatory properties in Chile}}
\newacronym{BCE} {BCE} {Before Common Era}
\newglossaryentry{Butler} {name={Butler}, description={A middleware component for persisting and retrieving image datasets (raw or processed), calibration reference data, and catalogs}}
\newacronym{CADC} {CADC} {Canadian Astronomy Data Centre}
\newacronym{CCD} {CCD} {Charge-Coupled Device}
\newacronym{CNRS} {CNRS} {Centre national de la recherche scientifique}
\newacronym{CPU} {CPU} {Central Processing Unit}
\newacronym{CTI} {CTI} {Charge Transfer Inefficiency}
\newacronym{CVMFS} {CVMFS} {CernVM File System}
\newglossaryentry{Camera} {name={Camera}, description={The LSST subsystem responsible for the 3.2-gigapixel LSST camera, which will take more than 800 panoramic images of the sky every night. SLAC leads a consortium of Department of Energy laboratories to design and build the camera sensors, optics, electronics, cryostat, filters and filter exchange mechanism, and camera control system}}
\newglossaryentry{Center} {name={Center}, description={An entity managed by AURA that is responsible for execution of a federally funded project}}
\newglossaryentry{Charge-Coupled Device} {name={Charge-Coupled Device}, description={a particular kind of solid-state sensor for detecting optical-band photons. It is composed of a 2-D array of pixels, and one or more read-out amplifiers}}
\newglossaryentry{Coadd Image} {name={Coadd Image}, description={An image that is the combination of multiple input images. The inputs are aligned to a common projection and pixel grid, corrected to the same photometric scale and zero-point, with bad pixels and artifacts rejected. (Image PSFs may also be matched prior to co-addition.) Coadd Images have had non-astrophysical background removed}}
\newglossaryentry{ComCam} {name={ComCam}, description={The commissioning camera is a single-raft, 9-CCD camera that will be installed in LSST during commissioning, before the final camera is ready.}}
\newglossaryentry{Commissioning} {name={Commissioning}, description={A two-year phase at the end of the Construction project during which a technical team a) integrates the various technical components of the three subsystems; b) shows their compliance with ICDs and system-level requirements as detailed in the LSST Observatory System Specifications document (OSS, LSE-30); and c) performs science verification to show compliance with the survey performance specifications as detailed in the LSST Science Requirements Document (SRD, LPM-17)}}
\newglossaryentry{Construction} {name={Construction}, description={The period during which LSST observatory facilities, components, hardware, and software are built, tested, integrated, and commissioned. Construction follows design and development and precedes operations. The LSST construction phase is funded through the NSF MREFC account}}
\newglossaryentry{Contract} {name={Contract}, description={A binding legal agreement between parties obligating the one (typically the  'seller') to furnish certain supplies or services and the other (typically, the buyer) to compensate the seller for the supplies or services with some form of consideration, (typically money). The term, 'contract' is used interchangeably with 'sub-award' 'agreement' 'memorandum of understanding and/or agreement' and 'purchase order' Each is a term used to differentiate between a purchase-order-format type document and a complex purchase in a subcontract/sub-award-format type document. These also include awards and notices of awards; job orders or task letters issued under basic ordering agreements; letter contracts; orders, such as purchase orders and subcontracts under which the order becomes effective by written acceptance or performance; and bilateral contract modifications}}
\newacronym{DAC} {DAC} {Data Access Center}
\newacronym{DC2} {DC2} {Data Challenge 2}
\newacronym{DCR} {DCR} {Differential Chromatic Refraction}
\newacronym{DE-AC02} {DE-AC02} {\gls{Department of Energy} contract number prefix}
\newacronym{DES} {DES} {Dark Energy Survey}
\newacronym{DESC} {DESC} {Dark Energy Science Collaboration}
\newacronym{DIA} {DIA} {Difference Image Analysis}
\newglossaryentry{DIAObject} {name={DIAObject}, description={A DIAObject is the association of DIASources, by coordinate, that have been detected with signal-to-noise ratio greater than 5 in at least one difference image. It is distinguished from a regular Object in that its brightness varies in time, and from a SSObject in that it is stationary (non-moving)}}
\newglossaryentry{DIASource} {name={DIASource}, description={A DIASource is a detection with signal-to-noise ratio greater than 5 in a difference image}}
\newacronym{DIMM} {DIMM} {Differential Image Motion Monitor}
\newacronym{DM} {DM} {Data Management}
\newacronym{DMS} {DMS} {Data Management Subsystem}
\newacronym{DMTN} {DMTN} {DM Technical Note}
\newacronym{DOE} {DOE} {Department of Energy}
\newacronym{DP0} {DP0} {Data Preview 0}
\newacronym{DP1} {DP1} {Data Preview 1}
\newacronym{DP2} {DP2} {Data Preview 2}
\newacronym{DPDD} {DPDD} {Data Product Definition Document}
\newacronym{DR} {DR} {Data Release}
\newacronym{DR1} {DR1} {Data Release 1}
\newacronym{DR3} {DR3} {Data Release 3}
\newacronym{DRP} {DRP} {Data Release Processing}
\newglossaryentry{Data Access Center} {name={Data Access Center}, description={Part of the LSST Data Management System, the US and Chilean DACs will provide authorized access to the released LSST data products, software such as the Science Platform, and computational resources for data analysis. The US DAC also includes a service for distributing bulk data on daily and annual (Data Release) timescales to partner institutions, collaborations, and LSST Education and Public Outreach (EPO). }}
\newglossaryentry{Data Management} {name={Data Management}, description={The LSST Subsystem responsible for the Data Management System (DMS), which will capture, store, catalog, and serve the LSST dataset to the scientific community and public. The DM team is responsible for the DMS architecture, applications, middleware, infrastructure, algorithms, and Observatory Network Design. DM is a distributed team working at LSST and partner institutions, with the DM Subsystem Manager located at LSST headquarters in Tucson}}
\newglossaryentry{Data Management Subsystem} {name={Data Management Subsystem}, description={The Data Management Subsystem is one of the four subsystems which constitute the LSST Construction Project. The Data Management Subsystem is responsible for developing and delivering the LSST Data Management System to the LSST Operations Project}}
\newglossaryentry{Data Management System} {name={Data Management System}, description={The computing infrastructure, middleware, and applications that process, store, and enable information extraction from the LSST dataset; the DMS will process peta-scale data volume, convert raw images into a faithful representation of the universe, and archive the results in a useful form. The infrastructure layer consists of the computing, storage, networking hardware, and system software. The middleware layer handles distributed processing, data access, user interface, and system operations services. The applications layer includes the data pipelines and the science data archives' products and services}}
\newglossaryentry{Data Product} {name={Data Product}, description={The LSST survey will produce three categories of Data Products. Prompt, Data Release, User Generated. Previously referred to as Levels 1, 2, and 3}}
\newglossaryentry{Data Release} {name={Data Release}, description={The approximately annual reprocessing of all LSST data, and the installation of the resulting data products in the LSST Data Access Centers, which marks the start of the two-year proprietary period}}
\newglossaryentry{Data Release Processing} {name={Data Release Processing}, description={Deprecated term; see Data Release Production}}
\newglossaryentry{Data Release Production} {name={Data Release Production}, description={An episode of (re)processing all of the accumulated LSST images, during which all output DR data products are generated. These episodes are planned to occur annually during the LSST survey, and the processing will be executed at the Archive Center. This includes Difference Imaging Analysis, generating deep Coadd Images, Source detection and association, creating Object and Solar System Object catalogs, and related metadata}}
\newglossaryentry{Department of Energy} {name={Department of Energy}, description={cabinet department of the United States federal government; the DOE has assumed technical and financial responsibility for providing the LSST camera. The DOE's responsibilities are executed by a collaboration led by SLAC National Accelerator Laboratory}}
\newglossaryentry{Difference Image} {name={Difference Image}, description={Refers to the result formed from the pixel-by-pixel difference of two images of the sky, after warping to the same pixel grid, scaling to the same photometric response, matching to the same PSF shape, and applying a correction for Differential Chromatic Refraction. The pixels in a difference thus formed should be zero (apart from noise) except for sources that are new, or have changed in brightness or position. In the LSST context, the difference is generally taken between a visit image and template. }}
\newglossaryentry{Difference Image Analysis} {name={Difference Image Analysis}, description={The detection and characterization of sources in the Difference Image that are above a configurable threshold, done as part of Alert Generation Pipeline}}
\newglossaryentry{Differential Chromatic Refraction} {name={Differential Chromatic Refraction}, description={The refraction of incident light by Earth's atmosphere causes the apparent position of objects to be shifted, and the size of this shift depends on both the wavelength of the source and its airmass at the time of observation. DCR corrections are done as a part of DIA}}
\newglossaryentry{Director} {name={Director}, description={The person responsible for the overall conduct of the project; the LSST director is charged with ensuring that both the scientific goals and management constraints on the project are met. S/he is the principal public spokesperson for the project in all matters and represents the project to the scientific community, AURA, the member institutions of LSST-DA, and the funding agencies}}
\newglossaryentry{DocuShare} {name={DocuShare}, description={The trade name for the enterprise management software used by LSST to archive and manage documents}}
\newglossaryentry{Document} {name={Document}, description={Any object (in any application supported by DocuShare or design archives such as PDMWorks or GIT) that supports project management or records milestones and deliverables of the LSST Project}}
\newacronym{E2V} {E2V} {Teledyne}
\newacronym{ECDFS} {ECDFS} {Extended Chandra Deep Field-South Survey}
\newacronym{EDFS} {EDFS} {Euclid Deep Field South}
\newacronym{EFD} {EFD} {Engineering and Facility Database}
\newacronym{EPO} {EPO} {Education and Public Outreach}
\newacronym{ESNet} {ESNet} {Energy Sciences Network}
\newacronym{ESO} {ESO} {European Southern Observatory}
\newglossaryentry{Education and Public Outreach} {name={Education and Public Outreach}, description={The LSST subsystem responsible for the cyberinfrastructure, user interfaces, and outreach programs necessary to connect educators, planetaria, citizen scientists, amateur astronomers, and the general public to the transformative LSST dataset}}
\newacronym{FAA} {FAA} {U.S. Federal Aviation Administration}
\newacronym{FBS} {FBS} {Feature-Based Scheduler}
\newacronym{FGCM} {FGCM} {Forward Global Calibration Method}
\newacronym{FITS} {FITS} {Flexible Image Transport System}
\newacronym{FOV} {FOV} {field of view}
\newacronym{FTS3} {FTS3} {File Transfer Service 3}
\newacronym{FWHM} {FWHM} {Full Width at Half-Maximum}
\newglossaryentry{Filter} {name={Filter}, description={A filter in astronomy is an optical element used to restrict the passband of light reaching the focal plane, it transmits a selected range of wavelengths. Filters elements are often named after standard photometric passbands, such as those used in the SDSS survey: u, g, r, i, z}}
\newglossaryentry{Firefly} {name={Firefly}, description={A framework of software components written by IPAC for building web-based user interfaces to astronomical archives, through which data may be searched and retrieved, and viewed as FITS images, catalogs, and/or plots. Firefly tools will be integrated into the Science Platform}}
\newglossaryentry{Flexible Image Transport System} {name={Flexible Image Transport System}, description={an international standard in astronomy for storing images, tables, and metadata in disk files. See the IAU FITS Standard for details}}
\newglossaryentry{ForcedSource} {name={ForcedSource}, description={DRP table resulting from forced photometry}}
\newacronym{FrDF} {FrDF} {French Data Facility}
\newacronym{GAaP} {GAaP} {Gaussian Aperture and PSF}
\newacronym{GBDES} {GBDES} {Gary Bernstein Dark Energy Survey}
\newacronym{GP} {GP} {Gaussian Process}
\newacronym{GPS} {GPS} {Global Positioning System}
\newacronym{GPU} {GPU} {Graphics Processing Unit}
\newglossaryentry{Gaia} {name={Gaia}, description={a space observatory of the European Space Agency, launched in 2013 and expected to operate until 2025. The spacecraft is designed for astrometry: measuring the positions, distances and motions of stars with unprecedented precision}}
\newglossaryentry{Gaussian Aperture and PSF} {name={Gaussian Aperture and PSF}, description={ involves Gaussianizing the PSFs and then using a Gaussian aperture (instead of top-hat) for measuring photometry. The aperture+PSF is designed to be the same across all bands, so that you measure consistent colors.}}
\newacronym{HEALPix} {HEALPix} {Hierarchical Equal-Area iso-Latitude Pixelisation}
\newglossaryentry{HSM} {name={HSM}, description={Shape measurement algorithm from Hirata \& Seljak (2003) and Mandelbaum et al. (2005)}}
\newacronym{HST} {HST} {Hubble Space Telescope}
\newglossaryentry{Handle} {name={Handle}, description={The unique identifier assigned to a document uploaded to DocuShare}}
\newacronym{HiPS} {HiPS} {Hierarchical Progressive Survey}
\newacronym{IAU} {IAU} {International Astronomical Union}
\newacronym{IN2P3} {IN2P3} {Institut National de Physique Nucléaire et de Physique des Particules}
\newacronym{IPAC} {IPAC} {No longer an acronym; science and data center at Caltech}
\newacronym{ISR} {ISR} {Instrument Signal Removal}
\newacronym{ITL} {ITL} {Imaging Technology Laboratory,  University of Arizona (UA))}
\newacronym{IVOA} {IVOA} {International Virtual Observatory Alliance}
\newglossaryentry{Instrument Signature Removal} {name={Instrument Signature Removal}, description={Instrument Signature Removal is a pipeline that applies calibration reference data in the course of raw data processing, to remove artifacts of the instrument or detector electronics, such as removal of overscan pixels, bias correction, and the application of a flat-field to correct for pixel-to-pixel variations in sensitivity}}
\newglossaryentry{J2000} {name={J2000}, description={Julian Date referring to the instant of 12 noon (midday) on January 1, 2000. IAU standard equinox.}}
\newacronym{JD} {JD} {\gls{Julian Date}}
\newglossaryentry{Julian Date} {name={Julian Date}, description={The Julian Date (JD) of any instant is the Julian day number for the preceding noon (UTC), plus the fraction of the day elapsed since that instant. The Julian day number is a running sequence of integral days, starting at noon, since the beginning of the Julian Period; JD 0.0 corresponds to noon on 1 January 4713 BCE. Various Julian Date converters are available on the Web. For example, 18h 00m 00.0s UT on 2014-July-01 (near the start of LSST construction) corresponds to JD 2456840.25}}
\newacronym{LDM} {LDM} {LSST Data Management (Document \gls{Handle})}
\newacronym{LPM} {LPM} {LSST Project Management (Document \gls{Handle})}
\newacronym{LSE} {LSE} {LSST \gls{Systems Engineering} (Document Handle)}
\newacronym{LSST} {LSST} {Legacy Survey of Space and Time}
\newglossaryentry{LSST Camera} {name={LSST Camera}, description={3.2 Gigapixel camera and lens system build by SLAC to perform the Legacy Survey of Space and Time.}}
\newglossaryentry{LSST Project Office} {name={LSST Project Office}, description={Official name of the stand-alone AURA operating center responsible for execution of the LSST construction project under the NSF MREFC account}}
\newglossaryentry{LSST Science Pipelines} {name={LSST Science Pipelines}, description={software used to perform the LSST data reduction pipelines.lsst.io}}
\newacronym{LSST-DA} {LSST-DA} {LSST Discovery Alliance}
\newacronym{LSSTCam} {LSSTCam} {LSST Science Camera}
\newacronym{LSSTComCam} {LSSTComCam} {Rubin Commissioning Camera}
\newacronym{LSSTPO} {LSSTPO} {\gls{LSST Project Office}}
\newacronym{M1M3} {M1M3} {Primary Mirror Tertiary Mirror}
\newacronym{M2} {M2} {Secondary Mirror}
\newacronym{MC} {MC} {Monte-Carlo (simulation/process)}
\newacronym{MJD} {MJD} {Modified Julian Date}
\newacronym{MOC} {MOC} {Multi-Order Coverage (IVOA standard)}
\newacronym{MODTRAN} {MODTRAN} {MODerate resolution TRANsmission model}
\newacronym{MPC} {MPC} {Minor Planet Center}
\newacronym{MPCORB} {MPCORB} {Minor Planet Center Orbit database}
\newacronym{MREFC} {MREFC} {\gls{Major Research Equipment and Facility Construction}}
\newglossaryentry{Major Research Equipment and Facility Construction} {name={Major Research Equipment and Facility Construction}, description={the NSF account through which large facilities construction projects such as LSST are funded}}
\newacronym{NCSA} {NCSA} {National \gls{Center} for Supercomputing Applications}
\newacronym{NEO} {NEO} {Near-Earth Object}
\newacronym{NES} {NES} {North Ecliptic Spur}
\newacronym{NOIRLab} {NOIRLab} {NSF's National Optical-Infrared Astronomy Research Laboratory; \url{https://noirlab.edu}}
\newacronym{NSF} {NSF} {National Science Foundation}
\newacronym{NTP} {NTP} {Network Time Protocol}
\newglossaryentry{National Science Foundation} {name={National Science Foundation}, description={primary federal agency supporting research in all fields of fundamental science and engineering; NSF selects and funds projects through competitive, merit-based review}}
\newglossaryentry{Non-Standard Visit} {name={Non-Standard Visit}, description={Any single observation of a LSST field that is not comprised of either two 15 second 'Snap' exposures (a standard visit) or one 30 second exposure (an alternative standard visit). For example, exposure times for Special Programs might be significantly shorter or longer than a standard visit (or of random length)}}
\newacronym{OSS} {OSS} {Observatory System Specifications; \gls{LSE}-30}
\newglossaryentry{Object} {name={Object}, description={In LSST nomenclature this refers to an astronomical object, such as a star, galaxy, or other physical entity. E.g., comets, asteroids are also Objects but typically called a Moving Object or a Solar System Object (SSObject). One of the DRP data products is a table of Objects detected by LSST which can be static, or change brightness or position with time}}
\newacronym{ObsCore} {ObsCore} {Observation Data Model Core Components (\gls{IVOA} standard)}
\newacronym{ObsTAP} {ObsTAP} {Observation (\gls{metadata}) Table Access Protocol (part of IVOA ObsCore standard)}
\newglossaryentry{Operations} {name={Operations}, description={The 10-year period following construction and commissioning during which the LSST Observatory conducts its survey}}
\newacronym{PIFF} {PIFF} {PSFs In the Full FOV}
\newacronym{PNG} {PNG} {Portable Network Graphics}
\newacronym{POSIX} {POSIX} {Portable Operating System Interface}
\newacronym{PSF} {PSF} {Point Spread Function}
\newacronym{PSTN} {PSTN} {Project Science Technical Note}
\newacronym{PTC} {PTC} {Photon Transfer Curve}
\newacronym{Pan-STARRS} {Pan-STARRS} {Panoramic Survey Telescope and Rapid Response System}
\newglossaryentry{Project Manager} {name={Project Manager}, description={The person responsible for exercising leadership and oversight over the entire Rubin project; he or she controls schedule, budget, and all contingency funds}}
\newglossaryentry{Prompt Processing} {name={Prompt Processing}, description={The data processing which occurs at the Archive Center based on the stream of images coming from the telescope. This includes both Alert Production, which scans the image stream to identify and send alerts on transient and variable sources, and Solar System Processing, which identifies and characterizes objects in our solar system. It also includes specialized rapid calibration and Commissioning processing. Prompt Processing generates the Prompt Data Products.}}
\newacronym{Q1} {Q1} {Quarter one}
\newglossaryentry{Qserv} {name={Qserv}, description={LSST's distributed parallel database. This database system is used for collecting, storing, and serving LSST Data Release Catalogs and Project metadata, and is part of the Software Stack}}
\newacronym{RA} {RA} {Right Ascension}
\newacronym{REB} {REB} {Readout Electronics Board}
\newacronym{REST} {REST} {REpresentational State Transfer}
\newacronym{RINGSS} {RINGSS} {Ring-Image Next Generation Scintillation Sensor}
\newacronym{RMS} {RMS} {Root-Mean-Square}
\newacronym{RSP} {RSP} {Rubin Science Platform}
\newacronym{RTN} {RTN} {Rubin Technical Note}
\newglossaryentry{raft} {name={raft}, description={The sensors in the LSST camera are packaged into replaceable electronic assemblies, called rafts, consisting of 9 butted sensors (CCDs) in a 3x3 mosaic. Each raft is a replaceable unit in the LSST camera. There are 21 science rafts in the camera plus 4 additional corner rafts with specialized, non-science sensors, making for a total of 189 CCDs per focal plane image. The 21 science rafts are numbered from "0,1" through "0,3", "1,0" through "3,4", and "4,1" through "4,3". (In other words, the 25 combinations from "0,0" through "4,4" minus the four corners which are non-science.)}}
\newglossaryentry{Release} {name={Release}, description={Publication of a new version of a document, software, or data product. Depending on context, releases may require approval from Project- or DM-level change control boards, and then form part of the formal project baseline}}
\newglossaryentry{Rubin Operations} {name={Rubin Operations}, description={operations phase of Vera C. Rubin Observatory}}
\newacronym{S3} {S3} {(Amazon) Simple Storage Service}
\newacronym{S3DF} {S3DF} {SLAC Shared Scientific Data Facility}
\newacronym{SCP} {SCP} {South Celestial Pole}
\newacronym{SDSS} {SDSS} {Sloan Digital Sky Survey}
\newacronym{SLAC} {SLAC} {SLAC National Accelerator Laboratory}
\newglossaryentry{SLAC National Accelerator Laboratory} {name={SLAC National Accelerator Laboratory}, description={A national laboratory funded by the US Department of Energy (DOE); SLAC leads a consortium of DOE laboratories that has assumed responsibility for providing the LSST camera. Although the Camera project manages its own schedule and budget, including contingency, the Camera team’s schedule and requirements are integrated with the larger Project.  The camera effort is accountable to the LSSTPO.}}
\newacronym{SLR} {SLR} {Single Lens Reflex}
\newacronym{SMC} {SMC} {Small Magellanic Cloud}
\newacronym{SNR} {SNR} {Signal to Noise Ratio}
\newacronym{SOAR} {SOAR} {Southern Astrophysical Research Telescope}
\newacronym{SODA} {SODA} {Server-side Operations for Data Access (IVOA standard)}
\newacronym{SQL} {SQL} {Structured Query Language}
\newacronym{SQR} {SQR} {SQuARE document handle}
\newacronym{SRD} {SRD} {LSST Science Requirements; LPM-17}
\newacronym{STFC} {STFC} {UK Science and Technology Facilities Council}
\newglossaryentry{Science Collaboration} {name={Science Collaboration}, description={An autonomous body of scientists interested in a particular area of science enabled by the LSST dataset, which through precursor studies, simulations, and algorithm development lays the groundwork for the large-scale science projects the LSST will enable.  In addition to preparing their members to take full advantage of LSST early in its operations phase, the science collaborations have helped to define the system's science requirements, refine and promote the science case, and quality check design and development work}}
\newglossaryentry{Science Pipelines} {name={Science Pipelines}, description={The library of software components and the algorithms and processing pipelines assembled from them that are being developed by DM to generate science-ready data products from LSST images. The Pipelines may be executed at scale as part of LSST Prompt or Data Release processing, or pieces of them may be used in a standalone mode or executed through the Rubin Science Platform. The Science Pipelines are one component of the LSST Software Stack}}
\newglossaryentry{Science Platform} {name={Science Platform}, description={A set of integrated web applications and services deployed at the LSST Data Access Centers (DACs) through which the scientific community will access, visualize, and perform next-to-the-data analysis of the LSST data products}}
\newglossaryentry{Sensor} {name={Sensor}, description={A sensor is a generic term for a light-sensitive detector, such as a CCD. For LSST, sensors consist of a 2-D array of roughly 4K x 4K pixels, which are mounted on a raft in a 3x3 mosaic. Each sensor is divided into 16 channels or amplifiers. The 9 sensors that make up a raft are numbered from "0,0" through "2,2"}}
\newglossaryentry{Simonyi Survey Telescope} {name={Simonyi Survey Telescope}, description={The telescope at the Rubin Observatory that will perform the LSST (this refers to all physical components: the mirror, the mount assembly, etc.).}}
\newglossaryentry{Sloan Digital Sky Survey} {name={Sloan Digital Sky Survey}, description={is a digital survey of roughly 10,000 square degrees of sky around the north Galactic pole, plus a ~300 square degree stripe along the celestial equator}}
\newglossaryentry{Snap} {name={Snap}, description={One 15 second exposure within a Standard Visit in the LSST cadence}}
\newglossaryentry{Software Stack} {name={Software Stack}, description={Often referred to as the LSST Stack, or just The Stack, it is the collection of software written by the LSST Data Management Team to process, generate, and serve LSST images, transient alerts, and catalogs. The Stack includes the LSST Science Pipelines, as well as packages upon which the DM software depends. It is open source and publicly available}}
\newglossaryentry{Solar System Object} {name={Solar System Object}, description={A solar system object is an astrophysical object that is identified as part of the Solar System: planets and their satellites, asteroids, comets, etc. This class of object had historically been referred to within the LSST Project as Moving Objects}}
\newglossaryentry{Solar System Processing} {name={Solar System Processing}, description={A component of the Prompt Processing system, Solar System Processing identifies new SSObjects using unassociated DIASources.}}
\newglossaryentry{Source} {name={Source}, description={A single detection of an astrophysical object in an image, the characteristics for which are stored in the Source Catalog of the DRP database. The association of Sources that are non-moving lead to Objects; the association of moving Sources leads to Solar System Objects. (Note that in non-LSST usage "source" is often used for what LSST calls an Object.)}}
\newglossaryentry{Standard Visit} {name={Standard Visit}, description={A single observation of a LSST field comprised of two 15 second 'Snap' exposures that are immediately combined. An 'Alternate Standard Visit' is a single observation of a LSST field comprised of one 30 second exposure}}
\newglossaryentry{Subsystem} {name={Subsystem}, description={A set of elements comprising a system within the larger LSST system that is responsible for a key technical deliverable of the project}}
\newglossaryentry{Subsystem Manager} {name={Subsystem Manager}, description={responsible manager for an LSST subsystem; he or she exercises authority, within prescribed limits and under scrutiny of the Project Manager, over the relevant subsystem's cost, schedule, and work plans}}
\newglossaryentry{Summit} {name={Summit}, description={The site on the Cerro Pach\'{o}n, Chile mountaintop where the LSST observatory, support facilities, and infrastructure will be built}}
\newglossaryentry{Summit Facility} {name={Summit Facility}, description={The main Observatory and Auxiliary Telescope buildings at the Summit Site on Cerro Pach\'{o}n, Chile}}
\newglossaryentry{Systems Engineering} {name={Systems Engineering}, description={an interdisciplinary field of engineering that focuses on how to design and manage complex engineering systems over their life cycles. Issues such as requirements engineering, reliability, logistics, coordination of different teams, testing and evaluation, maintainability and many other disciplines necessary for successful system development, design, implementation, and ultimate decommission become more difficult when dealing with large or complex projects. Systems engineering deals with work-processes, optimization methods, and risk management tools in such projects. It overlaps technical and human-centered disciplines such as industrial engineering, control engineering, software engineering, organizational studies, and project management. Systems engineering ensures that all likely aspects of a project or system are considered, and integrated into a whole}}
\newacronym{TAP} {TAP} {Table Access Protocol (\gls{IVOA} standard)}
\newacronym{TOPCAT} {TOPCAT} {Tool for OPerations on Catalogues And Tables}
\newglossaryentry{Task} {name={Task}, description={Tasks are the basic unit of code re-use in the LSST Stack. They perform a well defined, logically contained piece of functionality. Tasks come standard with configuration, logging, processing metadata, and debugging features. For further details, see How to Write a Task in the source code documentation.  Tasks can be nested, providing a natural way to structure - and configure - high level algorithms that delegate work to lower-level algorithms}}
\newacronym{UA} {UA} {University of Arizona}
\newacronym{UK} {UK} {United Kingdom}
\newacronym{UKDF} {UKDF} {United Kingdom Data Facility}
\newacronym{URL} {URL} {Universal Resource Locator}
\newacronym{US} {US} {United States}
\newacronym{USDAC} {USDAC} {United States \gls{Data Access Center}}
\newacronym{USDF} {USDF} {United States Data Facility}
\newacronym{UT} {UT} {Universal Time}
\newacronym{UTC} {UTC} {Coordinated Universal Time}
\newacronym{VLT} {VLT} {Very Large Telescope (\gls{ESO})}
\newacronym{VO} {VO} {Virtual Observatory}
\newacronym{VST} {VST} {VLT Survey Telescope}
\newglossaryentry{Validation} {name={Validation}, description={A process of confirming that the delivered system will provide its desired functionality; overall, a validation process includes the evaluation, integration, and test activities carried out at the system level to ensure that the final developed system satisfies the intent and performance of that system in operations}}
\newglossaryentry{Visit} {name={Visit}, description={A sequence of one or more consecutive exposures at a given position, orientation, and filter within the LSST cadence. See Standard Visit, Alternate Standard Visit, and Non-Standard Visit}}
\newacronym{WAAS} {WAAS} {Wide Area Augmentation System}
\newacronym{WBS} {WBS} {\gls{Work Breakdown Structure}}
\newacronym{WCS} {WCS} {World Coordinate System}
\newacronym{WFD} {WFD} {Wide-Fast-Deep}
\newacronym{WebDav} {WebDav} {Web Distributed Authoring and Versioning}
\newglossaryentry{Work Breakdown Structure} {name={Work Breakdown Structure}, description={a tool that defines and organizes the LSST project's total work scope through the enumeration and grouping of the project's discrete work elements}}
\newglossaryentry{World Coordinate System} {name={World Coordinate System}, description={a mapping from image pixel coordinates to physical coordinates; in the case of images the mapping is to sky coordinates, generally in an equatorial (RA, Dec) system. The WCS is expressed in FITS file extensions as a collection of header keyword=value pairs (basically, the values of parameters for a selected functional representation of the mapping) that are specified in the FITS Standard}}
\newacronym{XP} {XP} {B or R Photometry (Gaia)}
\newglossaryentry{airmass} {name={airmass}, description={The pathlength of light from an astrophysical source through the Earth's atmosphere. It is given approximately by sec z, where z is the angular distance from the zenith (the point directly overhead, where airmass = 1.0) to the source}}
\newglossaryentry{algorithm} {name={algorithm}, description={A computational implementation of a calculation or some method of processing}}
\newglossaryentry{arcmin} {name={arcmin}, description={arcminute minute of arc (unit of angle)}}
\newglossaryentry{arcsec} {name={arcsec}, description={arcsecond second of arc (unit of angle)}}
\newglossaryentry{astrometry} {name={astrometry}, description={In astronomy, the sub-discipline of astrometry concerns precision measurement of positions (at a reference epoch), and real and apparent motions of astrophysical objects. Real motion means 3-D motions of the object with respect to an inertial reference frame; apparent motions are an artifact of the motion of the Earth. Astrometry per se is sometimes confused with the act of determining a World Coordinate System (WCS), which is a functional characterization of the mapping from pixels in an image or spectrum to world coordinate such as (RA, Dec) or wavelength}}
\newglossaryentry{astronomical object} {name={astronomical object}, description={A star, galaxy, asteroid, or other physical object of astronomical interest. Beware: in non-LSST usage, these are often known as sources}}
\newacronym{au} {au} {astronomical unit}
\newglossaryentry{background} {name={background}, description={In an image, the background consists of contributions from the sky (e.g., clouds or scattered moonlight), and from the telescope and camera optics, which must be distinguished from the astrophysical background. The sky and instrumental backgrounds are characterized and removed by the LSST processing software using a low-order spatial function whose coefficients are recorded in the image metadata}}
\newglossaryentry{brighter-fatter effect} {name={brighter-fatter effect}, description={The common term used to refer to one of the photometric qualities of the LSST camera: sources with a higher flux have a broader PSF. This is accounted for during calibration}}
\newglossaryentry{cadence} {name={cadence}, description={The sequence of pointings, visit exposures, and exposure durations performed over the course of a survey}}
\newglossaryentry{calibration} {name={calibration}, description={The process of translating signals produced by a measuring instrument such as a telescope and camera into physical units such as flux, which are used for scientific analysis. Calibration removes most of the contributions to the signal from environmental and instrumental factors, such that only the astronomical component remains}}
\newglossaryentry{cloud} {name={cloud}, description={A visible mass of condensed water vapor floating in the atmosphere, typically high above the ground or in interstellar space acting as the birthplace for stars.  Also a way of computing (on other peoples computers leveraging their services and availability).}}
\newglossaryentry{configuration} {name={configuration}, description={A task-specific set of configuration parameters, also called a 'config'. The config is read-only; once a task is constructed, the same configuration will be used to process all data. This makes the data processing more predictable: it does not depend on the order in which items of data are processed. This is distinct from arguments or options, which are allowed to vary from one task invocation to the next}}
\newglossaryentry{deblend} {name={deblend}, description={Deblending is the act of inferring the intensity profiles of two or more overlapping sources from a single footprint within an image. Source footprints may overlap in crowded fields, or where the astrophysical phenomena intrinsically overlap (e.g., a supernova embedded in an external galaxy), or by spatial co-incidence (e.g., an asteroid passing in front of a star). Deblending may make use of a priori information from images (e.g., deep CoAdds or visit images obtained in good seeing), from catalogs, or from models. A 'deblend' is commonly referred to in terms of 'parent' (total) and 'child' (component) objects}}
\newglossaryentry{deepCoadd} {name={deepCoadd}, description={A Coadd Image designed to produce detections as maximum depth. Produced by AssembleCoaddTask}}
\newglossaryentry{deg} {name={deg}, description={degree; unit of angle}}
\newglossaryentry{element} {name={element}, description={A node in the hierarchical project WBS}}
\newglossaryentry{epoch} {name={epoch}, description={Sky coordinate reference frame, e.g., J2000. Alternatively refers to a single observation (usually photometric, can be multi-band) of a variable source}}
\newglossaryentry{flux} {name={flux}, description={Shorthand for radiative flux, it is a measure of the transport of radiant energy per unit area per unit time. In astronomy this is usually expressed in cgs units: erg/cm2/s}}
\newglossaryentry{footprint} {name={footprint}, description={See 'source footprint', 'instrumental footprint', or 'survey footprint', `Footprint` is a Python class representing a source footprint}}
\newglossaryentry{forced photometry} {name={forced photometry}, description={A measurement of the photometric properties of a source, or expected source, with one or more parameters held fixed. Most often this means fixing the location of the center of the brightness profile (which may be known or predicted in advance), and measuring other properties such as total brightness, shape, and orientation. Forced photometry will be done for all Objects in the Data Release Production}}
\newglossaryentry{instrumental footprint} {name={instrumental footprint}, description={The size and shape of a region on the sky that is covered by the field of view of an instrument, or part of an instrument, e.g., the LSST Camera, or ComCam, or a single LSST CCD.  Often represented by a geometric region defined in field-angle space}}
\newglossaryentry{interoperability} {name={interoperability}, description={the ability of systems or software to exchange and make use of information between them.}}
\newglossaryentry{metadata} {name={metadata}, description={General term for data about data, e.g., attributes of astronomical objects (e.g. images, sources, astroObjects, etc.) that are characteristics of the objects themselves, and facilitate the organization, preservation, and query of data sets. (E.g., a FITS header contains metadata)}}
\newglossaryentry{middleware} {name={middleware}, description={Software that acts as a bridge between other systems or software usually a database or network. Specifically in the Data Management System this refers to Butler for data access and Workflow management for distributed processing.}}
\newglossaryentry{passband} {name={passband}, description={The window of wavelength or the energy range admitted by an optical system; specifically the transmission as a function of wavelength or energy. Typically the passband is limited by a filter. The width of the passband may be characterized in a variety of ways, including the width of the half-power points of the transmission curve, or by the equivalent width of a filter with 100\% transmission within the passband, and zero elsewhere}}
\newglossaryentry{patch} {name={patch}, description={An quadrilateral sub-region of a sky tract, with a size in pixels chosen to fit easily into memory on desktop computers}}
\newglossaryentry{pipeline} {name={pipeline}, description={A configured sequence of software tasks (Stages) to process data and generate data products. Example: Association Pipeline}}
\newglossaryentry{provenance} {name={provenance}, description={Information about how LSST images, Sources, and Objects were created (e.g., versions of pipelines, algorithmic components, or templates) and how to recreate them}}
\newglossaryentry{schema} {name={schema}, description={The definition of the metadata and linkages between datasets and metadata entities in a collection of data or archive.}}
\newglossaryentry{seeing} {name={seeing}, description={An astronomical term for characterizing the stability of the atmosphere, as measured by the width of the point-spread function on images. The PSF width is also affected by a number of other factors, including the airmass, passband, and the telescope and camera optics}}
\newglossaryentry{shape} {name={shape}, description={In reference to a Source or Object, the shape is a functional characterization of its spatial intensity distribution, and the integral of the shape is the flux. Shape characterizations are a data product in the DIASource, DIAObject, Source, and Object catalogs}}
\newglossaryentry{sky map} {name={sky map}, description={A sky tessellation for LSST. The Stack includes software to define a geometric mapping from the representation of World Coordinates in input images to the LSST sky map. This tessellation is comprised of individual tracts which are, in turn, comprised of patches}}
\newglossaryentry{software} {name={software}, description={The programs and other operating information used by a computer.}}
\newglossaryentry{source footprint} {name={source footprint}, description={A set of pixels that are determined to be part of a Source (or DIASource). It is implemented as a list of spans. A span contains coordinates of a stripe of pixels: row (y) given span belongs to, and a section of a column (xStart, xEnd). In DM code, the term 'footprint' refers to a 'source footprint'}}
\newglossaryentry{survey footprint} {name={survey footprint}, description={The portion of the sky covered by data from an astronomical survey, e.g., the main wide-fast-deep LSST 10-year survey, the LSST deep drilling fields, or the Science Validation data taken during commissioning.  Sometimes represented by Boolean maps or other summary statistics in an all-sky representation, e.g., the IVOA MOC standard}}
\newglossaryentry{tracklet} {name={tracklet}, description={Links between unassociated DIASources within one night to identify moving objects}}
\newglossaryentry{tract} {name={tract}, description={A portion of sky, a spherical convex polygon, within the LSST all-sky tessellation (sky map). Each tract is subdivided into sky patches}}
\newglossaryentry{transient} {name={transient}, description={A transient source is one that has been detected on a difference image, but has not been associated with either an astronomical object or a solar system body}}
 \makeglossaries

\newcommand{\campaignstartdate}{2024-10-24\xspace}

\newcommand{\nnightscomcam}{48\xspace}
\newcommand{\sizeinbytes}{3.5\xspace TB\xspace}
\newcommand{\tenyearcatalogsize}{15\xspace PB\xspace}

\newcommand{\totalarea}{$\sim$15\xspace deg$^2$\xspace}

\newcommand{\nnewasteroiddiscoveries}{93\xspace}
\newcommand{\nexposuresaoscommissioning}{10000\xspace}
\newcommand{\nexposurescalibcommissioning}{2000\xspace}
\newcommand{\nexposuresspcommissioning}{2000\xspace}
\newcommand{\sciencepipelinesversion}{v29.1\xspace}
\newcommand{\sciencepipelinesurl}{\nolinkurl{https://pipelines.lsst.io/v/v29\_1\_1}\xspace}
\newcommand{\maxdepthshalloweranyband}{2.2\xspace magnitudes\xspace}
\newcommand{\nvisits}{1792\xspace}
\newcommand{\nexposures}{1792\xspace}
\newcommand{\nfields}{seven\xspace}

\newcommand{\dponeenddate}{2024-12-11\xspace}
\newcommand{\exposuretime}{30\xspace s\xspace}

\newcommand{\bestimagequality}{0\farcs58\xspace}
\newcommand{\medianimagequalityallbands}{1\farcs14\xspace}

\newcommand{\nraws}{16125\xspace}
\newcommand{\rawhdd}{18\xspace MB\xspace}
\newcommand{\nrawpixx}{4608\xspace}
\newcommand{\nrawpixy}{4096\xspace}
\newcommand{\rawplatescale}{0\farcs2 per pixel\xspace}

\newcommand{\rawfov}{0.058\xspace deg$^2$\xspace}
\newcommand{\nvisitimages}{15972\xspace}
\newcommand{\visitimagehdd}{110\xspace MB\xspace}
\newcommand{\nvisitimagepixx}{4072\xspace}
\newcommand{\nvisitimagepixy}{4000\xspace}

\newcommand{\visitimagefovx}{0\fdg23\xspace}
\newcommand{\visitimagefovy}{0\fdg22\xspace}
\newcommand{\visitimagefov}{0.051\xspace deg$^2$\xspace}
\newcommand{\ndeepcoadds}{2644\xspace}

\newcommand{\ndeepcoaddpixx}{3400\xspace}
\newcommand{\ndeepcoaddpixy}{3400\xspace}

\newcommand{\ntemplatecoadds}{2730\xspace}

\newcommand{\ndifferenceimages}{15972\xspace}

\newcommand{\ntotaltracts}{18938\xspace}

\newcommand{\tractarea}{2.8\xspace deg$^2$\xspace}
\newcommand{\tractoverlap}{1\farcm0\xspace}
\newcommand{\npatchx}{10\xspace}
\newcommand{\npatchy}{10\xspace}

\newcommand{\innerpatcharea}{0.028\xspace deg$^2$\xspace}

\newcommand{\outerpatcharea}{0.036\xspace deg$^2$\xspace}
\newcommand{\patchoverlap}{80\farcs0\xspace}
\newcommand{\deepcoaddmaxfwhm}{1\farcs7\xspace}
\newcommand{\nsurveypropertymaps}{14\xspace}

\newcommand{\nvisitsummaries}{1786\xspace}
\newcommand{\nvisitdetectorsummaries}{16071\xspace}
\newcommand{\nsolarsystemsources}{5988\xspace}
\newcommand{\nsolarsystemobjects}{431\xspace}
\newcommand{\nsfpfails}{153\xspace}
\newcommand{\nobjects}{2.3\xspace million\xspace}
\newcommand{\nsources}{46\xspace million\xspace}
\newcommand{\ndiaobjects}{1.1\xspace million\xspace}
\newcommand{\ndiasources}{3.1\xspace million\xspace}
\newcommand{\nforcedsources}{269\xspace million\xspace}
\newcommand{\nforcedobjects}{2.3\xspace million\xspace}
\newcommand{\ndiaforcedsources}{197\xspace million\xspace}
\newcommand{\ndiaforcedobjects}{1.1\xspace million\xspace}
\newcommand{\nssobjects}{431\xspace}
\newcommand{\nextendedobjects}{1.6\xspace million\xspace}
\newcommand{\ndeepcoaddvisitimages}{15375\xspace}
\newcommand{\ntemplatecoaddvisitimages}{13113\xspace}
\newcommand{\udepth}{24.55\xspace}
\newcommand{\gdepth}{26.18\xspace}
\newcommand{\rdepth}{25.96\xspace}
\newcommand{\idepth}{25.71\xspace}
\newcommand{\zdepth}{25.07\xspace}
\newcommand{\ydepth}{23.1\xspace}
 
\begin{document}
\setcounter{footnote}{0}
\doi{10.71929/rubin/2570536}
\date{\today}
\title{The Vera C. Rubin Observatory Data Preview 1}
\shorttitle{Rubin DP1}

\hypersetup{
    pdftitle={The Vera C. Rubin Observatory Data Preview 1},
    pdfauthor={},
    pdfkeywords={}
}

\author[sname='Vera C. Rubin Observatory Team']{Vera~C.~Rubin~Observatory~Team}
\affiliation{Vera C.\ Rubin Observatory Project Office, 950 N.\ Cherry Ave., Tucson, AZ  85719, USA}
\email{pubboard@lists.lsst.org}

\author[0000-0002-5947-2454,gname='Tatiana',sname='Acero-Cuellar']{Tatiana~Acero-Cuellar}
\affiliation{Department of Physics and Astronomy, University of Delaware, Newark, DE 19716-2570, USA}
\email{taceroc@udel.edu}

\author[0009-0006-1601-3246,gname='Emily',sname='Acosta']{Emily~Acosta}
\affiliation{Vera C.\ Rubin Observatory Project Office, 950 N.\ Cherry Ave., Tucson, AZ  85719, USA}
\email{eacosta@lsst.org}

\author[0009-0008-8623-871X,gname='Christina L.',sname='Adair']{Christina~L.~Adair}
\affiliation{SLAC National Accelerator Laboratory, 2575 Sand Hill Rd., Menlo Park, CA 94025, USA}
\email{cadair@slac.stanford.edu}

\author[0000-0001-9431-3806,gname='Prakruth',sname='Adari']{Prakruth~Adari}
\affiliation{Department of Physics and Astronomy, Stony Brook University, Stony Brook, NY 11794, USA}
\email{prakruth.adari@stonybrook.edu}

\author[0000-0002-4100-8928,gname='Jennifer K.',sname='Adelman-McCarthy']{Jennifer~K.~Adelman-McCarthy}
\affiliation{Fermi National Accelerator Laboratory, P. O. Box 500, Batavia, IL 60510, USA}
\email{jen_a@fnal.gov}

\author[0009-0000-7835-3963,gname='Anastasia',sname='Alexov']{Anastasia~Alexov}
\affiliation{Vera C.\ Rubin Observatory Project Office, 950 N.\ Cherry Ave., Tucson, AZ  85719, USA}
\email{aalexov@lsst.org}

\author[0009-0009-9491-8923,gname='Russ',sname='Allbery']{Russ~Allbery}
\affiliation{Vera C.\ Rubin Observatory Project Office, 950 N.\ Cherry Ave., Tucson, AZ  85719, USA}
\email{rra@lsst.org}

\author[gname='Robyn',sname='Allsman']{Robyn~Allsman}
\affiliation{Vera C.\ Rubin Observatory Project Office, 950 N.\ Cherry Ave., Tucson, AZ  85719, USA}
\email{robynallsman@gmail.com}

\author[0009-0008-9216-7516,gname='Yusra',sname='AlSayyad']{Yusra~AlSayyad}
\affiliation{Department of Astrophysical Sciences, Princeton University, Princeton, NJ 08544, USA}
\email{yusra@astro.princeton.edu}

% ...

\begin{abstract}
We present Rubin Data Preview 1 (DP1), the first data from the NSF-DOE Vera C. Rubin Observatory, comprising raw and calibrated single-epoch images, coadds, difference images, detection catalogs, and ancillary data products.
DP1 is based on \nexposures optical/near-infrared exposures acquired over \nnightscomcam distinct nights by the Rubin Commissioning Camera, LSSTComCam, on the Simonyi Survey Telescope at the Summit Facility on Cerro Pach\'on, Chile in late 2024.
DP1 covers \totalarea distributed across \nfields roughly equal-sized non-contiguous fields, each independently observed in six broad photometric bands, $ugrizy$.
The median FWHM of the point-spread function across all bands is approximately \medianimagequalityallbands, with the sharpest images reaching about \bestimagequality.
The 5$\sigma$  point source  depths for coadded  images in the deepest field, the Extended Chandra Deep Field South, are: $u = \udepth,  g = \gdepth, r = \rdepth, i = \idepth, z = \zdepth, y = \ydepth$.
Other fields are no more than \maxdepthshalloweranyband shallower in any band, where they have nonzero coverage.
DP1 contains approximately \nobjects distinct astrophysical objects, of which \nextendedobjects are extended in at least one band in coadds, and \nsolarsystemobjects solar system objects, of which \nnewasteroiddiscoveries are new discoveries.
DP1 is approximately \sizeinbytes in size and is available to Rubin data rights holders via the Rubin Science Platform, a cloud-based environment for the analysis of petascale astronomical data.
While small compared to future LSST releases, its high quality and diversity of data support a broad range of early science investigations ahead of full operations in 2026.
\end{abstract} 
\section{Introduction}
\label{sec:intro}
The \gls{NSF}–\gls{DOE} Vera C. Rubin Observatory is a ground-based, wide-field optical/near-infrared facility located on Cerro Pach\'on in northern Chile.
Named in honor of Vera C. Rubin, a pioneering astronomer whose groundbreaking work in the 20th century provided the first convincing evidence for the existence of dark matter \citep[][]{1970ApJ...159..379R, 1980ApJ...238..471R}, the observatory’s prime mission is to carry out the \gls{LSST} \citep{2019ApJ...873..111I}.
This 10-year survey is designed to obtain rapid-cadence, multi-band imaging of the entire visible southern sky approximately every 3–4 nights. 
Over its main 18,000 deg$^2$ footprint, the LSST is expected to reach a depth of $\sim$27 magnitude in the r-band, with $\sim$800 visits per pointing in all filters \citep[][]{2022ApJS..258....1B}.

The Rubin Observatory system consists of four main components: the \gls{Simonyi Survey Telescope}, featuring an 8.4 m diameter (6.5 m effective aperture) primary mirror that delivers a wide field of view; the 3.2-gigapixel \gls{LSSTCam}, capable of imaging 9.6 square degrees per exposure
\footnote{We define an ``exposure" as the process of exposing all  detectors in the focal plane. It is synonymous with the term ``visit" in 	DP1. By contrast, an ``image" is the output of a single detector following an exposure.} 
with seeing-limited quality in six broadband filters, \textit{ugrizy} (320--1050 nm); an automated \gls{Data Management System} that processes and archives tens of terabytes of data per night, generating science-ready data products within minutes for a global community of scientists; and an \gls{EPO} program that provides real-time data access, interactive tools, and educational content to engage the public.
The integrated system's \'etendue\footnote{The product of the primary mirror area and the angular area of its field of view for a given set of observing conditions.} of 319\,$\text{m}^2\,\text{deg}^2$, is over an order of magnitude larger than that of any previous optical observatory, enabling a fast, large-scale survey with exceptional depth in a fraction of the time compared to other observatories.

The observatory's design is driven by four key science themes: probing dark energy and dark matter; taking an inventory of the solar system; exploring the transient and variable optical sky; and mapping the Milky Way \citep{2019ApJ...873..111I}.
These themes inform the optimization of a range of system parameters, including image quality; photometric and astrometric accuracy; single-visit depth; coadded survey depth; the filter complement; the total number of visits per pointing and their distribution on the sky; and total sky coverage.
Additionally, they inform the design of the data processing and access systems.
By optimizing the system parameters to support a wide range of scientific goals, we maximize the observatory's scientific output across all areas, making Rubin a powerful discovery machine capable of addressing a broad range of astrophysical questions.

Throughout the duration of the \gls{LSST},  Rubin Observatory will issue a series of Data Releases, each representing a complete reprocessing of all \gls{LSST} data collected up to that point.
Prior to the start of the \gls{LSST} survey, commissioning activities generated a significant volume of science-grade data.
To make this early data available to the community, the Rubin Early Science Program \citep{RTN-011} was established.
One key component of this program is a series of Data Previews; early versions of the \gls{LSST} Data Releases.
These previews include preliminary data products derived from both simulated and commissioning data, which, together with early versions of the data access services, are intended to support high-impact early science, facilitate community readiness, and inform the development of Rubin’s operational capabilities ahead of the start of full survey operations.
All data and services provided through the Rubin Early Science Program are offered on a shared-risk basis\footnote{Shared risk means early access with caveats: the community benefits from getting a head start on science, preparing analyses, and providing feedback, while also accepting that the system may not work as well as it will during full operations.}.

This paper describes Rubin's second of three planned Data Previews: \gls{DP1} \citep[][]{10.71929/rubin/2570308}.
The first, \gls{DP0}, contained data products produced from the processing of simulated \gls{LSST}-like data sets.
These were released together with a very early version of the \gls{RSP} \citep{LSE-319}, which provided the data access services.
\gls{DP0} was released in multiple phases; DP0.1, DP0.2, and DP0.3,  each building upon the previous and incorporating new data and functionalities.
DP0.1 and DP0.2 uses data from the cosmoDC2 simulations \citep[][]{2021ApJS..253...31L} prepared by the Dark Energy Science Collaboration (DESC), whereas DP0.3 is based on simulated datasets from the Solar System Science Collaboration (SSSC).
Online documentation for \gls{DP0} is available at \url{https://dp0.lsst.io}.
 
\gls{DP1} contains data products derived from the reprocessing of science-grade exposures acquired by the \gls{LSSTComCam} in late 2024.
The third and final Data Preview, \gls{DP2}, is planned to be based on a reprocessing of all science-grade data taken with Rubin's \gls{LSSTCam} during commissioning.

All Rubin Data Releases and Previews are subject to a two-year proprietary period, with immediate access granted exclusively to LSST data rights holders\footnote{Individuals or institutions with formal authorization to access proprietary data collected by the Vera C. Rubin Observatory. See \nolinkurl{https://www.lsst.org/scientists/international-drh-list}}\citep[][]{rdo-013}.
After the two-year proprietary period, \gls{DP1} will be made public. 
However, even once the data become public, access for individuals without data rights will not be provided through Rubin Data Access Centers  in the US and Chile \citep[][]{rdo-013}. 
Alternative access mechanisms are still under discussion and have not yet been finalized.

In this paper, we describe the contents and validation of Rubin \gls{DP1}, the first Data Preview to deliver data derived from observations conducted by the Vera C. Rubin Observatory, as well as the data-access mechanisms and community-support services that accompany it. 
\gls{DP1} is based on the reprocessing of \nexposures science-grade exposures acquired during the first on-sky commissioning campaign, conducted over \nnightscomcam nights between \campaignstartdate and \dponeenddate.
It covers a  total area of approximately \totalarea distributed across \nfields distinct non-contiguous fields.
The data products include raw and calibrated single-\gls{epoch} images, coadded images, difference images, detection catalogs, and other derived data products.
\gls{DP1} is about \sizeinbytes in size and contains around \nobjects distinct astronomical objects, detected in \ndeepcoadds coadded images.
Full \gls{DP1} release documentation is available at \url{https://dp1.lsst.io}.
Despite Rubin Observatory still being in commissioning and not yet complete at the time the observations were acquired, Rubin \gls{DP1} provides an important first look at the data, showcasing its characteristics and capabilities.

The structure of this paper is as follows.
In section \ref{sec:on_sky_campaign} we describe the observatory system and overall construction and commissioning status at the time of data acquisition, the \nfields fields included in \gls{DP1}, and the observing strategy used.
Section \ref{sec:data_products} summarizes the contents of \gls{DP1} and the data products contained in the release.
The data processing pipelines are described in section \ref{sec:drp}, followed by a description of the data validation and performance assessment in section \ref{sec:performance}.
Section \ref{sec:data_services} describes the \gls{RSP}, a \gls{cloud}-based data science infrastructure that provides tools and services to Rubin data rights holders to access, visualize and analyze peta-scale data generated by the \gls{LSST}.
Section \ref{sec:community_science} presents Rubin Observatory's model for community support, which emphasizes self-help via documentation and tutorials, and employs an open platform for issue reporting that enables crowd-sourced solutions.
Finally, a summary of the \gls{DP1} release and information on expected future releases of data is given in section \ref{sec:summary}.
The appendix contains a useful glossary of terms used throughout this paper.

All magnitudes quoted are in the AB system \citep{1983ApJ...266..713O}, unless otherwise specified. \section{On-Sky Commissioning Campaign}
\label{sec:on_sky_campaign}
The primary objective of the first Rubin on-sky commissioning campaign was to optically align the Simonyi Survey Telescope and verify its ability to deliver acceptable image quality using the Commissioning Camera, \gls{LSSTComCam}.
Additionally, the campaign provided valuable operational experience to support commissioning the LSST Science Camera, \gls{LSSTCam} \citep{2024SPIE13096E..1OL,2024SPIE13096E..1SR}.
We note that commissioning \gls{LSSTComCam} was not an objective of the campaign; rather \gls{LSSTComCam} was used as a tool to support broader observatory commissioning, including early testing of the \gls{AOS} (\secref{ssec:simonyi}) and the LSST Science Pipelines (\secref{ssec:pipelines_commissioning}).
As a result, many artifacts present in the data are specific to \gls{LSSTComCam} and will be addressed only if they persist with \gls{LSSTCam}.
Accordingly, the image quality achieved during this campaign, and in the \gls{DP1} data, may not reflect the performance ultimately expected from \gls{LSSTCam}.

Approximately 16,000 exposures\footnote{We define an exposure as the process of exposing all \gls{LSSTComCam} detectors. It is synonymous with visit in \gls{DP1}. By contrast, an image is the output of a single \gls{LSSTComCam} detector following an exposure.} were collected during this campaign, the majority in support of \gls{AOS} commissioning, system-level verification, and end-to-end testing of the telescope’s hardware and software.
This included over \nexposuresaoscommissioning exposures for \gls{AOS} commissioning, more than \nexposurescalibcommissioning bias and dark calibration frames, and over \nexposuresspcommissioning exposures dedicated to commissioning the LSST Science Pipelines.
For \gls{DP1}, we have selected a subset of \nexposures science-grade exposures from this campaign that are most useful for the community to begin preparing for early science.

At the time of the campaign, the observatory was still under construction, with several key components, such as dome thermal control, full mirror control, and the final \gls{AOS} configuration either incomplete or still undergoing commissioning.
As a result, image quality varied widely throughout the campaign and exhibited a broader distribution than is expected with \gls{LSSTCam}.
Despite these limitations, the campaign successfully demonstrated system integration and established a functional observatory.

\subsection{Simonyi Survey Telescope}
\label{ssec:simonyi}
The Simonyi Survey Telescope \citep{2024SPIE13094E..09S} features a unique three-mirror design, including an 8.4-meter \gls{M1M3} fabricated from a single substrate and a 3.5-meter \gls{M2}.
This compact \gls{configuration} supports a wide 3.5-degree field of view while enabling exceptional stability, allowing the telescope to slew and settle in under five seconds.
To achieve the scientific goals of the 10-year \gls{LSST}, the Observatory must maintain high image quality across its wide field of view \citep{2008arXiv0805.2366I}.
This is accomplished through the \gls{AOS} \citep{2015ApOpt..54.9045X,MegiasHomar_2024}, which corrects, between successive exposures, wavefront distortions caused by optical misalignments and mirror surface deformations, primarily due to the effect of gravitational and thermal loads.

The \gls{AOS}, which comprises an open-loop component and a closed-loop component, optimizes image quality by aligning the camera and \gls{M2} relative to \gls{M1M3}, as well as adjusting the shapes of all three mirrors to nanometer precision.
The \gls{AOS} open-loop component corrects for predictable distortions and misalignments, while the closed-loop component addresses unpredictable or slowly varying aberrations using feedback from the corner wavefront sensors.
The closed-loop wavefront sensing technique is curvature wavefront sensing, which infers wavefront errors in the optical system by analyzing extra- and intra-focal star images \citep{2023aoel.confE..67T}.
Since \gls{LSSTComCam} lacks dedicated wavefront sensors, wavefront errors were instead estimated by defocusing the telescope $\pm$1.5 mm on either side of focus and applying the curvature wavefront sensing pipeline to the resulting images.
Each night began with an initial alignment correction using a laser tracker to position the system within the capture range of the closed-loop \gls{algorithm} \citep{10.1117/12.3019031}.
Once this coarse alignment was complete, the \gls{AOS} refined the optical alignment and applied mirror surfaces corrections to optimize the image quality across the \gls{LSSTComCam} field of view.

During LSST \gls{Science Pipelines} commissioning (\secref{ssec:pipelines_commissioning}),  observations were conducted using the AOS in open-loop mode only, without closed-loop corrections between exposures. 
Closed-loop operation, which requires additional intra- and extra-focal images with LSSTComCam, was not compatible with the continuous data acquisition needed by the pipelines.
The image quality for these data was monitored by measuring the \gls{PSF} at \gls{FWHM}, and closed-loop sequences were periodically run when image quality degradation was observed.

\subsection{The LSST Commissioning Camera}
\label{ssec:comcam}
\gls{LSSTComCam} \citep{2022SPIE12184E..0JS,2020SPIE11447E..0LS,2018SPIE10700E..3DH, 10.71929/rubin/2561361} is a 144-megapixel version of the 3.2-gigapixel \gls{LSSTCam}.
It covers approximately 5\% of the \gls{LSSTCam} focal plane area, with a field of view of $\sim$0.5\,deg$^2$   (40\arcmin$\times$40\arcmin), compared to LSSTCam's 9.6\,deg$^2$. 
It was developed to validate camera interfaces with other observatory components and evaluate overall system performance prior to the start of \gls{LSSTCam} commissioning. 
Although \gls{LSSTComCam} has a smaller imaging area, it shares the same plate scale of \rawplatescale and is housed in a support structure that replicates the mass, center of gravity, and physical dimensions of \gls{LSSTCam}.
All mechanical and utility interfaces to the telescope are implemented identically, enabling full end-to-end testing of observatory systems, including readout electronics, image acquisition, and data pipelines.
Although the \gls{LSSTComCam} cryostat employs a different cooling system (Cryotels) to that of \gls{LSSTCam}, it included a refrigeration pathfinder to validate the cryogenic system intended for \gls{LSSTCam}.

\begin{figure}[htb!]
\includegraphics[width=\linewidth]{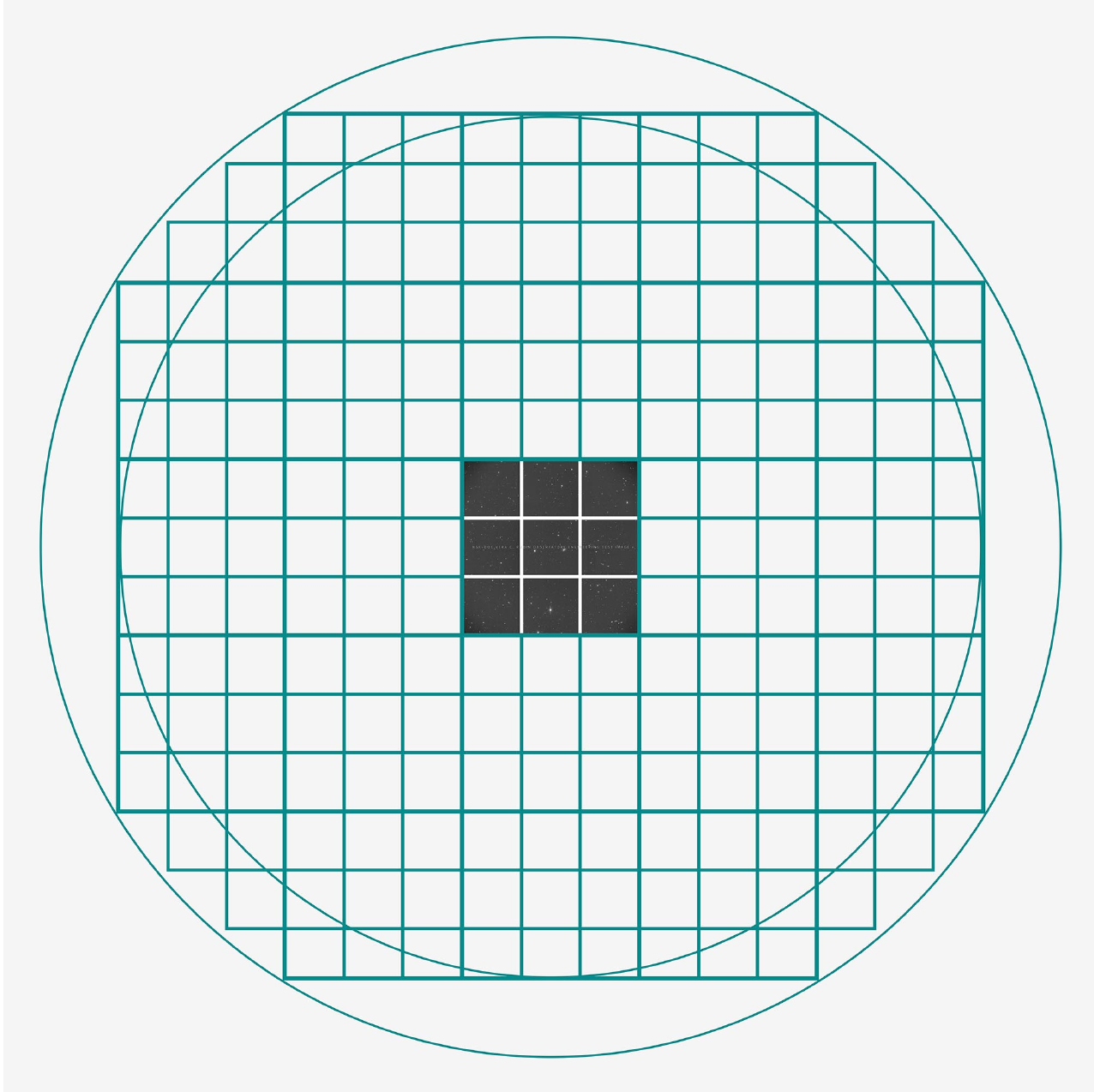}
\caption{LSSTComCam focal plane layout illustrating the placement of its nine sensors, shown in gray, which form a raft. The view is looking down from above the focal plane through the \gls{LSSTComCam} lenses. \gls{LSSTComCam} is Raft 22 (R22). We also indicate the location of the LSSTCam sensors (open squares) to highlight the field-of-view of LSSTComCam in relation to that of LSSTCam.}
\label{fig:comcam_raft_in_lsstcam_focal_plane}
\end{figure}

The \gls{LSSTCam} focal plane comprises 25 modular rafts arranged in a 5$\times$5 grid, of which 21 are science rafts dedicated to imaging and 4 are corner rafts used for guiding and wavefront sensing.
\gls{LSSTCam} employs CCD sensors from two vendors: \gls{ITL} and \gls{E2V}.
In contrast, \gls{LSSTComCam} contains only a single science raft equipped exclusively with \gls{ITL} sensors.
\figref{fig:comcam_raft_in_lsstcam_focal_plane} presents a schematic of the \gls{LSSTCam} focal plane, with the \gls{LSSTComCam} raft positioned at the center, corresponding to the \gls{LSSTCam} central science raft location.
The perspective is from above, looking down through the \gls{LSSTComCam} lenses onto the focal plane. 

Each science raft is a self-contained unit comprising nine 4K$\times$4K \gls{CCD} \citep{RevModPhys.82.2307} sensors arranged in a 3$\times$3 mosaic, complete with integrated readout electronics and cooling systems.
Each sensor is subdivided into 16 segments arranged in a 2$\times$8 layout, with each segment consisting of 512$\times$2048 pixels and read out in parallel using individual amplifiers.
This design is identical across all science rafts. 
To maintain uniform performance and \gls{calibration}, each raft is populated exclusively with sensors from a single vendor.

\begin{figure}[htb!]
\plotone{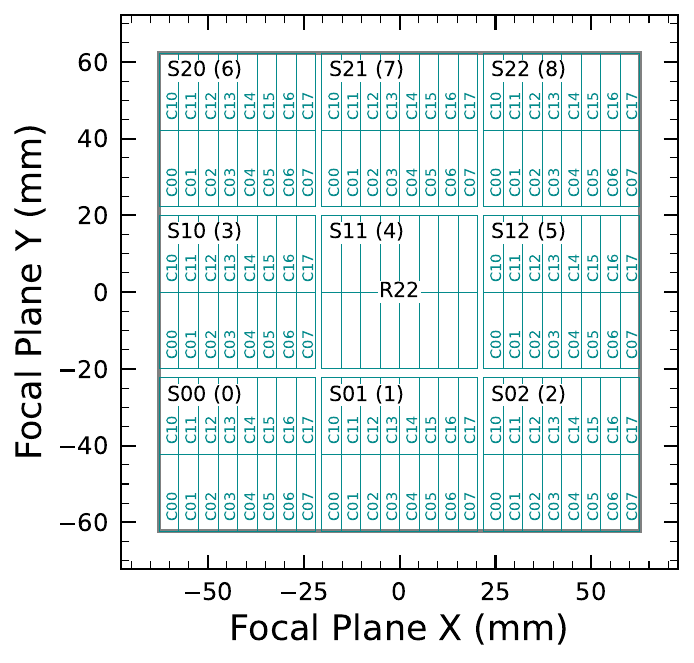}
\caption{\gls{LSSTComCam} focal plane layout, showing Raft 22 (R22) and the placement and numbering scheme of sensors (S) and amplifiers (C). The view is from above, looking through the \gls{LSSTComCam} lenses onto the focal plane. Each sensor contains 16 amplifiers, and the raft is composed of a 3$\times$3 array of sensors. The detector number for each sensor is indicated in parentheses.}
\label{fig:raft_schematic}
\end{figure}

\gls{LSSTComCam} consists of a single science raft, designated Raft 22 (R22), equipped solely with \gls{ITL} sensors.
These sensors were selected from the best-performing  remaining \gls{ITL} devices after the \gls{LSSTCam} rafts were fully populated. 
Some exhibit known issues such as high readout noise (e.g., Detector 8) and elevated \gls{CTI} (e.g., Detector 5).
Consequently, certain image artifacts present in the \gls{DP1} dataset may be specific to \gls{LSSTComCam}.
\figref{fig:raft_schematic} shows the \gls{LSSTComCam} R22 focal plane layout  and the placement and numbering scheme of sensors (S) and amplifiers (C). 
This configuration is identical across all science rafts in \gls{LSSTCam}.
The \gls{LSSTCam} and \gls{LSSTComCam} focal planes are described in detail in \cite{CTN-001}.

\subsubsection{Filter Complement}
\label{sssec:comcam_filters}
\begin{figure}[htb!]
\plotone{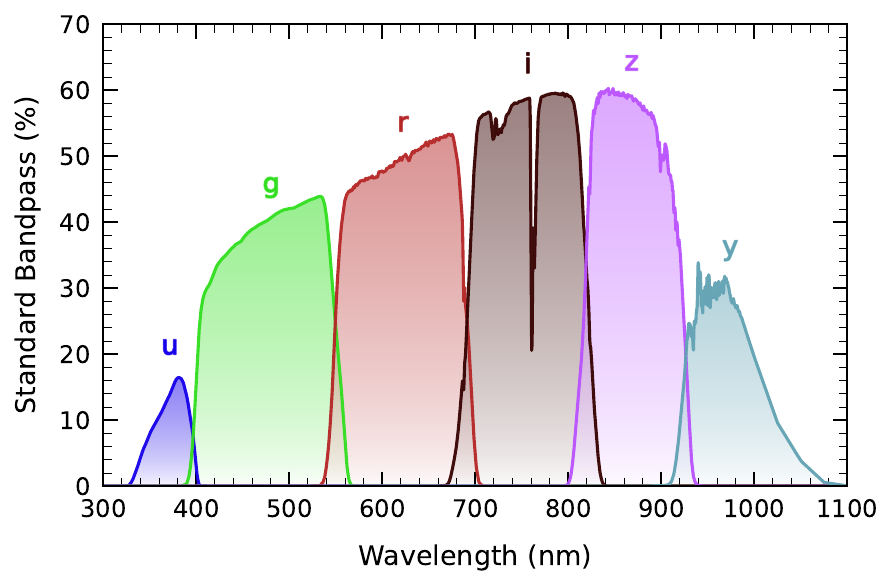}
\caption{\gls{LSSTComCam} standard bandpasses, illustrating full system throughput. The bandpasses include a standard atmosphere at airmass 1.2, telescope optics, camera surfaces, and mean \gls{ITL} detector quantum efficiency. The corresponding transmission curves are provided as a \gls{DP1} data product.}
\label{fig:comcam_standard_bandpasses}
\end{figure}

\gls{LSSTComCam} supports imaging with six broadband filters $ugrizy$ spanning 320--1050 nm, identical in design to \gls{LSSTCam}.
However, its filter exchanger can hold only three filters at a time, comapared to five with \gls{LSSTCam}.
The full-system throughput of the six \gls{LSSTComCam} filters, which encompasses contributions from a standard atmosphere at airmass 1.2, telescope optics, camera surfaces, and the mean \gls{ITL} detector quantum efficiency is shown in \figref{fig:comcam_standard_bandpasses}.
The corresponding transmission curves are provided as a \gls{DP1} data product (\secref{sssec:transmission_curves}).

\subsubsection{Timing Calibration}
\label{ssec:comcam_timing}
The absolute time accuracy of data taken with \gls{LSSTComCam} relies on the \gls{NTP} for clock synchronization, which should be accurate to approximately 1 millisecond. 
In order to evaluate the absolute timing accuracy of the entire system we observed the geosynchronous satellite EUTELSAT 117 West B with a set of 10 usable 10-second exposures over two nights.  
EUTELSAT 117 West B is part the \gls{GPS} system and serves as one of the \gls{WAAS} satellites operated for the \gls{FAA} and used to broadcast \gls{GPS} corrections to air traffic.

As these satellites are part of the \gls{GPS} system, their positions are tracked very precisely and the record of their locations is published after the fact and can be downloaded.  
Following the technique previously employed by other surveys, \citep{2018PASP..130f4505T},  we observed the satellite while tracking the sky and then downloaded the data-files with its precise locations from the National Satellite Test Bed web site\footnote{\nolinkurl{https://www.nstb.tc.faa.gov/nstbarchive.html}}.  
By comparing the measured and predicted locations of the start of the satellite track on the sky, we determined that (relative to the start of integration-time recorded in the \gls{FITS} headers) our time was accurate to 53.6 $\pm$ 11.0 milliseconds.

This work continues to be an area of ongoing study, with the exact timing of when the shutter open command is issued, and the complete profile of the shutter movement not yet determined. 
However the open command is on average near 29 milliseconds later. Incorporating the delays into the fit reduces the offset to 24.8 $\pm$ 11.0 milliseconds.

The full shutter takes approximately 396 milliseconds to completely open.  
As the \gls{LSSTComCam} sensors are centered in the aperture, the center of the focal plane should be exposed about half-way through the shutter open procedure, 198 milliseconds after the open command. 
There are uncertainties on the full motion profile, and the blade direction motions are currently not known, but the fraction of the shutter aperture subtended by the focal plane is 52\%. 
This implies that that the shutter will pass any pixel between 198 +/- 103 milliseconds.  
Subtracting this from the fitted delay of 24.8 milliseconds and adding the fitted error of 11.0 milliseconds in quadrature, results in a current conservative estimate of the delay of -173.2 $\pm$ 104.1 milliseconds, consistent with and smaller than the constraints on the timing offset determined using astrometric residuals from known asteroid associations presented in \secref{ssec:asteroid_association}.

\subsection{Flat Field System}
\label{ssec:flat_field_system}
During the on-sky campaign, key components of the Rubin calibration system ~\citep{2022SPIE12182E..0RI}, including the flat field screen, had not yet been installed.
As a result, flat fielding for \gls{DP1} relied entirely on twilight flats.
While twilight flats pose challenges such as non-uniform illumination and star print-through, they were the only available option during \gls{LSSTComCam} commissioning and for \gls{DP1} processing.
To mitigate these limitations, dithered, tracked exposures were taken over a broad range of azimuth and rotator angles to construct combined flat \gls{calibration} frames.
Exposure times were dynamically adjusted to reach target signal levels of between 10,000 and 20,000 electrons.
Future campaigns with \gls{LSSTCam} will benefit from more stable and uniform flat fielding using the Rubin flat field system, described in \citet{SITCOMTN-086}.
%
\subsection{LSST Science Pipelines Commissioning}
\label{ssec:pipelines_commissioning}
\begin{figure*}[bt!]
\centering
\plotone{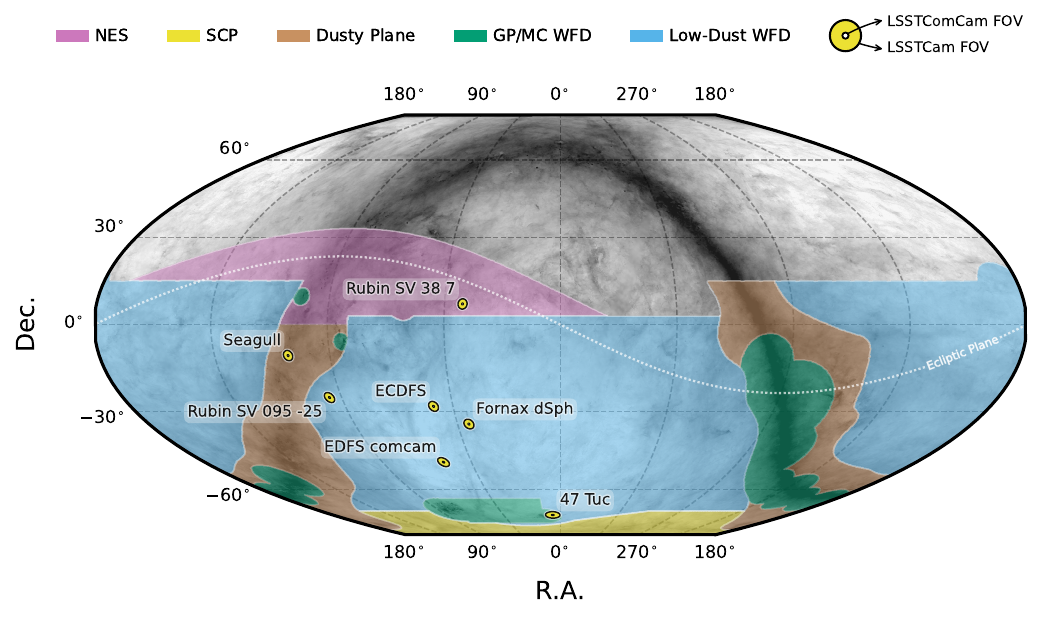}
\caption{Locations of the seven \gls{DP1} fields overlaid on the \gls{LSST} baseline survey footprint. NES: North Ecliptic Spur; SCP: South Celestial Pole; Low-Dust \gls{WFD}: regions away from the Galactic Plane (GP) observed with a \gls{WFD} cadence; GP/MC \gls{WFD}: Galactic Plane and Magellanic Clouds regions observed with a \gls{WFD} cadence. The fields of view of the \gls{LSSTCam} and \gls{LSSTComCam} focal planes are shown as concentric yellow circles about the pointing center of each field. The background Milky Way dust map is based on the FDS model \citep{1999ApJ...524..867F}.}
\label{fig:dp1_fields_on_sky}
\end{figure*}

Commissioning of the LSST Science Pipelines, \citep{PSTN-019},  began once the telescope was able to routinely deliver sub-arcsecond image quality.
The goals included testing the internal astrometric and photometric calibration across a range of observing conditions, 
validating the difference image analysis and prompt processing \citep{dmtn-219} framework, and accumulating over 200 visits per band to evaluate 
deep coadded images with integrated exposure times roughly equivalent to those of the planned LSST \gls{WFD} 10-year depth.
To support these goals, \nfields target fields were selected that span a range of stellar densities, overlap with external reference datasets, and collectively span the full breadth of the four primary \gls{LSST} science themes.
These \nfields fields form the basis of the \gls{DP1} dataset.
\figref{fig:dp1_fields_on_sky} shows the locations of these \nfields fields on the sky, overlaid on the LSST baseline survey footprint \citep{PSTN-051, PSTN-052, PSTN-053, PSTN-055, PSTN-056}, along with the sky coverage of both the \gls{LSSTCam} and \gls{LSSTComCam} focal planes.
Each of the \nfields target fields was observed repeatedly in multiple bands over many nights.
A typical observing \gls{epoch} on a given target field consisted of 5-20 visits in each of the three loaded filters.
Only images taken as 1x30 second exposures have been included in \gls{DP1}.
All images were acquired using the Rubin \gls{FBS}, version 3.0 \citep{Naghib_2019, peter_yoachim_2024_13985198}.
\tabref{tab:dp1_fields} lists the \nfields \gls{DP1} fields and their pointing centers, and provides a summary of the band coverage in each.

\begin{figure}[htb!]
\centering
\plotone{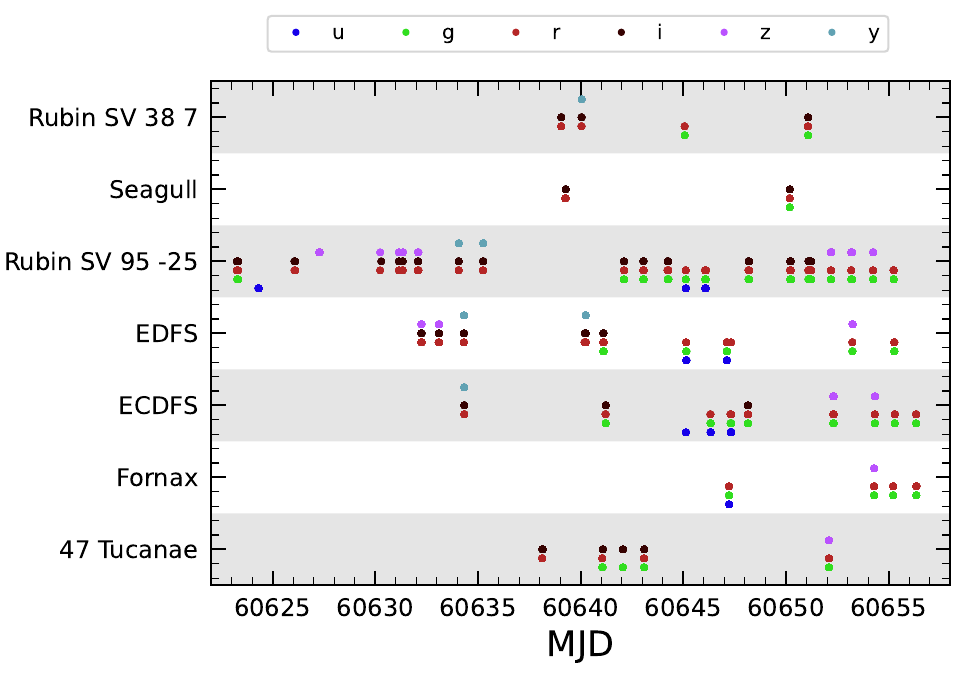}
\caption{Temporal distribution of \gls{DP1} observations, grouped by field as a function of \gls{MJD} and color-coded by filter. Each point indicates that a given field was observed at least once in the corresponding filter on that date.}
\label{fig:target_fields_temporal_sampling}
\end{figure}

\begin{deluxetable*}{llcccccccccr}
\caption{DP1 fields and pointing centers with the number of exposures in each band per field. ICRS coordinates are in units of decimal degrees, and are specified as J2000.}
\label{tab:dp1_fields}
\tablehead{
  \colhead{\textbf{Field Code}} & 
  \colhead{\textbf{Field Name}} &
  \colhead{\textbf{RA}} & 
  \colhead{\textbf{Dec}} & 
  \colhead{} &  
  \multicolumn{6}{c}{\textbf{Band}} & 
  \colhead{\textbf{Total}} \\
  \cline{3-4} \cline{6-11}
  & & \colhead{deg} & \colhead{deg} & & 
  \colhead{$u$} & \colhead{$g$} & \colhead{$r$} & 
  \colhead{$i$} & \colhead{$z$} & \colhead{$y$} & \colhead{}
}
\startdata
47\_Tuc & \parbox[t]{6cm}{47 Tucanae Globular Cluster} & 6.128 & -72.090 & & 
\parbox{0.3cm}{6} & \parbox{0.3cm}{10} & \parbox{0.3cm}{32} & \parbox{0.3cm}{19} & \parbox{0.3cm}{0} & \parbox{0.3cm}{5} & 72 \\
ECDFS & \parbox[t]{6cm}{Extended Chandra Deep Field South} & 53.160 & -28.100 & & 
\parbox{0.3cm}{43} & \parbox{0.3cm}{230} & \parbox{0.3cm}{237} & \parbox{0.3cm}{162} & \parbox{0.3cm}{153} & \parbox{0.3cm}{30} & 855 \\
EDFS\_comcam & \parbox[t]{6cm}{Rubin SV Euclid Deep Field South} & 59.150 & -48.730 & & 
\parbox{0.3cm}{20} & \parbox{0.3cm}{61} & \parbox{0.3cm}{87} & \parbox{0.3cm}{42} & \parbox{0.3cm}{42} & \parbox{0.3cm}{20} & 272 \\
Fornax\_dSph & \parbox[t]{6cm}{Fornax Dwarf Spheroidal Galaxy} & 40.080 & -34.450 & & 
\parbox{0.3cm}{0} & \parbox{0.3cm}{5} & \parbox{0.3cm}{25} & \parbox{0.3cm}{12} & \parbox{0.3cm}{0} & \parbox{0.3cm}{0} & 42 \\
Rubin\_SV\_095\_-25 & \parbox[t]{6cm}{Rubin SV Low Galactic Latitude Field} & 95.040 & -25.000 & & 
\parbox{0.3cm}{33} & \parbox{0.3cm}{82} & \parbox{0.3cm}{84} & \parbox{0.3cm}{23} & \parbox{0.3cm}{60} & \parbox{0.3cm}{10} & 292 \\
Rubin\_SV\_38\_7 & \parbox[t]{6cm}{Rubin SV Low Ecliptic Latitude Field} & 37.980 & 7.015 & & 
\parbox{0.3cm}{0} & \parbox{0.3cm}{44} & \parbox{0.3cm}{40} & \parbox{0.3cm}{55} & \parbox{0.3cm}{20} & \parbox{0.3cm}{0} & 159 \\
Seagull & \parbox[t]{6cm}{Seagull Nebula} & 106.300 & -10.510 & & 
\parbox{0.3cm}{10} & \parbox{0.3cm}{37} & \parbox{0.3cm}{43} & \parbox{0.3cm}{0} & \parbox{0.3cm}{10} & \parbox{0.3cm}{0} & 100 \\
\hline
Total & & & & & 
\parbox{0.3cm}{112} & \parbox{0.3cm}{469} & \parbox{0.3cm}{548} & \parbox{0.3cm}{313} & \parbox{0.3cm}{285} & \parbox{0.3cm}{65} & 1792
\enddata
\end{deluxetable*}
 %
Figure~\ref{fig:target_fields_temporal_sampling} shows the temporal sampling of observations by filter and by night.
The figure indicates the dates on which each field was observed in a given band but does not convey the total number of observations obtained per filter on any individual night. 
Gaps in coverage across some bands arise from the fact that \gls{LSSTComCam} can only accommodate three filters at a time (see \secref{ssec:comcam}).
As the campaign progressed, the temporal sampling became denser across all fields, reflecting improved efficiency and increased time allocated for science observations.
The \gls{ECDFS} field received the most consistent and densest temporal sampling.
It is important to note that the time sampling in the \gls{DP1} dataset differs significantly from what will be seen in the final \gls{LSST} data.
%
All fields except for the low ecliptic latitude field, Rubin\_SV\_38\_7, used a small random dithering pattern.
The random translational dithers of the telescope boresight were applied for each visit, with offsets of up to 0.2 degrees around the pointing center. 
The rotational dithers of the camera rotator were typically approximately 1 degree per visit, with larger random offsets at each filter change, which worked to keep operational efficiency high. 
The Rubin\_SV\_38\_7 field used a different dither pattern to optimize coverage of Solar System Objects and test Solar System Object linking across multiple nights. 
These observations used a 2$\times$2 grid of \gls{LSSTComCam} pointings to cover an area of about 1.3 degree$\times$1.3 degrees.
The visits cycled between the grid's four pointing centers, each separated by 0.65  degrees, and used small random translational dithers to fill chip gaps with the goal of acquiring 3-4 visits per pointing center per band in each observing epoch.
The RA and Dec values provided in \tabref{tab:dp1_fields} for this field represent approximately the center of the four fields.

\begin{figure*}[ht]
\centering
 \includegraphics[width=\linewidth]{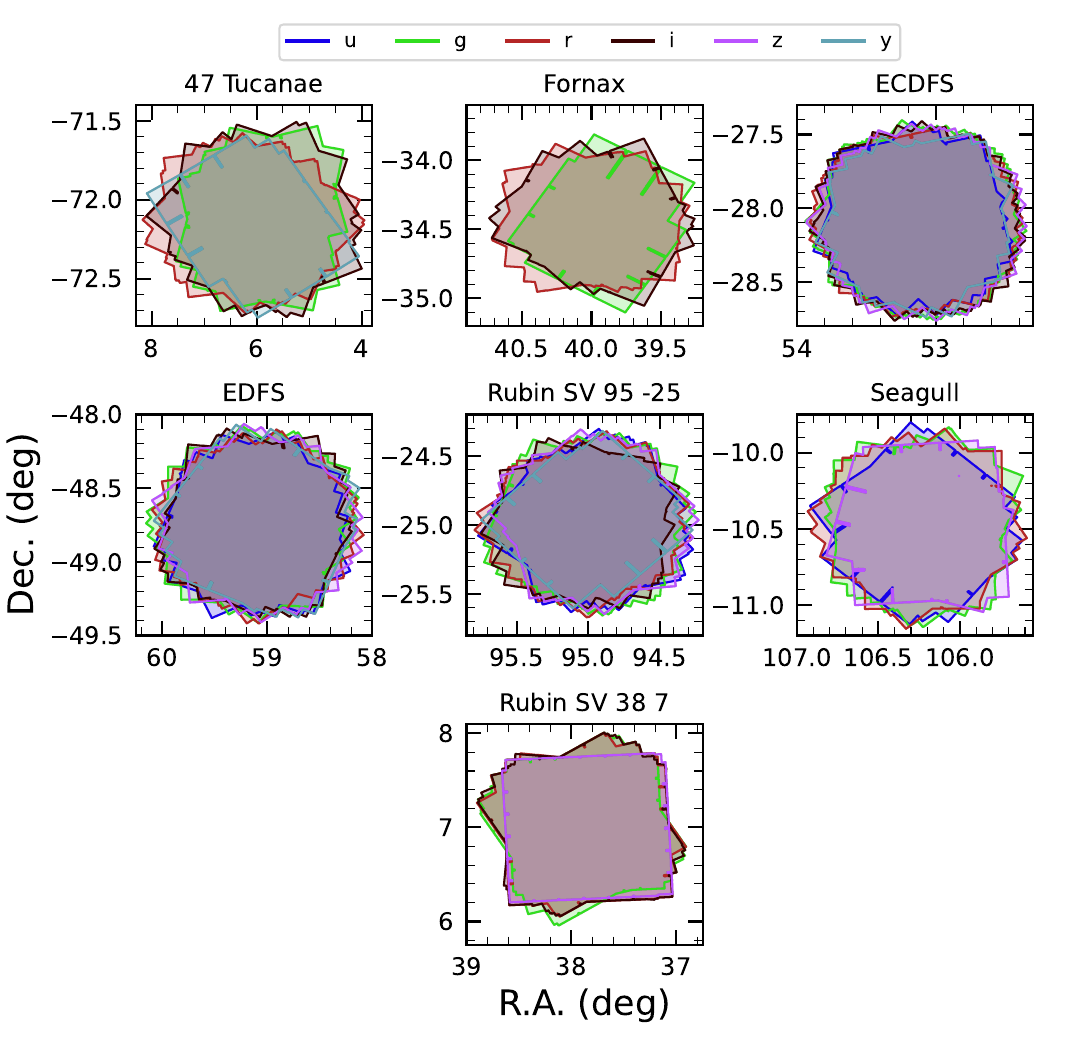}
 \caption{Sky coverage maps showing the distribution of visits in each field, color coded by band. The images clearly show the focal plane chip gaps and dithering pattern. Only the detectors for which single frame processing succeeded are included in the plots, which explains why the central region of 47\_Tuc looks thinner than the other fields. }
\label{fig:dp1_fields_coverage_maps}
\end{figure*}

\figref{fig:dp1_fields_coverage_maps} shows sky coverage maps showing the distribution of visits in each of the \nfields \gls{DP1} fields, color coded by band. 
The images clearly show the focal plane chip gaps and dithering pattern.
Only the detectors for which single frame processing succeeded are included in the plots, which explains why the central region of 47\_Tuc looks thinner than the other fields (see \secref{ssec:crowded_fields}). 
\tabref{tab:dp1_m5_depths} reports the 5$\sigma$ point source depths for coadded images per field and per band, where coverage in a band is non-zero, together with the expected 10-year LSST depths derived from the baseline simulated survey \citep[][]{2022ApJS..258....1B}.

\setlength{\tabcolsep}{3pt}  
\begin{deluxetable}{lcccccc}
\caption{DP1 median 5$\sigma$ coadded point-source detection limits per field and band, expressed in magnitudes, compared with the expected 10-year LSST values derived from the baseline simulated survey \citep{2022ApJS..258....1B}. } 
\label{tab:dp1_m5_depths}
\tablehead{
  \colhead{\textbf{Field Code}} & \multicolumn{6}{c}{\textbf{Band}} \\
  \cline{2-7}
   &$u$&$g$&$r$&$i$&$z$&$y$
}
\startdata
47\_Tuc & - & 24.03 & 24.24 & 23.90 & - & 21.79 \\
ECDFS & 24.55 & 26.18 & 25.96 & 25.71 & 25.07 & 23.10 \\
EDFS\_comcam & 23.42 & 25.77 & 25.72 & 25.17 & 24.47 & 23.14 \\
Fornax\_dSph & - & 24.53 & 25.07 & 24.64 & - & - \\
Rubin\_SV\_095\_-25 & 24.29 & 25.46 & 24.95 & 24.86 & 24.32 & 22.68 \\
Rubin\_SV\_38\_7 & - & 25.46 & 25.15 & 24.86 & 23.52 & - \\
Seagull & 23.51 & 24.72 & 24.19 & - & 23.30 & - \\
\hline
LSST 10-year & 25.73 & 26.86 & 26.88 & 26.34 & 25.63 & 24.87 \\
\enddata
\end{deluxetable}
 
\subsection{Delivered Image Quality}
\label{ssec:image_quality}
The delivered image quality is influenced by contributions from both the observing system (i.e., dome, telescope and camera) and the atmosphere.
During the campaign, the Rubin \gls{DIMM} was not operational, so atmospheric seeing was estimated using live data from the \gls{SOAR} \gls{RINGSS} seeing monitor, also located on Cerro Pach\'on.
Although accelerometers mounted on the mirror cell and top-end assembly were available to track dynamic optics effects, such as mirror oscillations that can degrade optical alignment, this data was not used during the campaign.
Mount encoder data were used to measure the mount jitter in every image, with a measured median contribution of 0.004 arcseconds to image degradation.
As the pointing model was not fine tuned, tracking errors could range from 0.2 to 0.4 arcseconds per image, depending on RA and Dec.
Dome and mirror-induced \gls{seeing} were not measured during the campaign.

\begin{figure*}[hbt!]
  \centering
  \begin{subfigure}[t]{0.45\textwidth}
    \includegraphics[width=\linewidth]{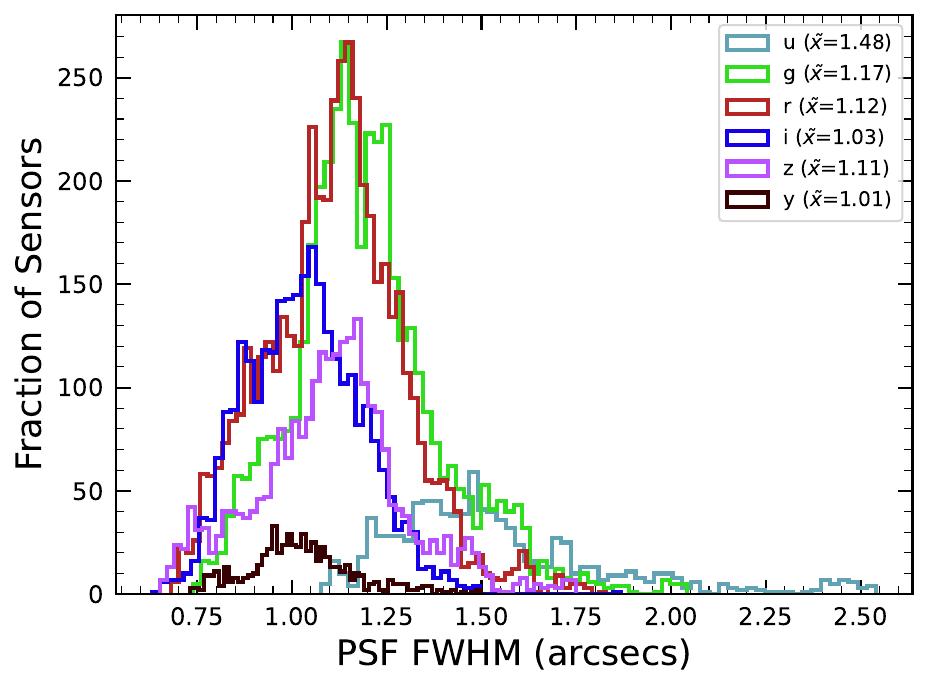}
    \caption{PSF FHWM (arcsecs) per passband across all \gls{DP1} target fields. } 
 \end{subfigure}\hfill
  \begin{subfigure}[t]{0.45\textwidth}
    \includegraphics[width=\linewidth]{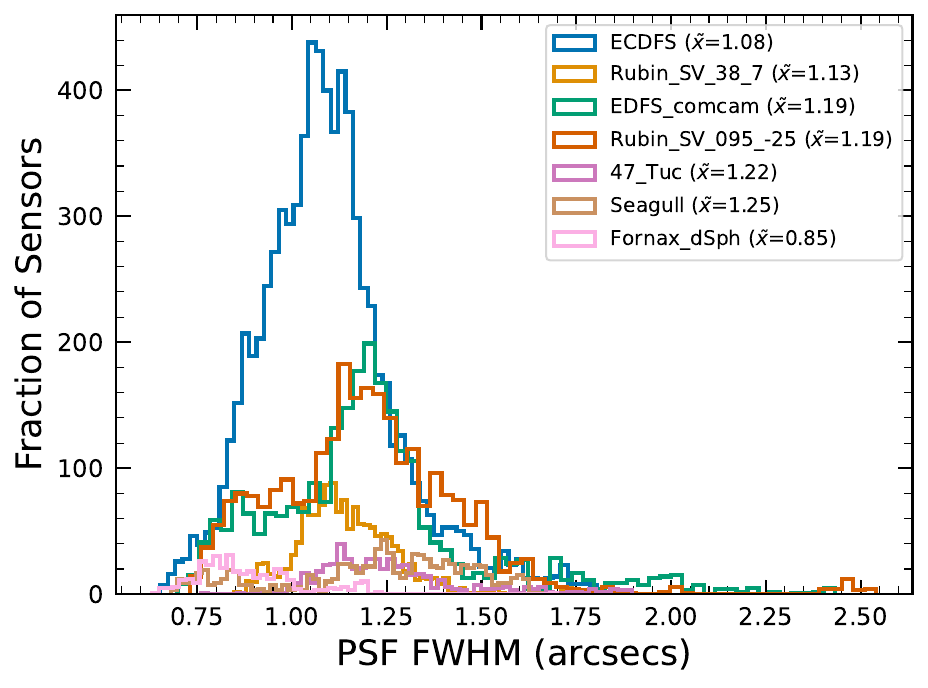}
    \caption{PSF FHWM (arcsecs) per \gls{DP1} target field across all passbands}
  \end{subfigure}\hfill
\caption{Histograms showing the  distribution of delivered image quality for all \nvisitdetectorsummaries single-epoch individual sensors in the \gls{DP1} dataset per passband (a) and per field (b). The median values are given in the legend. }
\label{fig:delivered_image_quality}
\end{figure*}

\setlength{\tabcolsep}{3pt}  
\begin{deluxetable}{lccccccc}
\caption{DP1 Median image quality per field and per band quantified as the PSF at FWHM in arcseconds.}
\label{tab:iq_band_field}
\tablehead{
  \colhead{\textbf{Field Code}} & \multicolumn{6}{c}{\textbf{Band}} & \textbf{All}\\
  \cline{2-7}
   &$u$ & $g$ & $r$ & $i$ & $z$ & $y$ &
}
\startdata
47\_Tuc & -- & 1.27 & 1.25 & 1.11 & -- & 1.33 & 1.22 \\
ECDFS & 1.40 & 1.14 & 1.08 & 1.00 & 1.00 & 1.07 & 1.08 \\
EDFS\_comcam & 1.88 & 1.25 & 1.20 & 1.10 & 1.18 & 0.99 & 1.19 \\
Fornax\_dSph & -- & 1.16 & 0.82 & 0.93 & -- & -- & 0.85 \\
Rubin\_SV\_095\_-25 & 1.40 & 1.25 & 1.14 & 0.97 & 1.17 & 0.82 & 1.19 \\
Rubin\_SV\_38\_7 & -- & 1.13 & 1.13 & 1.10 & 1.22 & -- & 1.13 \\
Seagull & 1.50 & 1.34 & 1.19 & -- & 1.19 & -- & 1.25 \\
\hline
All & 1.48 & 1.17 & 1.12 & 1.03 & 1.11 & 1.01 & 1.13 \\\enddata
\end{deluxetable}
 The \gls{DP1} median delivered image quality, quantified as the \gls{PSF} at \gls{FWHM} across all filters and target fields, is \medianimagequalityallbands. 
The best images achieve a \gls{PSF} \gls{FWHM} of approximately \bestimagequality. 
Both the per-sensor \gls{PSF} \gls{FWHM} and the overall median vary depending on the filter and the specific target field. 
The median delivered image quality per band and target field is provided in \tabref{tab:iq_band_field}.
\figref{fig:delivered_image_quality} shows the  distribution of PSF FWHM (in arcsec) over all \nvisitdetectorsummaries individual sensors images.
Ongoing efforts aim to quantify all sources of image degradation, including contributions from the camera system; static and dynamic optical components; telescope mount motion; observatory-induced seeing from the dome and primary mirror; and atmospheric conditions.
For the \gls{LSST}, the design specification for the median delivered image quality, referenced to the zenith and 550 nm, is 0\farcs7.  
This value corresponds to the measured median atmospheric seeing at the Cerro Pach\' on site and a system contribution to delivered image quality of 0\farcs35 added in quadrature.

 \section{Overview of the contents of Rubin \gls{DP1}}
\label{sec:data_products}
In this section we describe the Rubin \gls{DP1} data products and provide summary statistics for each.
For more detailed information, we refer the reader to the DOI-registered \gls{DP1} release documentation available at \url{https://dp1.lsst.io} and the catalog schemas available at \url{https://sdm-schemas.lsst.io}.\footnote{Searchable catalog schemas are also available to Data Rights Holders via the Rubin Science Platform at \nolinkurl{https://data.lsst.cloud}.}

The \gls{DP1} science data products are derived from the \nvisitimages individual \gls{CCD} images taken across \nexposures exposures in the \nfields \gls{LSSTComCam} commissioning fields (\secref{ssec:pipelines_commissioning}).
To aid legibility, we have separated the descriptions of the data products from the description of the data release processing pipeline (\secref{sec:drp}).
Similarly, as the \gls{DP1} data products can be accessed via one or both of \gls{IVOA} Services (\secref{sssec:ivoa_services}) or the Data \gls{Butler} (\secref{sssec:data_butler}), we describe them here in a manner that is agnostic to the means of access.

The data products that comprise \gls{DP1} provide an early preview of future LSST data releases and are strongly dependent on the type and quality of the data that was collected during the \gls{LSSTComCam} on-sky campaign (\secref{ssec:pipelines_commissioning}).
Consequently not all anticipated  \gls{LSST} data products, as described in the \gls{DPDD} \citep{LSE-163}, were produced for the \gls{DP1} dataset.

Rubin Observatory has adopted the convention by which single-epoch detections are referred to as ``Sources", and the astrophysical object associated with a given detection is referred to as an ``Object"
\footnote{We caution that this nomenclature is not universal; for example, some surveys use ``detections'' for what we call ``sources'', and ``sources'' for what we call ``objects''.}.
As such, a given Object will likely have multiple associated  Sources, since it will be observed in multiple epochs.

At the highest level, the \gls{DP1} data products fall into one of five types:
\begin{itemize}
\item \textbf{Science Images}, including single-\gls{epoch} images, deep and template coadded images, and difference images (\secref{ssec:science_images});
\item \textbf{Catalogs} of astrophysical Sources and Objects detected and measured in the aforementioned images. We also provide the astrometric and photometric reference catalog generated from external sources that was used during processing to generate the \gls{DP1} data products (\secref{ssec:catalogs});
\item \textbf{Maps}, which provide non-science-level visualizations of the data within the release. They include, for example, zoomable multi-band images and coverage maps (\secref{ssec:survey_property_maps});
\item \textbf{Ancillary data products}, including, for example, the parameters used to configure the data processing pipelines, log and processing performance files,  and \gls{calibration} data products (\secref{ssec:ancilliary});
\item \textbf{Metadata} in the form of tables containing information about each visit and processed image, such as pointing, exposure time, and a range of image quality summary statistics (\secref{ssec:metadata}).
\end{itemize}
While images and catalogs are expected to be the primary data products for scientific research, we also recognize the value of providing access to other data types to support investigations and ensure transparency.

To facilitate processing, Rubin \gls{DP1} uses a single skymap\footnote{A skymap is a tiling of the celestial sphere, organizing large-scale sky coverage into manageable sections for processing and analysis. While the skymap described here is specific to \gls{DP1}, we do not anticipate major changes to the skymap in future data releases.} that covers the entire sky area encompassing the seven \gls{DP1} fields.
The \gls{DP1} skymap divides the entire celestial sphere into \ntotaltracts \gls{tract}s, each covering approximately \tractarea.
The \gls{tract}s are arranged in rings of declination, ordered from south to north, then with increasing right ascension within a ring.
Each \gls{tract} is further subdivided into \npatchx$\times$\npatchy equally-sized patches.
Both \gls{tract}s and patches overlap with their neighboring regions.
The amount of overlap between \gls{tract}s changes with declination, with \gls{tract}s nearest the poles having the greatest degree of overlap; the minimum overlap between tracts is \tractoverlap.
By contrast, the amount of overlap between patches is constant, with each \gls{patch} overlapping each of its neighbouring patches by \patchoverlap.
Each patch covers \outerpatcharea which, due to the patch overlap, is slightly larger than the tract area divided by the number of patches in a tract.
The aerial coverage of a patch is comparable to, but somewhat smaller than, the \rawfov field-of-view of a single \gls{LSSTComCam} or \gls{LSSTCam} detector, meaning each detector image spans multiple patches.
The size of a tract is larger than the \gls{LSSTComCam} field of view. However, since each observed field extends across more than one tract, each field covers multiple tracts.

The skymap is integral to the production of co-added images.
To create a coadded image, the processing pipeline selects all calibrated science images in a given field that meet specific quality thresholds (\secref{ssec:science_images} and \secref{ssec:coaddition}) for a given \gls{patch}, warps them onto a single consistent pixel grid for that \gls{patch}, as defined by the skymap, then coadds them.
Each individual coadd image therefore covers a single \gls{patch}.

Throughout this section, the data product names are indicated using \texttt{monospace} font.
Data products are accessed via either the \gls{IVOA} Services ( \secref{sssec:ivoa_services}) or the Data \gls{Butler} (\secref{sssec:data_butler}).

\subsection{Science Images}
\label{ssec:science_images}
Science images are exposures of the night sky, as distinct from \gls{calibration} images (\secref{ssec:calibration_data}).
Although the release includes \gls{calibration} images, thereby allowing users to reprocess the raw images if needed, this is expected to be necessary only in rare cases.
Users are strongly encouraged to start from the \texttt{visit\_image} provided.
The data product names shown here are those used by the Data \gls{Butler}, but the names used in the \gls{IVOA} Services differ only slightly in that they are prepended by ``\texttt{lsst.}''.

\subsubsection{Raw Image}
\texttt{raw} images \citep{10.71929/rubin/2570310} are unprocessed data received directly from the camera.
Each \texttt{raw} corresponds to a single \gls{CCD} from a single \gls{LSSTComCam} exposure of \exposuretime duration.
Each \gls{LSSTComCam} exposure typically produces up to nine \texttt{raw}s, one per sensor in the focal plane.
However, a small number of exposures resulted in fewer than nine \texttt{raw} images due to temporary hardware issues or readout faults.

In total, \gls{DP1} includes \nraws raw images.
\tabref{tab:rawbreakdown} provides a summary by target and band.
A \texttt{raw} contains \nrawpixx $\times$ \nrawpixy pixels, including prescan and overscan, and occupies around \rawhdd of disk space.\footnote{Each amplifier image contains 3 and 64 columns of serial prescan and overscan pixels, respectively, and 48 rows of parallel overscan pixels, meaning a \texttt{raw} contains \nvisitimagepixx$\times$\nvisitimagepixy exposed pixels.}
The field of view of a single \texttt{raw}, excluding prescan and overscan regions, is roughly \visitimagefovx$\times$\visitimagefovy$\approx$\visitimagefov, corresponding to a plate scale of \rawplatescale.
\setlength{\tabcolsep}{6pt}  
\begin{deluxetable}{lccccccr}
\caption{Number of \texttt{raw} images per field and band.
Each raw image corresponds to a single 30-second LSSTComCam exposure on one CCD. Most exposures produce nine raw images, one per sensor in the focal plane, however some yield fewer due to occasional hardware or readout issues.}
\label{tab:rawbreakdown}
\tablehead{
  \colhead{\textbf{Field Code}} & \multicolumn{6}{c}{\textbf{Band}} & \colhead{\textbf{Total}} \\
  \cline{2-7}
  & $u$ & $g$ & $r$ & $i$ & $z$ & $y$ & 
}
\startdata
47\_Tuc & \parbox{0.3cm}{54} & \parbox{0.3cm}{90} & \parbox{0.3cm}{288} & \parbox{0.3cm}{171} & \parbox{0.3cm}{0} & \parbox{0.3cm}{45} & 648 \\
ECDFS & \parbox{0.3cm}{387} & \parbox{0.3cm}{2070} & \parbox{0.3cm}{2133} & \parbox{0.3cm}{1455} & \parbox{0.3cm}{1377} & \parbox{0.3cm}{270} & 7692 \\
EDFS\_comcam & \parbox{0.3cm}{180} & \parbox{0.3cm}{549} & \parbox{0.3cm}{783} & \parbox{0.3cm}{378} & \parbox{0.3cm}{378} & \parbox{0.3cm}{180} & 2448 \\
Fornax\_dSph & \parbox{0.3cm}{0} & \parbox{0.3cm}{45} & \parbox{0.3cm}{225} & \parbox{0.3cm}{108} & \parbox{0.3cm}{0} & \parbox{0.3cm}{0} & 378 \\
Rubin\_SV\_095\_-25 & \parbox{0.3cm}{297} & \parbox{0.3cm}{738} & \parbox{0.3cm}{756} & \parbox{0.3cm}{207} & \parbox{0.3cm}{540} & \parbox{0.3cm}{90} & 2628 \\
Rubin\_SV\_38\_7 & \parbox{0.3cm}{0} & \parbox{0.3cm}{396} & \parbox{0.3cm}{360} & \parbox{0.3cm}{495} & \parbox{0.3cm}{180} & \parbox{0.3cm}{0} & 1431 \\
Seagull & \parbox{0.3cm}{90} & \parbox{0.3cm}{333} & \parbox{0.3cm}{387} & \parbox{0.3cm}{0} & \parbox{0.3cm}{90} & \parbox{0.3cm}{0} & 900 \\
\cline{1-8}
Total & \parbox{0.3cm}{1008} & \parbox{0.3cm}{4221} & \parbox{0.3cm}{4932} & \parbox{0.3cm}{2814} & \parbox{0.3cm}{2565} & \parbox{0.3cm}{585} & 16125 \\
\enddata
\end{deluxetable} 
\subsubsection{Visit Image}
\texttt{visit\_image}s \citep{10.71929/rubin/2570311} are fully-calibrated processed images.
They have undergone instrument signature removal (\secref{ssec:isr}) and all the single frame processing steps described in \secref{ssec:single_frame_processing} which are, in summary: \gls{PSF} modeling, \gls{background} subtraction, and astrometric and photometric \gls{calibration}.
As with \texttt{raw}s, a \texttt{visit\_image} contains processed data from a single \gls{CCD} resulting from a single \exposuretime \gls{LSSTComCam} exposure.
As a consequence, a single \gls{LSSTComCam} exposure typically results in nine \texttt{visit\_image}s.
The handful of exposures with fewer than nine \texttt{raw} images also have fewer than nine \texttt{visit\_images}, but there are an additional \nsfpfails \texttt{raw} images that failed processing and for which there is thus no corresponding \texttt{visit\_image}.
The majority of failures -- 131 in total -- were due to challenges with astrometric fits or \gls{PSF} models in the 47\_Tuc crowded field. The other failures were in the Rubin\_SV\_095\_-25 (9 failures), ECDFS (8), Fornax\_dSph (3), and EDFS\_comcam (2) fields.

\begin{figure}[hbt!]
  \centering
  \includegraphics[width=\linewidth]{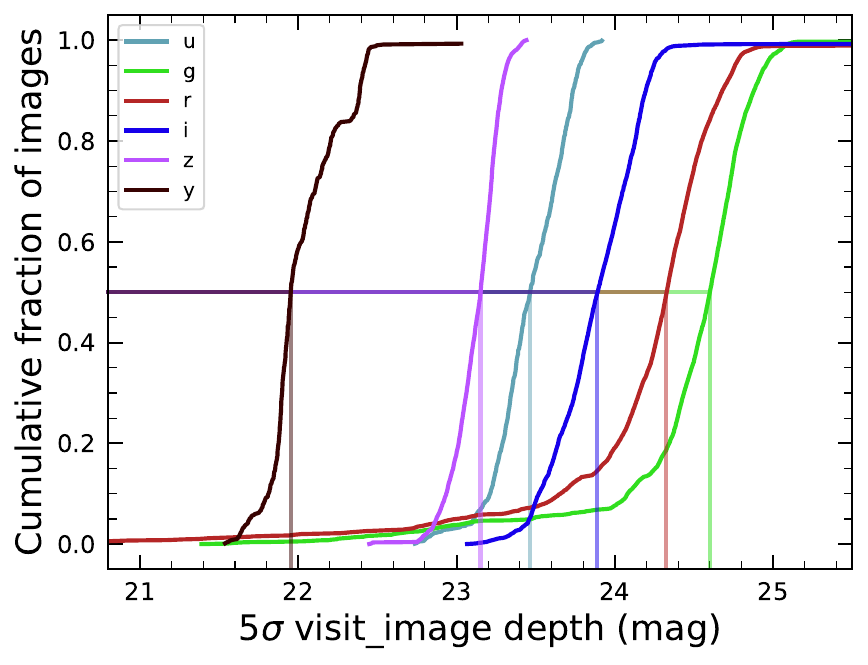}
  \caption{Normalized cumulative histograms of the $5\sigma$ depths of all \texttt{visit\_image}s in each band. The vertical lines indicate the 50th percentiles for each band (see legend).}
  \label{fig:visit_image_depths}
\end{figure}

In total, there are \nvisitimages \texttt{visit\_image}s in \gls{DP1}.
Each \texttt{visit\_image} comprises three images: a calibrated science image, a variance image, and a pixel-level bitmask that flags issues such as saturation, cosmic rays, or other artifacts.
Each \texttt{visit\_image} also contains a position-dependent \gls{PSF} model, \gls{WCS} information, and various \gls{metadata} providing information about the observation and processing.
The science and variance images and the pixel mask each contain \nvisitimagepixx$\times$ \nvisitimagepixy pixels.
In total, a single \texttt{visit\_image}, including all extensions and \gls{metadata}, occupies around \visitimagehdd of disk space.
A plot showing the normalized cumulative histogram of the $5\sigma$ depths of all the \texttt{visit\_image}s in \gls{DP1} is shown in \figref{fig:visit_image_depths}.

\subsubsection{Deep Coadd}
\texttt{deep\_coadd}s are created on a per-band basis, meaning only data from exposures taken with a common filter are coadded.
As such, there are up to six \texttt{deep\_coadd}s covering each \gls{patch} -- one for each of the six \gls{LSSTComCam} bands.
The process of producing \texttt{deep\_coadd}s is described in \secref{ssec:coadd_processing} but, to summarize, it involves the selection of suitable \texttt{visit\_image}s (both in terms of \gls{patch} coverage, band, and image quality), the warping of those \texttt{visit\_image}s onto a common pixel grid, and the co-adding of the warped \texttt{visit\_image}s.
To be included in a \gls{DP1} \texttt{deep\_coadd}, a \texttt{visit\_image} needed to have a \gls{PSF} \gls{FWHM} smaller than \deepcoaddmaxfwhm. Of the \nvisitimages \texttt{visit\_images}, \ndeepcoaddvisitimages satisfied this criterion and were therefore used to create \texttt{deep\_coadds}.

There are a total of \ndeepcoadds \texttt{deep\_coadd}s in \gls{DP1}.
As mentioned above, a single \texttt{deep\_coadd} covers one \gls{patch}, and includes a small amount of overlap with its neighboring \gls{patch}.
The skymap used for \gls{DP1} defines a \gls{patch} as having an on-sky area of \innerpatcharea excluding overlap, and \outerpatcharea including overlap.
A single \texttt{deep\_coadd} -- including overlap -- contains \ndeepcoaddpixx $\times$ \ndeepcoaddpixy equal-sized pixels, corresponding to a platescale of \rawplatescale.
Each \texttt{deep\_coadd} contains the science image (i.e., the coadd), a variance image, and a pixel mask; all three contain the same number of pixels.
Each \texttt{deep\_coadd} also contains a position-dependent \gls{PSF} model (which is the weighted sum of the \gls{PSF} models of the input \texttt{visit\_image}s), \gls{WCS} information, plus various \gls{metadata}.

The number of \texttt{visit\_image}s that contributed to a given \texttt{deep\_coadd} varies across the patch; the Survey Property Maps can be consulted to gain insights into the total exposure time at all locations covered by the survey.
Similarly, since coadds always cover an entire \gls{patch}, it is common for a \texttt{deep\_coadd} to contain regions that were not covered by any of the selected \texttt{visit\_image}s, particularly if the \gls{patch} is on the outskirts of a field and was thus not fully observed.
By the nature of how coadds are produced, such regions may contain seemingly valid \gls{flux} values (i.e., not necessarily zeros or \texttt{NaNs}), but will instead be flagged with the \texttt{NO\_DATA} flag in the pixel mask.
It is therefore crucial that the pixel mask be referred to when analyzing \texttt{deep\_coadds}.

\subsubsection{Template Coadd}
\texttt{template\_coadd}s \citep{10.71929/rubin/2570314} are those created to use as templates for difference imaging, i.e., the process of subtracting a template image from a \texttt{visit\_image} to identify either variable or \gls{transient} objects. It should be noted, however, that \texttt{template\_coadd}s are not themselves subtracted from \texttt{visit\_image}s but are, instead, warped to match the \gls{WCS} of a \texttt{visit\_image}.
It is this warped template that is subtracted from the \texttt{visit\_image} to create a difference image.
\footnote{For storage space reasons, warped templates are not retained for \gls{DP1}, as they can be readily and reliably recreated from the \texttt{template\_coadd}s.}
As with \texttt{deep\_coadd}s, \texttt{template\_coadd}s are produced by warping and co-adding multiple \texttt{visit\_image}s covering a given skymap-defined \gls{patch}.
The process of building \texttt{template\_coadd}s is the same as that for \texttt{deep\_coadd}s, but the selection criteria differ between the two types of coadd.
In the case of \texttt{template\_coadd}s, one third of \texttt{visit\_image}s covering the \gls{patch} in question with the narrowest \gls{PSF} \gls{FWHM} are selected.
If one third corresponds to fewer than twelve \texttt{visit\_image}s (i.e., there are fewer than 36 \texttt{visit\_image}s covering the \gls{patch}), then the twelve \texttt{visit\_images} with the narrowest \gls{PSF} \gls{FWHM} are selected.
Finally, if there are fewer than twelve \texttt{visit\_images} covering the \gls{patch}, then all \texttt{visit\_image}s are selected.
Of the \nvisitimages \texttt{visit\_image}s, \ntemplatecoaddvisitimages were used to create \texttt{template\_coadd}s.
This selection strategy is designed to optimize for \gls{seeing} when a \gls{patch} is well-covered by \texttt{visit\_image}s, yet still enable the production of \texttt{template\_coadd}s for poorly-covered patches.
As with \texttt{deep\_coadd}s, the number of \texttt{visit\_image}s that contributed to a \texttt{template\_coadd} varies across the patch.

\gls{DP1} contains a total of \ntemplatecoadds \texttt{template\_coadd}s.\footnote{The difference in the number of \texttt{deep\_coadd}s and \texttt{template\_coadd}s is due to the difference in the \texttt{visit\_image} selection criteria for each coadd.}
As with \texttt{deep\_coadd}s, a single \texttt{template\_coadd} covers a single \gls{patch}.
Since the same \texttt{skymap} is used when creating both \texttt{deep\_coadd} and \texttt{template\_coadd}s, the on-sky area and pixel count of \texttt{template\_coadd}s are the same as that of a \texttt{deep\_coadd} (see above).
Similarly, \texttt{template\_coadd}s contain the science image (i.e., the coadd), a variance image, and a pixel mask; all three contain the same number of pixels.
Also included are the \gls{PSF} model, \gls{WCS} information, and \gls{metadata}.
As is the case for \texttt{deep\_coadd}s, those pixels within \texttt{template\_coadd}s that are not covered by any of the selected \texttt{visit\_image}s may still have seemingly valid values, but are indicated with the \texttt{NO\_DATA} flag within the pixel mask.

\subsubsection{Difference Image}
\texttt{difference\_image}s \citep{10.71929/rubin/2570312} are generated by the subtraction of the warped, scaled, and \gls{PSF}-matched \texttt{template\_coadd} from the \texttt{visit\_image} (see \secref{ssec:diffim_analysis}). In principle, only those sources whose \gls{flux} has changed relative to the \texttt{template\_coadd} should be apparent (at a significant level) within a \texttt{difference\_image}. In practice, however, there are numerous spurious sources present in \texttt{difference\_image}s due to unavoidably imperfect template matching.

In total, there are \ndifferenceimages \texttt{difference\_image}s in \gls{DP1}, one for each \texttt{visit\_image}.

Like \texttt{visit\_image}s, \texttt{difference\_image}s contain the science (i.e., difference) image, a variance image, and a pixel mask; all three contain the same number of pixels, which is the same as that of the input \texttt{visit\_image}.
Also included is the \gls{PSF} model, \gls{WCS} information, and \gls{metadata}.

\subsubsection{Background Images}
Background images contain the model \gls{background} that has been generated and removed from a science image.
\texttt{visit\_image}s, \texttt{deep\_coadd}s and \texttt{template\_coadd}s all have associated \gls{background} images.\footnote{In future data releases, \gls{background} images may be included as part of their respective science image data product.} Background images contain the same number of pixels as their respective science image, and there is one \gls{background} image for each \texttt{visit\_image}, \texttt{deep\_coadd}, and \texttt{template\_coadd}.
Difference imaging analysis also measures and subtracts a \gls{background} model, but the \texttt{difference\_background} data product is not written out by default and is not part of \gls{DP1}.

Background images are not available via the \gls{IVOA} Service; they can only be accessed via the \gls{Butler} Data Service.

\subsection{Catalogs}
\label{ssec:catalogs}
In this section  we describe science-ready tables produced by the science pipelines.
All catalogs contain data for detections in the images described in \secref{ssec:science_images}, except  the  \texttt{Calibration} catalog, which contains reference data obtained from previous surveys.
Observatory-produced \gls{metadata} tables are described in \secref{ssec:metadata}.

The catalogs contains measurements for either Sources  detected in  \texttt{visit\_image}s and \texttt{difference\_image}s, or Objects detected in \texttt{deep\_coadd}s.
All catalogs store fluxes rather than magnitudes, with fluxes measured in nanojansky (1 nJy = $10^{-35}$ Wm$^{-2}$Hz$^{-1}$ ).
Fluxes are preferred for multi-epoch observations, as they can be averaged across epochs, unlike magnitudes.
Additionally, flux measurements on difference images (\secref{ssec:science_images}) are computed against a template, representing a flux difference.
As a result, flux measurements on difference images can be negative, particularly for faint sources in the presence of noise.

The \texttt{Source}, \texttt{Object}, \texttt{ForcedSource}, \texttt{DiaSource}, \texttt{DiaObject}, and \texttt{ForcedSourceOnDiaObject} catalogs described below each vary in terms of their specific columns but generally contain: one or more unique identification numbers, positional information, multiple types of \gls{flux} measurements (e.g., aperture fluxes, \gls{PSF} fluxes, Gaussian fluxes, etc.), and a series of boolean flags indicating characteristics such as saturation or cosmic ray contamination for each source/object.
The Solar System catalogs \texttt{SSObject} and \texttt{SSSource} deviate from this general structure in that they instead contain orbital parameters for all known asteroids.

Where applicable, quantities are prefixed with the band in which they were measured, and all measured properties are reported with their associated 1$\sigma$ uncertainties.
For example, \texttt{g\_ra} and \texttt{g\_raErr} refer to right ascension and its uncertainty, measured in the g-band.

Fluxes for various apertures are provided together with an uncertainty and  a flag, and named in the format \texttt{[band]\_ap[size]Flux}, where \texttt{[size]} is the aperture diameter in pixels.
For example, \texttt{g\_ap03Flux}, \texttt{g\_ap03FluxErr}, \texttt{g\_ap03Flux\_flag} provide the flux, uncertainty and flag measured within a 3.0-pixel aperture in the g-band.
Similarly for flux measurements using difference algorithms, e.g. \texttt{g\_psfFlux} provides the flux derived using the PSF model as a weight function, forced on g-band.

A complete list of columns  with description and units for all tables in DP1 is available at \nolinkurl{https://sdm-schemas.lsst.io/dp1.html}
Since \gls{DP1} is a preview release, it does not include all the catalogs expected in a full \gls{LSST} \gls{Data Release}.
Additionally, some catalogs may be missing columns, as not all quantities have been computed yet.
These quantities will be included in future releases, and, where it is known to be the case,  missing data are noted in the catalog descriptions that follow.

Catalog data are stored in the \gls{Qserv} database (\secref{sssec:qserv}) and are accessible via \gls{TAP}, and an online \gls{DP1} catalog \gls{schema} is available at \url{https://sdm-schemas.lsst.io/dp1.html}.
Catalog data are also accessible via the Data \gls{Butler} (see \secref{sssec:data_butler}).

\subsubsection{Source Catalog}
The \texttt{Source} catalog \citep{10.71929/rubin/2570323} contains data on all sources which are, prior to deblending (\secref{sssec:coadd_processing}), detected with a greater than 5$\sigma$ significance in each individual visit.
The detections reported in the \texttt{Source} catalog have undergone deblending; in the case of blended detections, only the deblended sources are included in the \texttt{Source} catalog.
It is important to note that while the criterion for inclusion in a \texttt{Source} catalog is a $>5\sigma$ detection in a \texttt{visit\_image} prior to deblending, the positions and fluxes are reported post-deblending.
Hence, it is possible for the \texttt{Source} catalog to contain sources whose \gls{flux}-to-error ratios -- potentially of all types (i.e., aperture \gls{flux}, \gls{PSF} \gls{flux}, etc.) -- are less than $5$.

In addition to the general information mentioned above (i.e., IDs, positions, fluxes, flags), the \texttt{Source} catalog also includes basic \gls{shape} and extendedness information.

The \texttt{Source} catalog contains data for \nsources \texttt{sources} in \gls{DP1}.

\begin{figure}[hbt!]
  \centering
  \includegraphics[width=\linewidth]{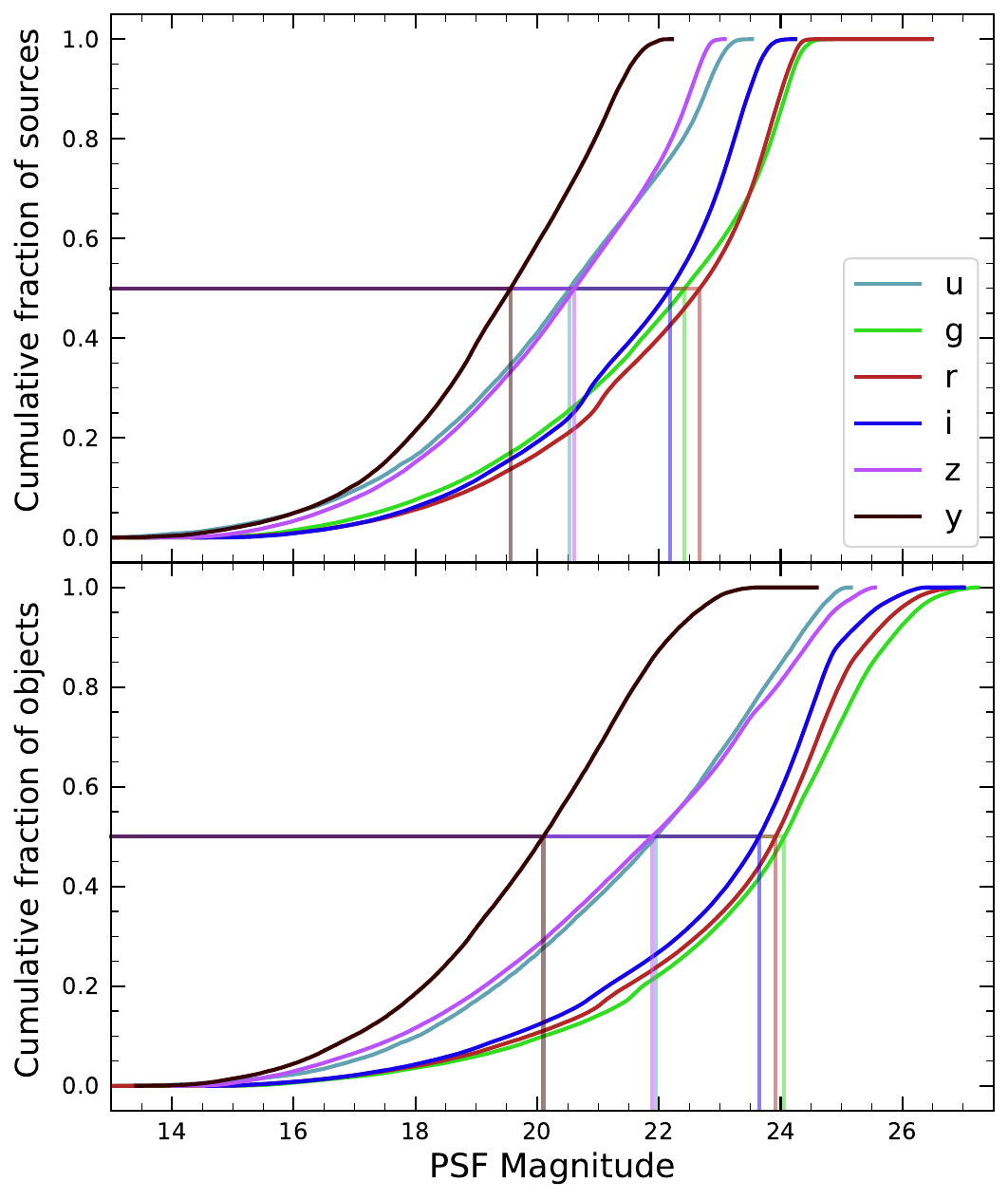}
  \caption{Normalized cumulative histograms of the PSF magnitudes of all $>5\sigma$-detected \texttt{sources} (top panel) and \texttt{objects} (bottom panel) contained in the \texttt{Source} and \texttt{Object} catalogs, respectively, separated according to band (see legend). The vertical lines indicate the 50th percentile for each band.}
  \label{fig:mag_hists}
\end{figure}

A cumulative histogram showing the PSF magnitudes of all \texttt{sources} contained within the \texttt{Source} catalogue is presented in the top panel of \figref{fig:mag_hists}

\subsubsection{Object Catalog}
The \texttt{Object} catalog \citep{10.71929/rubin/2570325} contains data on all objects detected with a greater than $5\sigma$ significance in the \texttt{deep\_coadd}s.
With coadd images produced on a per-band basis, a $>5\sigma$ detection in one or more of the bands will result in an object being included in the \texttt{Object} catalog.
For cases where an object is detected at $>5\sigma$ in more than one band, a cross-matching has been performed between bands to associate an object in one band with its counterpart(s) in the other bands.
As such, the \texttt{Object} catalog contains data from multiple bands.
The objects reported in the \texttt{Object} catalog have also undergone deblending; in the case of blended detections, only the deblended child objects are included in the catalog.
As with the \texttt{Source} catalog, the criterion for inclusion in the \texttt{Object} catalog is a $>5\sigma$ detection in one of the \texttt{deep\_coadd}s prior to deblending, yet the positions and fluxes of objects are reported post-deblending.
Hence, it is possible for \texttt{Object} catalog to contain \texttt{objects} whose \gls{flux}-to-error ratios --- potentially of all types and in all bands --- are less than $5$.

In addition to the general information mentioned above (i.e., IDs, positions, fluxes, flags), the \texttt{Object} catalog also includes basic \gls{shape} and extendedness information.
While they may be included in future data releases, no photometric redshifts, Petrosian magnitudes \citep{1976ApJ...209L...1P}, proper motions or periodicity information are included in the \gls{DP1} object catalogs.

The \texttt{Object} catalog contains data for \nobjects objects in \gls{DP1}.

\subsubsection{ForcedSource Catalog}
The \texttt{ForcedSource} catalog \citep{10.71929/rubin/2570327} contains forced \gls{PSF} photometry measurements performed on both \texttt{difference\_image}s (i.e., the \texttt{psfDiffFlux} column) and \texttt{visit\_image}s (i.e., the \texttt{psfFlux} column) at the positions of all the objects in the \texttt{Object} catalog, to allow assessment of the time variability of the fluxes.
We recommend using the \texttt{psfDiffFlux} column when generating light curves because this quantity is less sensitive to \gls{flux} from neighboring sources than \texttt{psfFlux}.
In addition to \gls{forced photometry} \gls{PSF} fluxes, a number of boolean flags are also included in the \texttt{ForcedSource} catalog.

The \texttt{ForcedSource} catalog contains a total of \nforcedsources entries across \nforcedobjects unique objects.

\subsubsection{DiaSource Catalog}
The \texttt{DiaSource} catalogs \citep{10.71929/rubin/2570317} contains data on all the sources detected at $>5\sigma$ significance --- including those associated with known Solar System objects --- in the \texttt{difference\_image}s.
Unlike sources detected in \texttt{visit\_image}s, sources detected in difference images (hereafter, ``DiaSource'') have gone through an association step in which an attempt has been made to associate them into underlying objects called ``DiaObject''.
The \texttt{DiaSource} catalog consolidates all this information across multiple visits and bands.
The detections reported in the \texttt{DiaSource} catalog have not undergone deblending.

The \texttt{DiaSource} catalog contains data for \ndiasources \texttt{DiaSources} in \gls{DP1}.

\subsubsection{DiaObject Catalog}
The \texttt{DiaObject} catalog \citep{10.71929/rubin/2570319} contains the astrophysical objects that DiaSources are associated with (i.e., the DiaObjects).
The \texttt{DiaObject} catalog contains only non-Solar System Objects; Solar System Objects are, instead, recorded in the \texttt{SSObject} catalog.
When a DiaSource is identified, the \texttt{DiaObject} and \texttt{SSObject} catalogs are searched for objects to associate it with.
If no association is found, a new DiaObject is created and the DiaSource is associated to it.
Along similar lines, an attempt has been made to associate DiaObjects across multiple bands, meaning the \texttt{DiaObject} catalog, like  the \texttt{Object} catalog, contains data from multiple bands.
Since DiaObjects are typically \gls{transient} or variable (by the nature of their means of detection), the \texttt{DiaObject} catalog contains summary statistics of their fluxes, such as the mean and standard deviation over multiple epochs; users must refer to the \texttt{ForcedSourceOnDiaObject} catalog (see below) or the \texttt{DiaSource} catalog for single \gls{epoch} \gls{flux} measurements of DiaObjects.

The \texttt{DIAObject} catalog contains data for \ndiaobjects DiaObjects in \gls{DP1}.

\subsubsection{ForcedSourceOnDiaObject Catalog}
The \texttt{ForcedSourceOnDiaObject} catalog \citep{10.71929/rubin/2570321} is equivalent to the \texttt{ForcedSource} catalog, but contains \gls{forced photometry} measurements obtained at the positions of all the DiaObjects in the \texttt{DiaObject} catalog.

The \texttt{ForcedSourceOnDiaObject} catalog contains a total of \ndiaforcedsources entries across \ndiaforcedobjects unique DiaObjects.

\subsection{SSObject Catalog}
The \texttt{SSObject} catalog \citep{10.71929/rubin/2570335} and the \gls{MPCORB} carry information about solar system objects.
The \gls{MPCORB} table provides the Minor Planet \gls{Center}-computed orbital elements for all known asteroids, including those that Rubin discovered.
For \gls{DP1}, the  \texttt{SSObject} catalog serves primarily to provide the mapping between the \gls{IAU} designation of an object (listed in \gls{MPCORB}), and the internal ssObjectId identifier, which is used as a key to find solar system object observations in the DiaSource and SSSource tables.
The \texttt{SSObject} catalog contains data for \nssobjects SSObjects in \gls{DP1}.

\subsubsection{SSSource Catalog}
The \texttt{SSSource} catalog \citep{10.71929/rubin/2570333} contains data on all DiaSources that are either associated with previously-known Solar System Objects, or have been confirmed as newly-discovered Solar System Objects by confirmation of their orbital properties.
As entries in the \texttt{SSSource} catalog stem from the \texttt{DiaSource} catalog, they have all been detected at $>5\sigma$ significance in at least one band.
The \texttt{SSSource} catalog contains data for \nsolarsystemsources Solar System Sources.

\subsubsection{CcdVisit Catalog}
The \texttt{CcdVisit} catalog \citep{10.71929/rubin/2570331} contains data for all \gls{CCD} images from a single visit. In princple, this means nine entries per visit, however due to a variety of technical reasons, not all \gls{CCD}s have data for each visit, and so the catalog may contain fewer than nine entries per visit.
In addition to technical information, such as the on-sky coordinates of the central pixel and measured pixel scale, the \texttt{CcdVisit} catalog contains a range of data quality measurements, such as whole-image summary statistics for the \gls{PSF} size, zeropoint, sky \gls{background}, sky noise, and quality of astrometric solution.
It provides an efficient method to access \texttt{visit\_image} properties without needing to access the image data.
When combined with the data contained in the \texttt{Visit} table described in \secref{ssec:metadata}, it provides a full picture of the telescope pointing and sky conditions at the time of observation.

The \texttt{CcdVisit} catalog contains \nvisitdetectorsummaries entries (nine entries for each of the \nvisitsummaries visits, minus three entries for one incomplete visit). This differs from the number of \texttt{visit\_image}s due to the more stringent requirements imposed to generate a science-ready image.

\subsubsection{Calibration Catalog}
\label{sssec:monster}
The \texttt{Calibration} catalog is the reference catalog that was used to perform astrometric and photometric \gls{calibration}.
It is a whole-sky catalog built specifically for \gls{LSST}, as no single prior reference catalog had both the depth and coverage needed to calibrate \gls{LSST} data.
It combines data from multiple previous reference catalogs and contains only stellar sources.
Full details on how the \texttt{Calibration} catalog was built are provided in \cite{DMTN-277}
\footnote{In \cite{DMTN-277},  the calibration reference catalog is referred to as ``The Monster". This terminology is also carried over to the \gls{DP1} Butler.}.
We provide a brief summary here.

For the \textit{grizy} bands, the input catalogs were (in order of decreasing priority): \gls{DES} Y6 Calibration Stars \citep{2023arXiv230501695R}; Gaia-\gls{XP} Synthetic Magnitudes \citep{2023A&A...674A..33G}; the \gls{Pan-STARRS}1 3PI Survey \citep{2016arXiv161205560C}; \gls{Data Release} 2 of the  SkyMapper survey \citep{2019PASA...36...33O}; and \gls{Data Release} 4 of the \gls{VST} \gls{ATLAS} survey \citep{2015MNRAS.451.4238S}.
For the \textit{u}-band, the input catalogs were (in order of decreasing priority): Standard Stars from \gls{SDSS} \gls{Data Release} 16 \citep{2020ApJS..249....3A}; Gaia-\gls{XP} Synthetic Magnitudes \citep{2023A&A...674A..33G}; and synthetic magnitudes generated using \gls{SLR}, which estimates the \textit{u}-band \gls{flux} from the \textit{g}-band \gls{flux} and \textit{g-r} colors.
This \gls{SLR} estimates were used to boost the number of \textit{u}-band reference sources, as otherwise the source density from the \textit{u}-band input catalogs is too low to be useful for the \gls{LSST}.

Only stellar sources were selected from each input catalog.
Throughout, the \texttt{Calibration} catalog uses the \gls{DES} bandpasses for the \textit{grizy} bands and the \gls{SDSS} bandpass for the \textit{u}-band; color transformations derived from high quality sources were used to convert fluxes from the various input catalogs (some of which did not use the \gls{DES}/SDSS bandpasses) to the respective bandpasses.
All sources from the input catalogs are matched to \textit{Gaia}-\gls{DR3} sources for robust astrometric information, selecting only isolated sources (i.e., no neighbors within 1\arcsec).

After collating the input catalogs and transforming the fluxes to the standard DES/SDSS bandpasses, the catalog was used to identify sources within a specific region of the sky.
This process generated a set of standard columns containing positional and flux information, along with their associated uncertainties.

\subsubsection{Source and Object Designations}\label{ssec:src_naming}
To refer to individual sources or objects from the \gls{DP1} catalogs, one should follow the LSST \gls{DP1} naming convention that has been registered with the International Astronomical Union.
Because the \texttt{Source}, \texttt{Object}, \texttt{DiaSource}, \texttt{DiaObject}, and \texttt{SSObject} tables each have their own unique IDs, their designations should differ.
In general, source and object designations should begin with the string ``LSST-\gls{DP1}'' (denoting the Legacy Survey of Space and Time, Data Preview 1), followed by a string specifying the table from which the source was obtained.
These strings should be ``O'' (for the \texttt{Object} table), ``S'' (\texttt{Source}), ``DO'' (\texttt{DiaObject}), ``DS'' (\texttt{DiaSource}), or ``SSO'' (\texttt{SSObject}).
Following the table identifier, the designation should contain the full unique numeric identifier from the specified table (i.e., the objectId, sourceId, diaObjectId, diaSourceId, or ssObjectId).
Each component of the identifier should be separated by dashes, resulting in a designation such as ``LSST-\gls{DP1}-TAB-123456789012345678''.
In summary, source designations should adhere to the formats listed below:

\begin{itemize}
\item Object: LSST-\gls{DP1}-O-609788942606161356 (for objectId 609788942606161356)
\item Source: LSST-\gls{DP1}-S-600408134082103129 (for sourceId 600408134082103129)
\item DiaObject: LSST-\gls{DP1}-DO-609788942606140532 (for diaObjectId 609788942606140532)
\item DiaSource: LSST-\gls{DP1}-DS-600359758253260853 (for diaSourceId 600359758253260853)
\item SSObject: LSST-\gls{DP1}-SSO-21163611375481943 (for ssObjectId 21163611375481943)
\end{itemize}

Tables that were not explicitly mentioned in the description above do not have their own unique IDs, but are instead linked to one of the five tables listed above via a unique ID.
For example, the \texttt{ForcedSource} table uses objectId, \texttt{ForcedSourceOnDiaObject} uses diaObjectId, \texttt{SSSource} uses diaSourceId and ssObjectId, and \texttt{MPCORB} uses ssObjectId.

\subsection{Maps}
Maps are two-dimensional visualizations of survey data.
In \gls{DP1}, these fall into two categories: Survey Property Maps and \gls{HiPS} Maps \citep{2015A&A...578A.114F}.

\subsubsection{Survey Property Maps}
\label{ssec:survey_property_maps}
Survey Property Maps \citep{10.71929/rubin/2570315} summarize how properties such as observing conditions or exposure time vary across the observed sky.
Each map provides the spatial distribution of a specific quantity at a defined sky position for each band by aggregating information from the images used to make the \texttt{deep\_coadd}.
Maps are initially created per-\gls{tract} and then combined to produce a final consolidated map.
At each sky location, represented by a spatial pixel in the \gls{HEALPix}\citep{2005ApJ...622..759G} grid, values are derived using statistical operations, such as minimum, maximum, mean, weighted mean, or sum, depending on the property.

\gls{DP1} contains \nsurveypropertymaps survey property maps.
The available maps describe total exposure times, observation epochs (one each for the earliest, mean, and latest observation epoch), \gls{PSF} size and \gls{shape} (one for each of the $e^1$ and $e^2$ shape parameters; see \secref{ssec:psf_models}), \gls{PSF} magnitude limits, sky \gls{background} and noise levels, as well as astrometric shifts (one each for right ascension and declination) and \gls{PSF} distortions (one for each of the $e^1$ and $e^2$ shape parameters) due to wavelength-dependent atmospheric \gls{DCR} effects.
They all use the dataset type  format \texttt{deep\_coadd\_<PROPERTY>\_consolidated\_map\_<STATISTIC>}.
For example, \texttt{deep\_coadd\_exposure\_time\_consolidated\_map\_sum} provides a spatial map of the total exposure time accumulated per
sky position in units of seconds.
All maps are stored in \texttt{HealSparse}\footnote{A sparse \gls{HEALPix} representation that efficiently encodes data values on the celestial sphere. \nolinkurl{https://healsparse.readthedocs.io}} format.
Survey property maps are only available via the Data \gls{Butler} (\secref{sssec:data_butler}).

\begin{figure*}[hbt!]
  \centering
  \begin{subfigure}[t]{0.31\textwidth}
  \includegraphics[width=\linewidth, height=5.8cm]{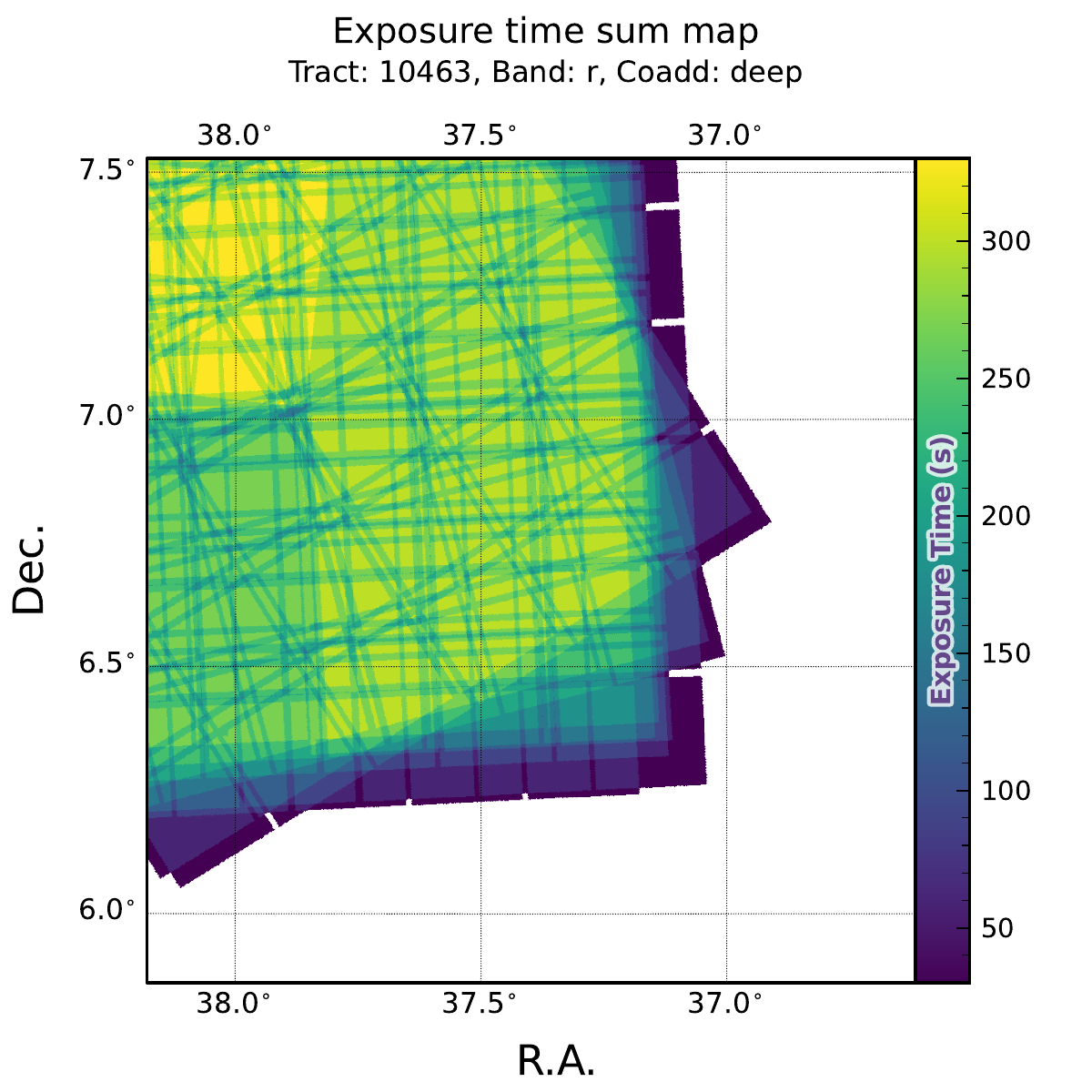}
  \caption{Exposure time sum map for \texttt{deep\_coadd} \gls{tract} 10463, $r$-band in field Rubin\_SV\_38\_7}
  \end{subfigure}\hfill
  \begin{subfigure}[t]{0.31\textwidth}
  \includegraphics[width=\linewidth, height=5.8cm]{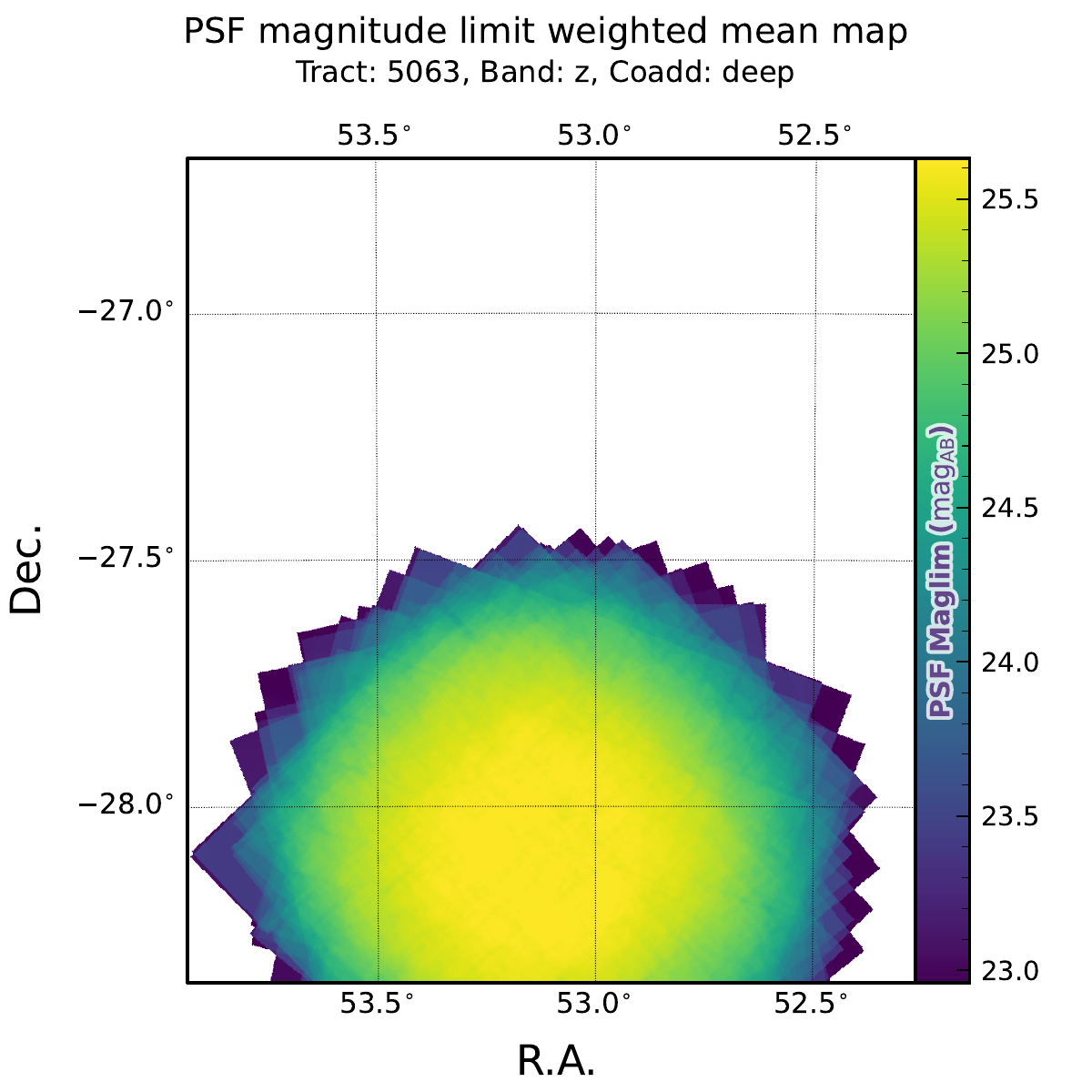}
  \caption{5$\sigma$ \gls{PSF} magnitude limit weighted mean map for \texttt{deep\_coadd} \gls{tract} 5063, $z$-band in field ECDFS}
  \end{subfigure}\hfill
    \begin{subfigure}[t]{0.31\textwidth}
  \includegraphics[width=\linewidth, height=5.8cm]{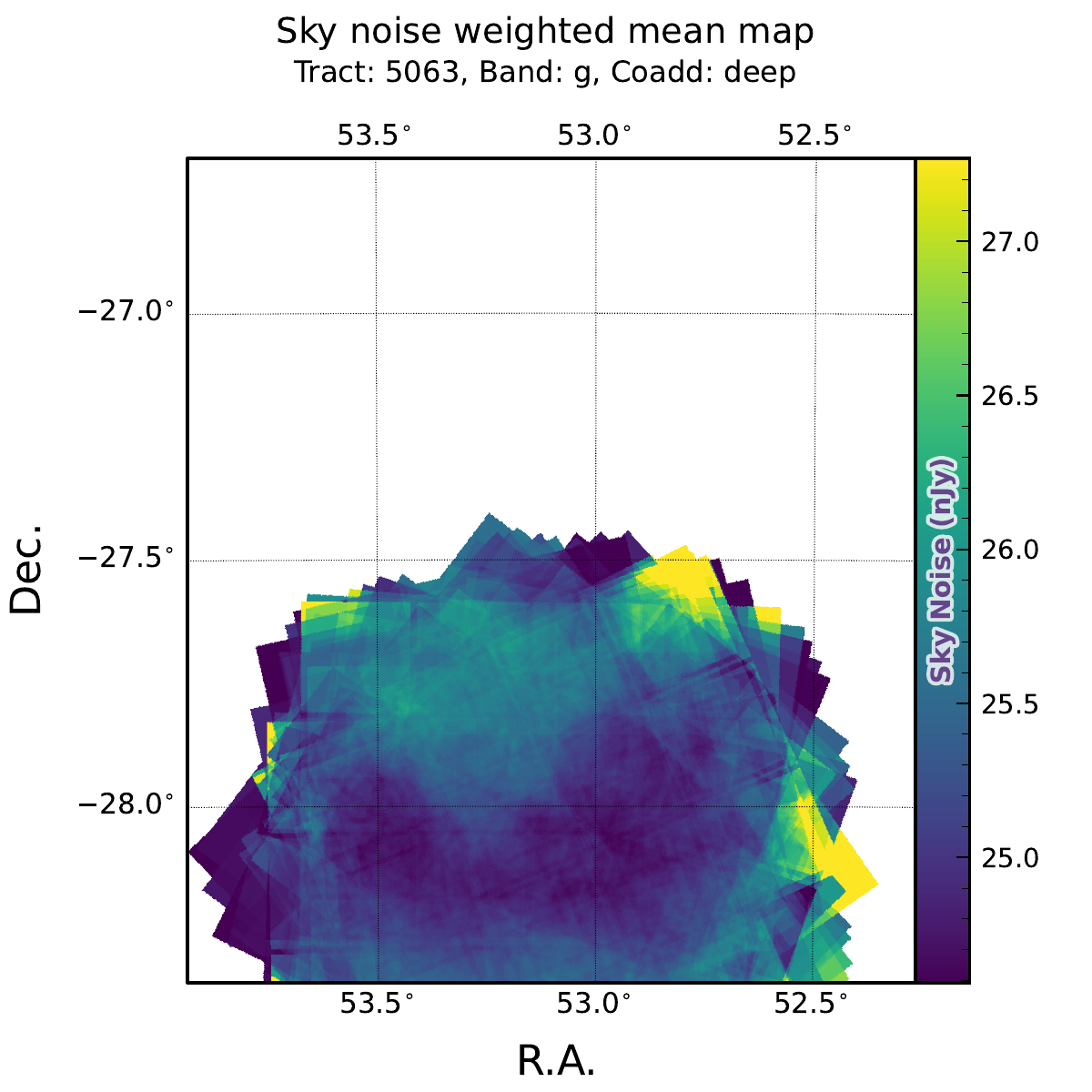}
  \caption{Sky noise weighted mean map for \texttt{deep\_coadd} \gls{tract} 5063, $z$-band in field ECDFS}
  \end{subfigure}\hfill
\caption{Examples of survey property maps from Rubin \gls{DP1} across different bands, clipped to the boundary of a single \gls{tract} for visual clarity.}
  \label{fig:survey_property_maps}
\end{figure*}

Figure \ref{fig:survey_property_maps} presents three survey property maps for exposure time, \gls{PSF} magnitude limit, and sky noise, computed for representative tracts and bands.
Because full consolidated maps cover widely separated tracts, we use clipped per-\gls{tract} views here to make the spatial patterns more discernible.

\subsubsection{HiPS Maps}
\gls{HiPS} Maps \citep{2015A&A...578A.114F}, offer an interactive way to explore seamless, multi-band tiles of the sky regions covered by \gls{DP1}, allowing for smooth panning and zooming.
\gls{DP1} provides multi-band \gls{HiPS} images created by combining data from individual bands of \texttt{deep\_coadd} and \texttt{template\_coadd} images, using an improved version (Lust et al. in prep) of the algorithm presented in \cite{2004PASP..116..133L}.
These images are false-color representations generated using various filter combinations for the red, green, and blue channels.

The available filter combinations include \textit{gri}, \textit{izy}, \textit{riz}, and \textit{ugr} for both \texttt{deep\_coadd} and \texttt{template\_coadd}.
Additionally, for \texttt{deep\_coadd} only, we provide color blends such as \textit{uug} and \textit{grz}.
Post-\gls{DP1}, we plan to also provide single-band HiPS images for all $ugrizy$ bands in both \gls{PNG} and \gls{FITS} formats.

\gls{HiPS} maps are only accessible through the \gls{HiPS} viewer in the \gls{RSP} Portal (\secref{ssec:rsp_portal}) and cannot be accessed via the Data \gls{Butler} (\secref{sssec:data_butler}).
All multi-band \gls{HiPS} images are provided in \gls{PNG} format.

\subsection{Metadata}
\label{ssec:metadata}
\gls{DP1} also includes \gls{metadata} about the observations, which are stored in the \texttt{Visit} table.
We distinguish it from a catalog as the data it contains was produced by the observatory directly, rather than the science pipelines.
The \texttt{Visit} table contains technical data for each visit, such as telescope pointing, camera rotation, \gls{airmass}, exposure start and end time, and total exposure time.
Some of the information contained within the \texttt{Visit} table is also contained in the \texttt{CCDVisit} catalogue described in \secref{ssec:catalogs} (e.g., exposure time), although the latter also includes information produced by the processing pipelines at a per-detector level, such as the PSF size and limiting magnitudes of a given \texttt{visit\_image}.

\subsection{Ancillary Data Products}
\label{ssec:ancilliary}
\gls{DP1} also includes several ancillary data products.
While we do not expect most users to need these, we describe them here for completeness.
All the Data Products described in this section can only be accessed via the Data Butler (\secref{sssec:data_butler}).

\subsubsection{Standard Bandpasses}
\label{sssec:transmission_curves}
\figref{fig:comcam_standard_bandpasses} shows the full-system throughput of the six \gls{LSSTComCam} filters.
The corresponding transmission curves are provided as a \gls{DP1} data product.
These datasets tabulate the full-system transmission of the six \gls{LSSTComCam} filters as a function of wavelength and were used as a reference for the \gls{LSSTComCam} \gls{DP1} photometry. The \texttt{standard\_passband} dataset is keyed by band and is stored in Astropy Table format.

\subsubsection{Task configuration, log, and metadata}
\gls{DP1} includes \gls{provenance}-related data products such as task logs, \gls{configuration} files, and task metadata.
Configuration files record the parameters used in each processing task, while logs and \gls{metadata} contain information output during processing.
These products help users understand the processing setup and investigate potential processing failures.

\subsubsection{Calibration Data Products}
\label{ssec:calibration_data}
Calibration data products include a variety of images and models that are used to characterize and correct the performance of the camera and other system components.
These include bias, dark, and flat-field images, \gls{PTC} gains, brighter-fatter kernels \citep{2014JInst...9C3048A}, charge transfer inefficiency (\gls{CTI}) models, linearizers, and illumination corrections.
For flat-field corrections, \gls{DP1} processing used combined flats, which are averaged from multiple individual flat-field exposures to provide a stable \gls{calibration}. These \gls{calibration} products are essential inputs to \gls{ISR} (\secref{ssec:isr}). While these products are included in \gls{DP1} for transparency and completeness, users should not need to rerun ISR for their science and are advised to start with the processed \texttt{visit\_image}.
 \section{Data Release Processing}
\label{sec:drp}

\gls{DRP} is the systematic processing of all Rubin Observatory data collected up to a certain date to produce the calibrated images, catalogs of detections, and derived data products described in Section \ref{sec:data_products}.
\gls{DP1} was processed entirely at the \gls{USDF} at SLAC using 17,024 CPU hours.\footnote{For future Data Releases, data processing will be distributed across the USDF, the French (FrDF)and UK (UKDF) data facilities.}

This section describes the pipeline algorithms used to produce \gls{DP1} and how they differ from those planned for full-scale LSST data releases.
Data Release Production consists of four major stages: (1) single-frame processing, (2) calibration, (3) coaddition, and (4) difference image analysis (DIA).

\subsection{LSST Science Pipelines Software}
\label{ssec:pipelines}
The  LSST Science Pipelines software \citep{PSTN-019, LDM-151} will be used to generate all Rubin Observatory and LSST data products.
They provide both the \gls{algorithm} and \gls{middleware} frameworks necessary to process raw data into science-ready data products, enabling analysis by the Rubin scientific community.
Version \sciencepipelinesversion of the pipelines was used to produce \gls{DP1}\footnote{Documentation for this version is available at \sciencepipelinesurl}.

\subsection{Single Frame Processing}
\label{ssec:single_frame_processing}

\subsubsection{Instrument Signature Removal}
\label{ssec:isr}
\begin{figure}[htb!]
  \centering
  \includegraphics[width=\linewidth]{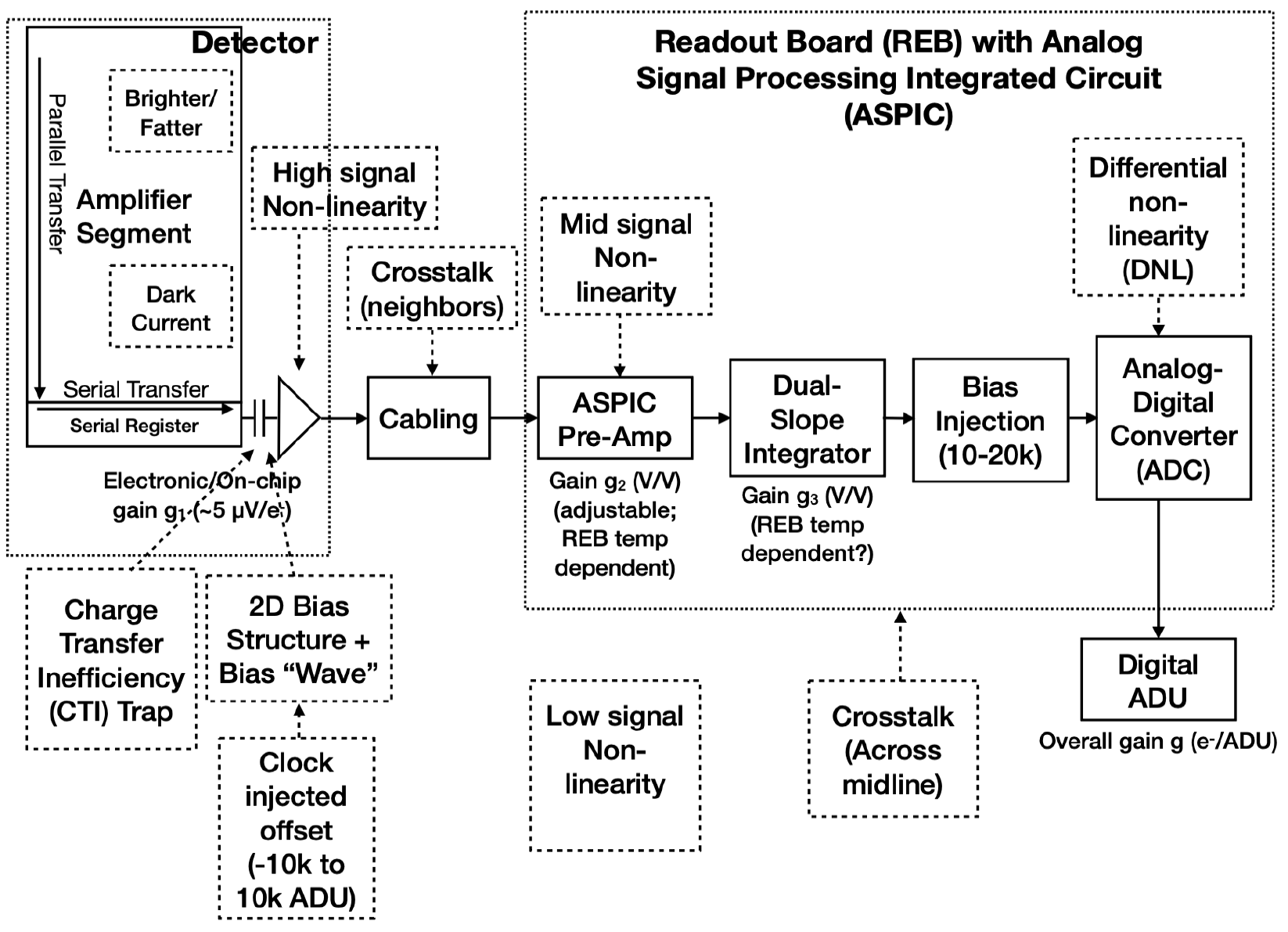}
  \caption{The model of the detector and REB components, labeled with the effects that they impart on signal.}
  \label{fig:isr_signal_chain}
\end{figure}

The first step in processing \gls{LSSTComCam} images is to correct for the effects introduced by the telescope and detector.
Each sensor and its readout amplifiers can vary slightly in performance, causing images of even a uniformly illuminated focal plane to exhibit discontinuities and shifts due to detector effects.
The \gls{ISR} pipeline aims to recover the original astrophysical signal as best as possible and produce science-ready single-epoch images for source detection and measurement.
A detailed description of the \gls{ISR} procedures can be found in \citet{SITCOMTN-086} and \citet{2025JATIS..11a1209P}.
\figref{fig:isr_signal_chain} illustrates the model of detector components and readout electronics and their impact on the signal, tracing the process from photons incident on the detector surface to the final quantized values\footnote{The images written to disk by the camera have values that are integers that come from the ADC converting an analog voltage.} recorded in the image files.
The \gls{ISR} \gls{pipeline} essentially ``works backward'' through the signal chain, correcting the integer analog-to-digital units (ADU) raw camera output back to a floating-point number of photoelectrons created in the silicon.
The physical detector, represented on the left in  \figref{fig:isr_signal_chain}, is the source of effects that arise from the silicon itself, such as the dark current and the brighter-fatter effect \citep{doi:10.1088/1538-3873/aab820,2024PASP..136d5003B}.
After the integration time has elapsed, the charge is shifted  to the serial register and read out, which can introduce charge transfer inefficiencies and a clock-injected offset level.
The signals for all amplifiers are transferred via cables to the \gls{REB}, during which crosstalk between the amplifiers may occur.
The \gls{ASPIC} on the \gls{REB} converts the analog signal from the detector into a digital signal, adding both quantization and a bias level to the image.
Although the signal chain is designed to be stable and linear, the presence of numerous sources of non-linearity indicates otherwise.

The \gls{ISR} processing pipeline for \gls{DP1} performs, in the following order: \gls{ADU} dithering to reduce quantization effects, serial overscan subtraction, saturation masking, gain normalization, crosstalk correction, parallel overscan subtraction, linearity correction, serial \gls{CTI} correction, image assembly, bias subtraction, dark subtraction, brighter-fatter correction, defect masking and interpolation, variance plane construction, flat fielding, and amplifier offset (amp-offset) correction\footnote{Amp-offset corrections are designed to address systematic discontinuities in background sky levels across amplifier boundaries. The implementation in the LSST Science Pipelines is based on the \texttt{Pan-STARRS} Pattern Continuity algorithm \citep{2020ApJS..251....4W}.}.
Flat fielding for \gls{DP1} was performed using combined flats produced from twilight flats acquired with sufficient rotational dithering to mitigate artifacts from print-through stars, as described in \secref{ssec:flat_field_system}.

\subsubsection{Background Subtraction}
\label{ssec:background_subtraction}

The background subtraction algorithms in the LSST Science Pipelines estimate and remove large-scale background signals from science imaging.
Such signals may include sky brightness from airglow, moonlight, scattered light instrumental effects, zodiacal light, and diffuse astrophysical emission.
In so doing, true astrophysical sources are isolated to allow for accurate detection and measurement.

To generate a \gls{background} model, each post-ISR image is divided into superpixels of $128\times128$ pixels.
Pixels with a mask flag set that indicates that they contain no useful science data or that they contain \gls{flux} from a preliminary source detection are masked.
The iterative $3\sigma$ clipped mean of the remaining pixels is calculated for each superpixel, constructing a \gls{background} statistics image.
A sixth-order Chebyshev polynomial is fit to these values on the scale of a single detector to allow for an extrapolation back to the native pixel resolution of the post-\gls{ISR} image.

\subsection{Calibration}
\label{ssec:drp_calibration}
Stars are detected in each post-\gls{ISR} image using a $5\sigma$ threshold.
Detections of the same star across multiple images are then associated to identify a consistent set of isolated stars with repeated observations suitable for use in PSF modeling, photometric \gls{calibration}, and astrometric \gls{calibration}.

Initial astrometric and photometric solutions are derived using only the calibration reference catalogs (see \secref{ssec:catalogs}), and an initial \gls{PSF} model is fit using PSFEx \citep{2011ASPC..442..435B}.
These preliminary solutions provide approximate source positions, fluxes, and \gls{PSF} shapes that serve as essential inputs to the \gls{calibration} process, enabling reliable source matching, selection of high-quality stars, and iterative refinement of the final astrometric, photometric, and \gls{PSF} models.
These preliminary solutions are subsequently replaced by more accurate fits, as described in the following sections.

\subsubsection{PSF Modeling}
\label{ssec:psf_modelling}
\gls{PSF} modeling in \gls{DP1} uses the Piff \citep{DES:2020vau} package.
Our configuration of Piff utilizes its \texttt{PixelGrid} model with a fourth-order polynomial interpolation per \gls{CCD}, except in the $u$-band, where star counts are insufficient to support a fourth-order fit.
In this case, a second-order polynomial is used instead.
Details on the choice of polynomial order, overall \gls{PSF} modeling performance, and known issues are discussed in \secref{ssec:psf_models}.

\subsubsection{Astrometric Calibration}
\label{ssec:global_astrometric_calibration}
Starting from the astrometric solution calculated in single frame processing (\secref{ssec:single_frame_processing}), the final astrometric solution is computed using the ensemble of visits in a given band that overlap a given \gls{tract}.
This allows the astrometric solution to be further refined by using all of the isolated point sources of sufficient signal-to-noise ratio in an image, rather than only those that appear in the reference catalog, as is done in single frame processing.
Using multiple whole visits rather than a single detector also allows us to account for effects that impact the full focal plane, and for the proper motion and parallax of the sources.

In order to perform the fit of the astrometric solution, isolated point sources are associated between overlapping visits and with the Gaia \gls{DR3} \citep{2023A&A...674A...1G} reference catalog where possible.
The model used for \gls{DP1} consists of a static map from pixel space to an intermediate frame (the per-detector model), followed by a per-visit map from the intermediate frame to the plane tangent to the telescope boresight (the per-visit model), then finally a deterministic mapping from the tangent plane to the sky.
The fit is done using the \texttt{gbdes} package \citep{2017PASP..129g4503B}, and a full description is given in \citet{DMTN-266}.

The per-detector model is intended to capture quasi-static characteristics of the telescope and camera.
During \gls{Rubin Operations}, the astrometric solution will allow for separate epochs with different per-detector models, to account for changes in the camera due to warming and cooling and other discrete events.
However, for \gls{DP1}, \gls{LSSTComCam} was assumed to be stable enough that all visits use the same per-detector model. The model itself is a separate two-dimensional polynomial for each detector.
For \gls{DP1}, a degree 4 polynomial was used; the degree of the polynomial mapping is tuned for each instrument and may be different for LSSTCam.
Further improvements may be made by including a pixel-based astrometric offset mapping, which would be fit from the ensemble of astrometric residuals, but this is not included in the \gls{DP1} processing.

The per-visit model attempts to account for the path of a photon from both atmospheric sources and those dependent on the telescope orientation.
This model is also a polynomial mapping, in this case a degree 6 two-dimensional polynomial.
Correction for \gls{DCR} (\secref{sec:differential_chromatic_refraction}) was not done for \gls{DP1}, but will be included in \gls{LSSTCam} processing during \gls{Rubin Operations}.
Future processing will also likely include a Gaussian Process fit to better account for atmospheric turbulence, as was demonstrated by \citet{2021AJ....162..106F} and \citet{2021A&A...650A..81L}.

The final component of the astrometric \gls{calibration} involves the positions of the isolated point sources included in the fit, which are described by five parameters: sky coordinates, proper motion, and parallax.
While proper motions and parallaxes are not released for \gls{DP1}, they are fitted for these sources in the astrometric solution to improve the astrometric calibration.

\subsubsection{Photometric Calibration}
\label{photometric_calibration}
Photometric \gls{calibration} of the \gls{DP1} dataset is based on the \gls{FGCM} ~\citep{2018AJ....155...41B}, adapted for the LSST Science Pipelines~\citep{2022PASJ...74..247A, SITCOMTN-086}.
We used the \gls{FGCM} to calibrate the full \gls{DP1} dataset with a forward model that uses a parameterized model of the atmosphere as a function of airmass along with a model of the instrument throughput as a function of wavelength.
The FGCM process typically begins with measurements of the instrumental throughput, including the mirrors, filters, and detectors.
However, because full scans of the \gls{LSSTComCam} as-built filters and individual detectors were not available, we instead used the nominal reference throughputs for the Simonyi Survey Telescope and LSSTCam.\footnote{Available at: \nolinkurl{https://github.com/lsst/throughputs/tree/1.9}}
These nominal throughputs were sufficient for the \gls{DP1} calibration, given the small and homogeneous focal plane consisting of only nine \gls{ITL} detectors.
The FGCM atmosphere model, provided by MODTRAN~\citep{1999SPIE.3756..348B}, was used to generate a look-up table for atmospheric throughput as a function of zenith distance at Cerro Pachón.
This model accounts for absorption and scattering by molecular constituents of the atmosphere, including $O_2$ and $O_3$; absorption by water vapor; and Mie scattering by airborne aerosol particulates.
Nightly variations in the atmosphere are modeled by minimizing the variance in repeated observations of stars with a \gls{SNR} greater than 10, measured using ``compensated aperture fluxes''.
These fluxes include a local \gls{background} subtraction (see \secref{ssec:background_subtraction}) to mitigate the impact of \gls{background} offsets.
The model fitting process incorporates all six bands ($ugrizy$) but does not include any gray (achromatic) terms, except for a linear assumption of mirror reflectance degradation, which is minimal over the short duration of the \gls{DP1} observation campaign.
As an additional constraint on the fit, we use a subset of stars from the reference catalog~\citep{DMTN-277}, primarily to constrain the system's overall throughput and establish the ``absolute'' calibration.

Photometric transformation relations between \gls{LSSTCam} and \gls{LSSTComCam} systems and other photometric systems are under development and are provided  in \citep{RTN-099}

\subsection{Visit Images and Source Catalogs}
\label{sssec:visit_images_source_catalogs}
With the final \gls{PSF} models, \gls{WCS} solutions, and photometric calibrations in place, we reprocess each single-epoch image to produce a final set of calibrated visit images and source catalogs.
Source detection is performed down to a $5\sigma$ threshold using the updated \gls{PSF} models, followed by measurement of \gls{PSF} and aperture fluxes.
These catalogs represent the best single-\gls{epoch} source characterization, but they are not intended for constructing light curves.
For time-domain analysis, we recommend using the \gls{forced photometry} tables described in \secref{sssec:lightcurves}.

\subsection{Coaddition Processing}
\label{ssec:coadd_processing}
\subsubsection{Coaddition}
\label{ssec:coaddition}
Only exposures with a \gls{seeing} better than 1.7 arcseconds FWHM are included in the deep coadded images. For the template coadds, typically only the top third of visits with the best \gls{seeing} are used (although see \secref{ssec:science_images} for more details), resulting in an even tighter image quality cutoff for the template coadds. Exposures with poor \gls{PSF} model quality, identified using internal diagnostics, are excluded to prevent contamination of the coadds with unreliable \gls{PSF} estimates. The remaining exposures are combined using an inverse-variance weighted mean stacking \gls{algorithm}.

To mitigate transient artifacts before coaddition, we apply the artifact rejection procedure described in \cite{DMTN-080} that identifies and masks features such as satellite trails, optical ghosts, and cosmic rays.
It operates on a time series of \gls{PSF}-matched images resampled onto a common pixel grid (``warps'') and leverages their temporal behavior to distinguish persistent astrophysical sources from transient artifacts.

Artifact rejection uses both direct (where no PSF-matching is performed) and PSF-matched warps, homogenized to a standard PSF of 1.8 arcseconds FWHM, broadly consistent with the 1.7 arcsecond FWHM \gls{seeing} threshold used in data screening.
A sigma-clipped mean of the \gls{PSF}-matched warps serves as a static sky model, against which individual warps are differenced to identify significant positive and negative residuals.
Candidate artifact regions are classified as \gls{transient} if they appear in less than a small percentage of the total number of exposures, with the threshold  based on the number of visits, $N$,  as follows:
\begin{itemize}
    \item $N=1$ or $2$: threshold $= 0$ (no clipping).
    \item $N=3$ or $4$: threshold $= 1$.
    \item $N=5$: threshold $= 2$.
    \item $N>5$: threshold $= 2+0.03N$.
\end{itemize}
Identified \gls{transient} regions are masked before coaddition, improving image quality and reducing contamination in derived catalogs.

\subsubsection{Detection, Deblending and Measurement}
\label{sssec:coadd_processing}
After constructing coadded images, sources are detected in each band, merged across bands, deblended, and measured to generate the final object catalogs  (\secref{ssec:catalogs}).
For each coadd in all six bands, we perform source detection at a $5\sigma$ detection threshold and then adjust the background with a per-patch constant (coadds are built from background-subtracted images, but the deeper detection on coadds redefines what is considered source versus background).
Detections across bands are merged in a fixed priority order, $irzygu$, to form a union detection catalog, which serves as input to deblending.

Deblending is performed using the Scarlet Lite algorithm, which implements the same model as Scarlet \citep{2018A&C....24..129M}, but operates on a single pixel grid.
This allows the use of analytic gradients, resulting in greater computational speed and memory efficiency.

\gls{Object} measurement is then performed on the deblended detection footprints in each band.
Measurements are conducted in three modes: independent per-band measurements, forced measurements in each band, and multiband measurements.

Most measurement algorithms operate through a single-band plugin system, largely as originally described in \citet{2018PASJ...70S...5B}.
The same plugins are run separately for each object on a deblended image, which uses the Scarlet model as a template to re-weight the original noisy coadded pixel values.
This effectively preserves the original image in regions where objects are not blended, while dampening the noise elsewhere.

A reference band is chosen for each object based on detection significance and measurement quality using the same priority order as detection merging ($irzygu$) and a second round of measurements is performed in forced mode using the shape and position from the reference band to ensure consistent colors \citep{2018PASJ...70S...5B}.

Measurement \gls{algorithm} outputs include object fluxes, centroids, and higher-order moments thereof like sizes and shapes. A variety of \gls{flux} measurements are provided, from aperture fluxes and forward modeling algorithms.

Composite model (CModel) magnitudes \citep{2004AJ....128..502A, 2018PASJ...70S...5B} are used to calculate the extendedness parameter, which functions as a star-galaxy classifier.
Extendedness is a binary classifier that is set to 1 if the PSF model flux is less than $98.5\%$ of the (free, not forced) CModel flux in a given band.
Additionally, the extendedness in the reference band is provided as a separate column for convenience as a multiband star-galaxy classification, and is recommended generally but also specifically for objects with low signal-to-noise ratio in some bands.

Gaussian-Aperture-and-PSF \citep[\gls{GAaP}][]{2008A&A...482.1053K, DMTN-190} fluxes are provided to ensure consistent galaxy colors across bands.
S\'ersic model \citep{1963BAAA....6...41S, 1968adga.book.....S} fits are run on all available bands simultaneously \cite[MultiProFit,][]{DMTN-312}.
The resulting S\'ersic model fluxes are provided as an alternative to CModel and are intended to represent total galaxy fluxes.
Like CModel, the S\'ersic model is a Gaussian mixture approximation to a true S\'ersic profile, convolved with a Gaussian mixture approximation to the \gls{PSF}.
S\'ersic model fits also include a free centroid, with all other structural parameters shared across all bands.
That is, the intrinsic model has no color gradients, but the convolved model may have color gradients if the \gls{PSF} parameters vary significantly between bands.

CModel measurements use a double ``shapelet'' \citep{2003ARA&A..41..645R} PSF model with a single shared shape.
The S\'ersic fits are intended to use a double Gaussian with independent shape parameters for each component.
Due to a pipeline misconfiguration, the S\'ersic fits actually used the shapelet PSF parameters, with the higher-order terms ignored (since MultiProFit does not support shapelet PSFs).
This bug is not expected to impact the galaxy fluxes significantly, since the higher-order shapelet PSF parameters tend to be small, and the fix will be applied in future campaigns.
Either way, the double Gaussian PSF parameters are included for each object.

Further details on the performance of these algorithms are found in \secref{ssec:fluxes}.

\subsection{Variability Measurement}
\subsubsection{Difference Imaging Analysis}
\label{ssec:diffim_analysis}
Difference Image Analysis (DIA) uses the decorrelated Alard \& Lupton image differencing algorithm \citep{DMTN-021}.
We detected both positive and negative \texttt{DIASource}s at $5\sigma$ in the difference image.
Sources with footprints containing both positive and negative peaks due to offsets from the template position or blending were fit with a dipole centroid code, which simultaneously fits offset positive and negative PSFs.
We filter the resulting \texttt{DIASource} catalog to remove detections with pixel flags indicative of artifacts, non-astrophysical trail lengths, or unphysically negative direct fluxes.
Finally, we perform a simple spatial association of \texttt{DIASource}s into \texttt{DIAObject}s using a one-arcsecond matching radius.

The Machine Learning reliability model applied to \gls{DP1} was developed with the aim to meet the latency requirements for Rubin \gls{Alert Production} when executed on CPUs.
Accordingly we developed a relatively simple model: a Convolutional Neural Network with three convolutional layers, and two fully connected layers.
The convolutional layers have a $5\times5$ kernel size, with 16, 32, and 64 filters, respectively.
A max-pooling layer of size 2 is applied at the end of each convolutional layer, followed by a dropout layer of 0.4 to reduce overfitting.
The last fully connected layers have sizes of 32 and 1.
The ReLU activation function is used for the convolutional layers and the first fully connected layer, while a sigmoid function is used for the output layer to provide a probabilistic interpretation.
The cutouts are generated by extracting postage stamps of $51\times51$ pixels centered on the detected sources.
The input data of the model consist of the template, science, and difference image stacked to have an array of \gls{shape} (3, 51, 51).
The model is implemented using PyTorch \citep{10.1145/3620665.3640366}.
The Binary Cross Entropy loss function was used, along with the \gls{Adam} optimizer with a fixed learning rate of $1\times10^{-4}$, weight decay of $3.6\times10^{-2}$, and a batch size of 128.
The final model uses the weights that achieved the best precision/purity for the test set.
Training was done on the \gls{S3DF} with an NVIDIA model L40S GPU.

The model was initially trained using simulated data from the second DESC Data Challenge (DC2; \citep{2021ApJS..253...31L}) plus randomly located injections of PSFs to increase the number of real sources, for a total of 89,066 real sources.
The same number of bogus sources were selected at random from non-injected DIASources.
Once the  \gls{LSSTComCam} data were available, the model was fine-tuned on a subset of the data containing 183,046 sources with PSF injections.
On the \gls{LSSTComCam} test set, the model achieved an accuracy of 98.06\%, purity of 97.87\%, and completeness of 98.27\%.
As discussed in \secref{ssec:performance_dia}, the injections used to train this model version do not capture all types of astrophysical variability, so performance on the test set will not be representative for variable stars, comets, and other types of variable objects.
The machine-learning reliability score, reported in the \texttt{reliability} column of the \texttt{DIASource} catalog, is a scalar value between 0 and 1 that quantifies the model’s confidence that a given detection is astrophysical.

\subsubsection{Light Curves}
\label{sssec:lightcurves}
To produce light curves, we perform multi-epoch \gls{forced photometry} on both the direct visit images and the difference images.
For light curves we recommend the \gls{forced photometry} on the difference images (\texttt{psfDiffFlux} on the ForcedSource Table), as it isolates the variable component of the flux and avoids contamination from static sources.
In contrast, \gls{forced photometry} on direct images includes flux from nearby or blended static objects, and this contamination can vary with seeing.
Centroids used in the multi-epoch \gls{forced photometry} stage are taken either from object positions measured on the coadds or from the DIAObjects (the associated DIASources detected on difference images).

\subsubsection{Solar System Processing}
\label{sec:drp:solsys}

Solar system processing in \gls{DP1} consists of two key components: the association of observations (sources) with known solar system objects, and the discovery of previously unknown objects by linking sets of
{\em tracklets}\footnote{A tracklet is defined as two or more detections of a moving object candidate taken in close succession in a single night.}.

The association component begins by generating expected positions for all objects in the Minor Planet Center orbit catalog, using ephemerides computed with the \texttt{Sorcha} survey simulation toolkit (Merritt et al., in press)\footnote{Available at \nolinkurl{https://github.com/dirac-institute/sorcha}}.
To enable fast lookup of objects potentially present in an observed visit, we use the \texttt{mpsky} package \citep{mpsky}.
In each image, the closest \texttt{DiaSource} within 1~arcsecond of a known solar system object's predicted position is associated to that object.
In \gls{DP1} we used a simple positional association to tag DiaSources that are likely observations of known asteroids.
The 1~arcsecond radius is intentionally generous;  we did not see evidence of mismatches at DP1 depth and volume.
This radius will be tuned for future processing campaigns.

The discovery component of Solar System processing uses the \texttt{heliolinx} package\footnote{\nolinkurl{https://github.com/heliolinx/heliolinx}}, which provides tools for asteroid identification and linking \citep{heliolinx}.
The repository contains code for the following tasks:
\begin{itemize}
    \item Tracklet creation with \texttt{make\_tracklets}
    \item Multi-night \gls{tracklet} linking with an algorithm
    \item Linkage post processing (orbit fitting, outlier rejection, and de-duplication) with \texttt{link\_purify}
\end{itemize}

The inputs to the discovery processing comprised all sources detected in difference images, regardless of whether they were tagged in the association step.
These inputs were produced by an early processing of \gls{LSSTComCam} commissioning data, some of which were later rejected during \gls{DP1} processing and therefore do not appear in the final \gls{DP1} data products.

About 10\% of all commissioning visits targeted the near-ecliptic field Rubin\_SV\_38\_7, chosen to facilitate asteroid discovery.
Rubin\_SV\_38\_7 produced the vast majority of asteroid discoveries in \gls{DP1}, as expected, but a few were found in off-ecliptic fields as well.

Tracklet creation with \texttt{make\_tracklets} used an upper limit angular velocity of 1.5 \gls{deg}/day, faster than any main belt asteroid and in the range of many \gls{NEO} discoveries.
While no formal minimum angular velocity was imposed, in practice it would be unlikely to detect objects moving slower  than about 0.01~deg~day$^{-1}$.
To minimize false \gls{tracklet}s from fields observed multiple times per night, the minimum \gls{tracklet} length was set to three detections, and a minimum on-sky motion of five arcseconds was required for a valid \gls{tracklet}.
To claim a discovery candidate, we required tracklets to be linked across at least three nights.

Multi-night \gls{tracklet} linking is the heart of Solar system discovery,  which connects (``links") tracklets belonging to the same object over a series of nights.
It employs the HelioLinC3D algorithm \citep{2020DPS....5221101E,2022DPS....5450404H}, a refinement of the original HelioLinC algorithm of \citet{2018AJ....156..135H}.
Each processing run tested each \gls{tracklet} with 324 different hypotheses spanning heliocentric distances from 1.5 to 9.8 \gls{au} and radial velocities spanning the full range of possible bound orbits (eccentricity 0.0 to nearly 1.0).
The upper limit of 10~\gls{au} was chosen because searches targeting more distant populations require different parameter choices.
This range of distance encompasses all main belt asteroids and Jupiter Trojans, as well as many comets and Mars-crossers and some \glspl{NEO}.
A dedicated search for objects at heliocentric distances out to 50~\gls{au} was also conducted; no distant objects were detected, consistent with expectations for the size of the \gls{DP1} data set.
Smaller heliocentric distances were not attempted here because nearby objects move rapidly across the sky and hence were not likely to remain long enough in an \gls{LSSTComCam} field to be discovered.

Candidate linkages, defined as groups of tracklets whose propagated orbits cluster within a radius of 1.33 $\times\,10^{3}$\,au at 1\,au,
are identified, then post-processed via \texttt{link\_purify} to yield a final, non‐overlapping set of high-confidence asteroid candidates, ranked by orbit-fit residuals and related metrics.
While heliolinx can produce false-positive or redundant raw linkages by design, these are filtered during post-processing by \texttt{link\_purify}, which applies a Rubin-specific, more stringent version of the MPC validation rules\footnote{\nolinkurl{https://minorplanetcenter.net/mpcops/documentation/identifications/additional/}}.
This step both rejects spurious linkages and deduplicates multiple hypotheses for the same object, ensuring that only the highest-quality, non-redundant linkages are carried forward for orbit determination and for distinguishing new discoveries from rediscoveries of known objects.
 \section{Performance Characterization and Known Issues}
\label{sec:performance}
In this section, we provide an assessment of the \gls{DP1} data quality and describe known issues.

\subsection{Sensor Anomalies and ISR}
\label{ssec:sensor_anomalies}
In addition to the known detector features identified before \gls{LSSTComCam} commissioning, most of which are handled by the ISR processing (see \secref{ssec:isr}), we discovered a number of new types of anomalies in the \gls{DP1} data. 
Since no corrections are currently available for these anomalies, they are masked and excluded from downstream data products.

\subsubsection{Vampire Pixels}
\begin{figure}[htb!]
  \centering
  \includegraphics[width=0.98\linewidth]{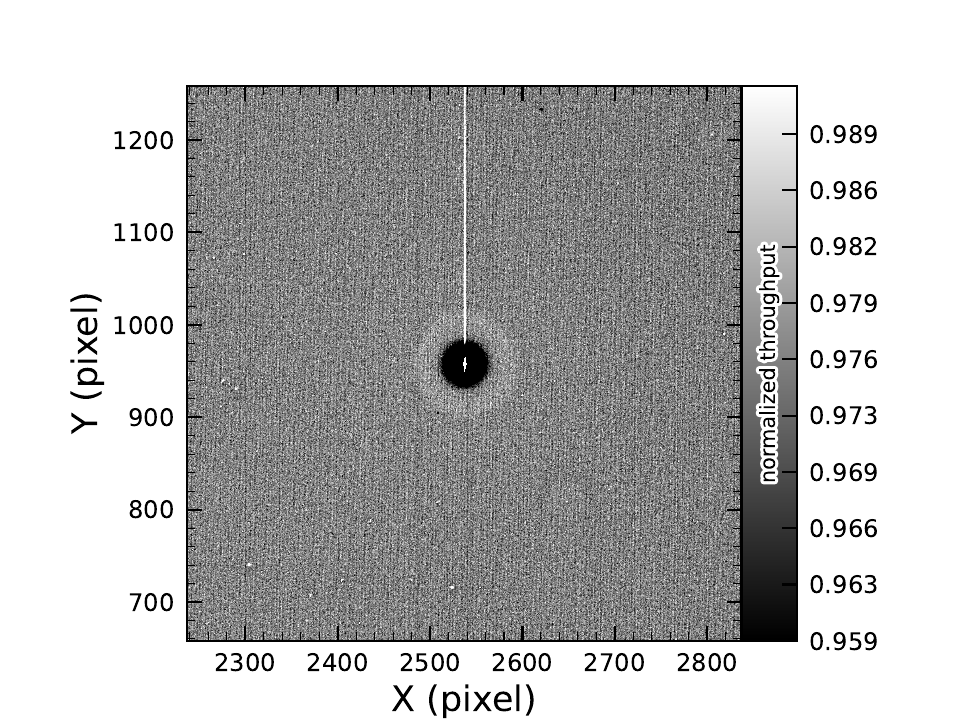}
  \caption{ A large vampire pixel near the center of R22\_S11, as seen on the $r$-band flat.  
  This clearly shows the central hot "vampire" pixels, surrounded by a region of depressed signal, with a brighter ring surrounding that caused by the local electric field effects.  
  The charge contained in the central pixels is incompletely shifted as the image is read, and that charge leaks out into subsequent rows as they are shifted through the remnant charge.  
  The columns that contain the hot pixels are masked as defects in all processing, as this feature cannot be otherwise corrected.}
   \label{fig:anomalies_vampire_pixels}
\end{figure}

``Vampire" pixels are visible on the images as a bright defect surrounded by a region of depressed flux, as though the defect is stealing charge from its neighboring pixels.
\figref{fig:anomalies_vampire_pixels} shows an example of a vampire pixel near the center of R22\_S11 on an $r$-band flat.

From studies on evenly illuminated images, vampires appear to conserve charge.
Unfortunately, no unique optimum way exists to redistribute this stolen flux so, following visual inspection, a defect mask was created to exclude them from processing.
We have found some similar features on the ITL detectors on LSSTCam, and will use the same approach to exclude them.

\subsubsection{Phosphorescence}
\begin{figure}[htb!]
  \centering
  \includegraphics[width=0.98\linewidth]{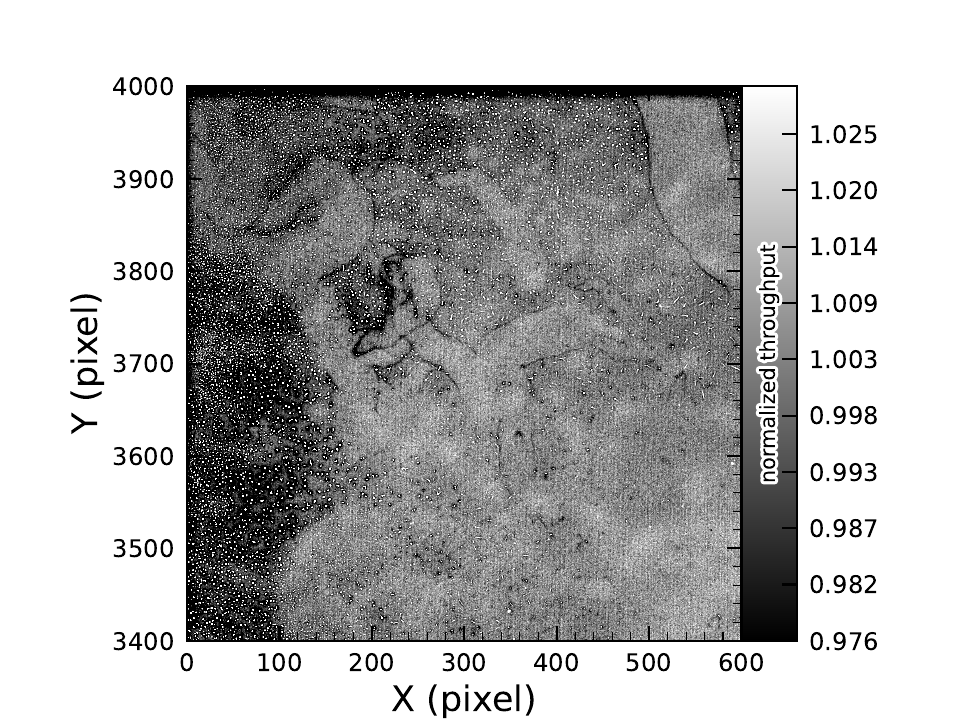}
  \caption{The top left corner of R22\_S01 in the g-band flat, showing the many small defect features that are caused by the remnant photoresist wax.
  A single large defect box masks this region from further analysis to prevent these features from contaminating measurements.}
  \label{fig:anomalies_phosphorescence}
\end{figure}

Some regions of the \gls{LSSTComCam} CCD raft were seen to contain large numbers of bright defects.
An example is shown in \figref{fig:anomalies_phosphorescence}  in a $g$-band flat. 
On further investigation, it appears that on some detectors a layer of photoresist wax was incompletely removed from the detector surface during production.
As this wax is now trapped below the surface coatings, there is no way to physically clean these surfaces.
If this wax responded to all wavelengths equally, then it would likely result in quantum efficiency dips, which might be removable during flat correction.
However, it appears that this wax is slightly phosphorescent, with a decay time on the order of minutes, resulting in the brightness of these defects being dependent on the illumination of prior exposures.
The worst of these regions were excluded with manual masks.

\subsubsection{Crosstalk}
\begin{figure}[htb!]
  \centering
  \includegraphics[width=0.98\linewidth]{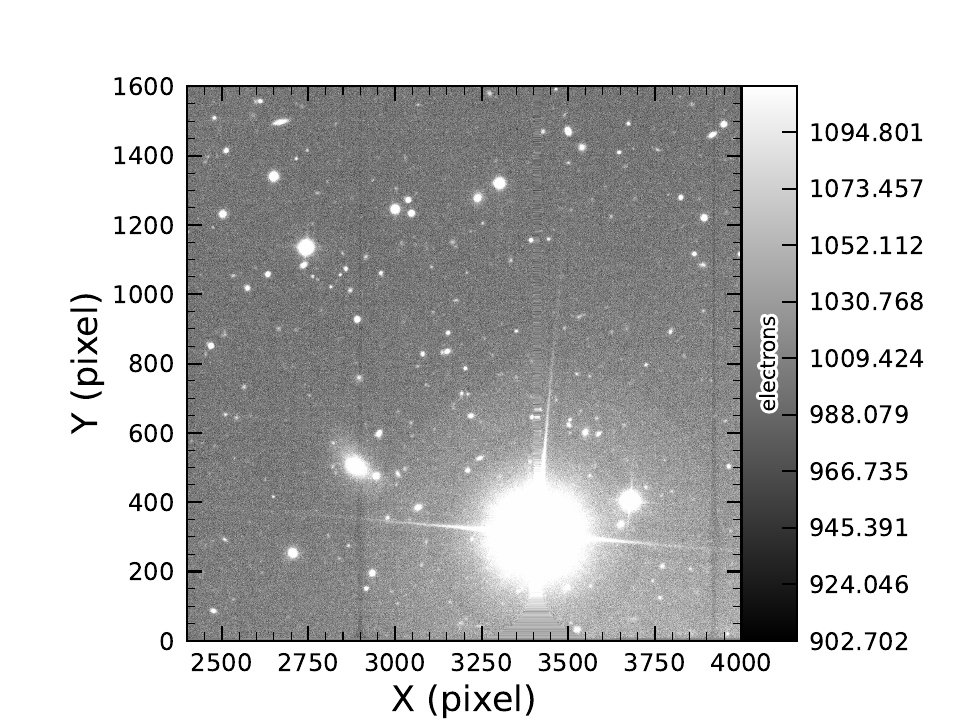}
  \caption{An example of a bright star with over-subtracted crosstalk residuals visible on neighboring amplifiers to both sides (exposure 2024120600239, detector R22\_S02).
  The horizontal banding stretching from the center of the star shows the interpolation pattern covering the saturated core and the ITL edge bleed near the serial register.}
  \label{fig:crosstalk_residual}
\end{figure}

Crosstalk refers to unwanted signal interference between adjacent pixels or amplifiers.
We use an average inter-amp crosstalk correction based on laboratory measurements with LSSTCam.
These average corrections proved satisfactory, and so have been used as-is for \gls{DP1} processing.
There are, however, some residual crosstalk features present post-correction, with a tendency towards over-subtraction.
\figref{fig:crosstalk_residual} shows an example  of a bright star with over-subtracted crosstalk residuals visible on neighboring amplifiers to both sides on exposure 2024120600239, detector R22\_S02.

\subsubsection{Bleed Trails}
Bleed trails are produced when charge from saturated pixels spills into adjacent pixels.
Bleed trails were anticipated on \gls{LSSTComCam} sensors, but they appear in more dramatic forms than had been expected.
As a bleed trail nears the serial register, it fans out into a ``trumpet'' shaped feature.
Although bright, these features do not have consistently saturated pixels.
In \gls{DP1} these ``edge bleeds'' were  identified and masked.

\begin{figure}[htb!]
  \centering
  \includegraphics[width=0.98\linewidth]{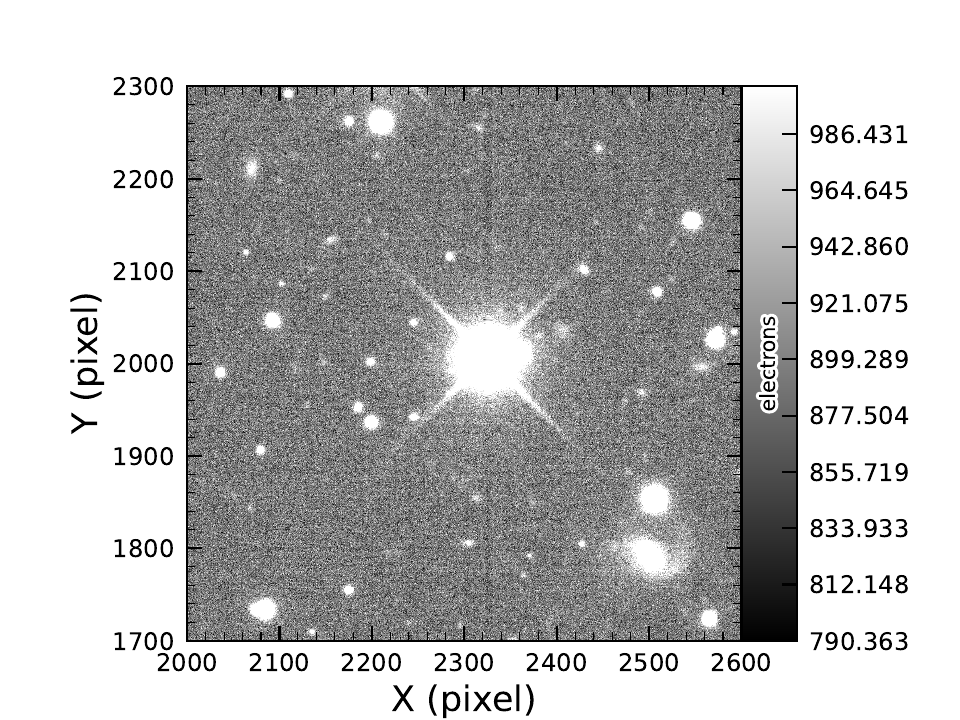}
  \caption{A bright star showing the ``ITL dip'' phenomenon, in which a dark trail extends out from the star to the top and bottom edges of the detector (exposure: 2024121000503, detector: R22\_S21).}
\label{fig:anomalies_itl_dip}
\end{figure}

Saturated sources can create a second type of bleed, where the central bleed drops below the background level.
The depressed columns along these trails extend across the entire readout column of the detector, crossing the detector mid-line.
We developed a model for these to identify which sources are sufficiently saturated to result in such a trail, which is then masked.  As this kind of trail appears only on the ITL detectors, we've named these features ``ITL dips''.
\figref{fig:anomalies_itl_dip} shows an example of a   bright star exhibiting the ``ITL dip'' phenomenon on exposure: 2024121000503, detector: R22\_S21.

\subsection{PSF Models}
\label{ssec:psf_models}
To characterize \gls{PSF}  performance, we use adaptive second moments \citep{2002AJ....123..583B} measured on \gls{PSF} stars and on the PSF model using the \gls{HSM} implementation \citep{2003MNRAS.343..459H, 2005MNRAS.361.1287M}.
All measurements are expressed in the pixel coordinate frame of each detector.
We characterize the performance of the PSF using the classical trace of the  second moment matrix $T$, along with the ellipticity parameters $e^1$ and $e^2$.
Measurements on the observed PSF stars are are denoted as  $T_{\text{PSF}}$, $e^1_{\text{PSF}}$,  $e^2_{\text{PSF}}$, while those from PSF models are denoted as  $T_{\text{model}}$, $e^1_{\text{model}}$, $e^2_{\text{model}}$.
We compare two PSF modeling approaches:
\begin{itemize}
\item Piff with second-order polynomial interpolation (Piff O2), the pipeline's default, and
\item Piff with fourth-order polynomial interpolation (Piff O4), which serves as the final \gls{DP1} PSF model.	
\end{itemize}
\tabref{tab:psf-1d_stats} summarizes each model’s ability to reconstruct the mean $T$, $e^1$, and $e^2$ on  \gls{LSSTComCam}. 
Both models exhibit a negative residual bias in the reconstructed PSF size, with Piff O4 providing improved performance over Piff O2.

\begin{figure}[htb!]
    \centering
    \includegraphics[width=0.8\columnwidth]{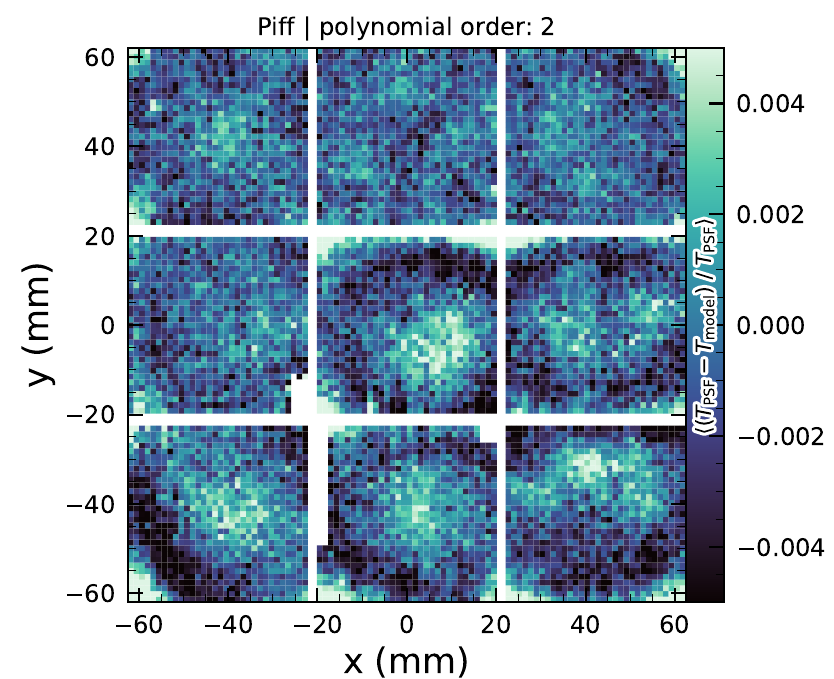}  
    \vspace{0.5cm}  
    \includegraphics[width=0.8\columnwidth]{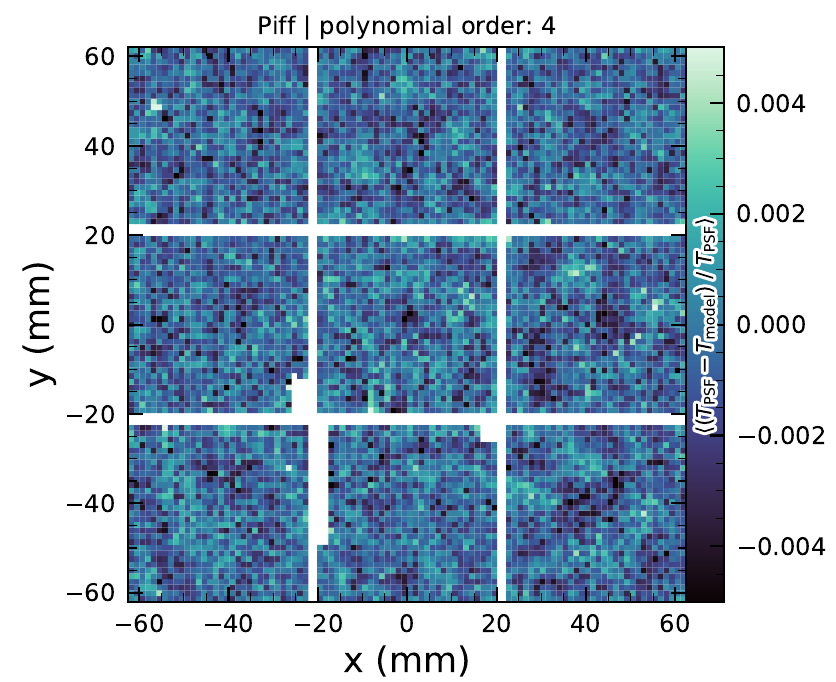}  
    \caption{Average across all visits of $\delta T/T$  for Piff O2 and Piff O4 modeling on LSSTComCam. Averages are computed using a 120$\times$120 binning.}
    \label{fig:psf_residuals_fov}
\end{figure}

\begin{deluxetable*}{lccc}
\caption{Observed mean values and comparison of model residuals, across all visits and filters}
\label{tab:psf-1d_stats}
\tablehead{
  \colhead{\textbf{Quantity}} & 
  \colhead{\textbf{Observed}} & 
  \colhead{\textbf{Piff O2}} & 
  \colhead{\textbf{Piff O4}} \\
  \colhead{} & 
  \colhead{} & 
  \colhead{$\times10^{-4}$} & 
  \colhead{$\times10^{-4}$} 
}
\startdata
$\langle T\rangle\ (\mathrm{pixel}^2)$ & $11.366 \pm 0.003$ & & \\
$\langle e^1\rangle$ & $(-6.07\pm0.05)\times10^{-3}$ & & \\
$\langle e^2\rangle$ & $(-4.57\pm0.05)\times10^{-3}$ & & \\
$\langle e\rangle$ & $(8.794\pm0.004)\times10^{-2}$ & & \\
$\langle \delta T / T\rangle$  & & $-4.0\pm0.2$ & $-5.0\pm0.2$ \\
$\langle \delta e^1\rangle$ & & $0.6\pm0.1$ & $0.5\pm0.1$ \\
$\langle \delta e^2\rangle$ & & $0.0\pm0.1$ & $0.0\pm0.1$ \\
\enddata
\end{deluxetable*} 
An alternative approach to evaluating the performance of the \gls{PSF} model is to examine the average $\delta T/T$, where $\delta T$ is $T_{\text{PSF}}$ - $T_{\text{model}}$, across visits, projected onto focal-plane coordinates, as shown in \figref{fig:psf_residuals_fov}. 
Piff reveals strong spatial correlations in the residuals, including a systematic offset consistent with the results presented in \tabref{tab:psf-1d_stats}. 
The presence of these spatial structures motivated the adoption of fourth-order polynomial interpolation in all bands except $u$-band. 
Although not shown in \figref{fig:psf_residuals_fov}, residual patterns persist even with third-order interpolation, indicating that it is insufficient to capture the complexity of the PSF variation. 
Increasing the interpolation order to five would nominally reduce the residuals further, but the limited number of stars available on some CCDs would not provide adequate constraints for such a model, while the resulting improvement would likely be minimal.
Preliminary analysis of \gls{LSSTCam} data in the laboratory at \gls{SLAC} shows that the \gls{ITL} sensors exhibit the same pattern as \gls{ITL} sensors on \gls{LSSTComCam}.

\begin{figure}[htb!]
    \centering
    \includegraphics[scale=0.33]{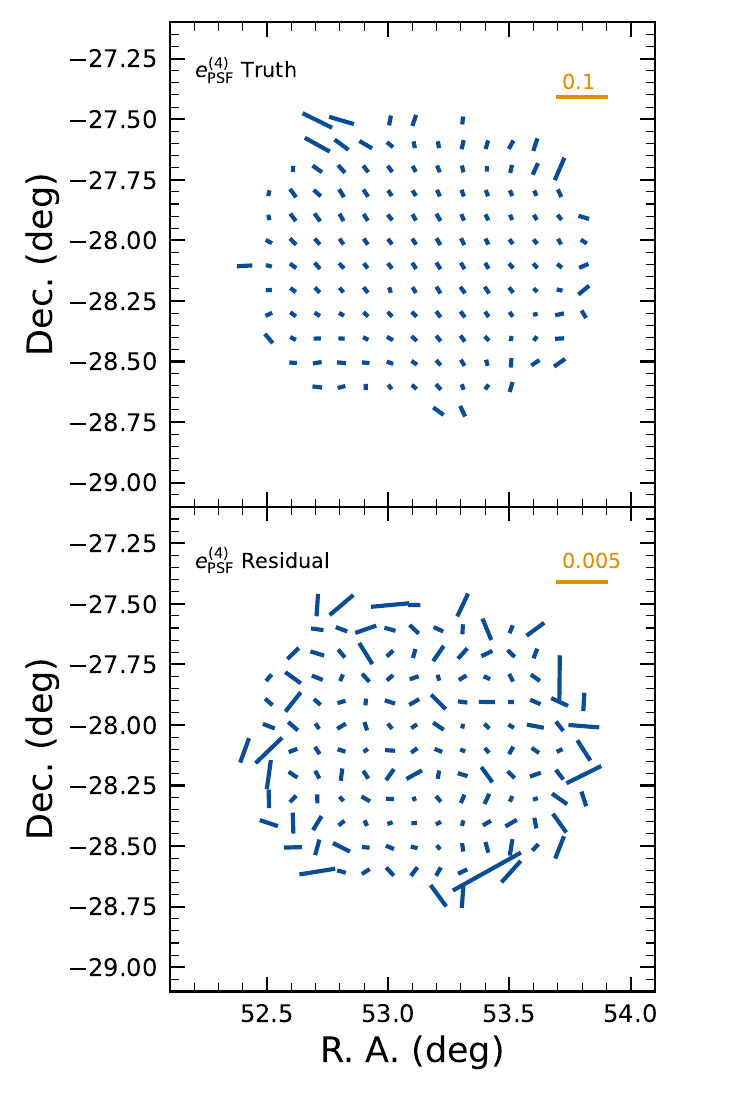}
    \includegraphics[scale=0.33]{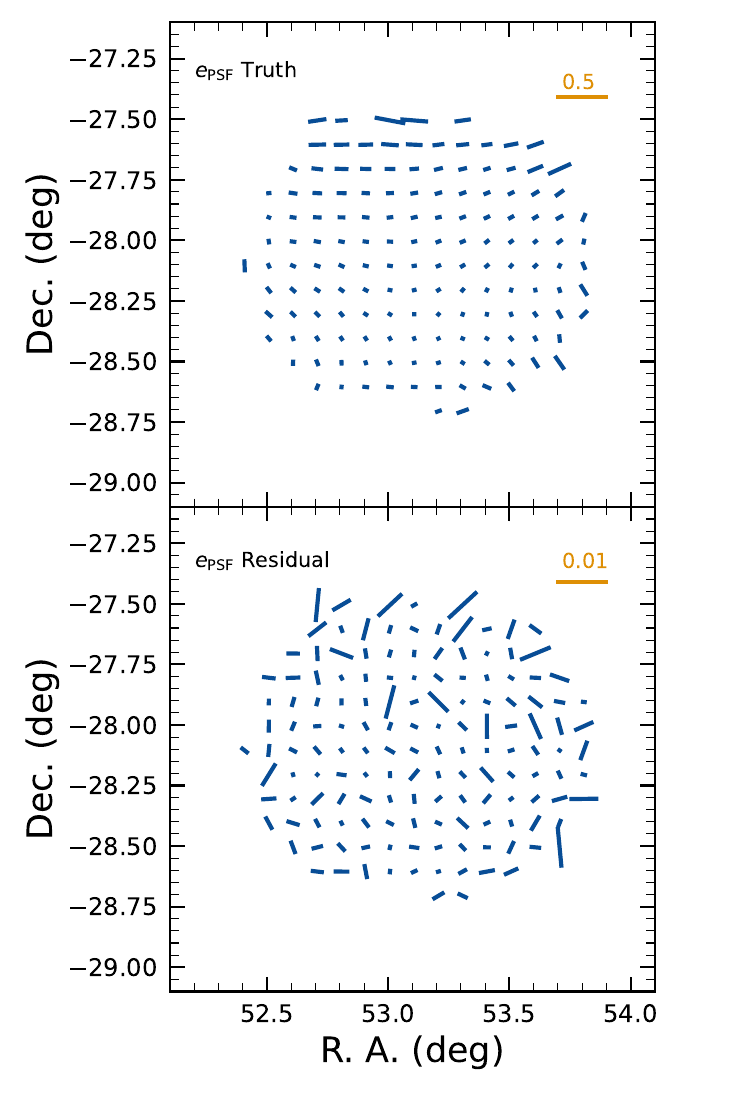}
    \caption{Whisker plots for the  \gls{ECDFS} field for $e$, $e^{(4)}$ and $\delta e$, $\delta e^{(4)}$.}
    \label{fig:psf_residuals_whisker_ECDFS}
\end{figure}

Another way to look at the \gls{PSF} modeling quality is via whisker plots of the \gls{PSF} second and fourth moments and their modeling residuals projected on a part of the sky.
In addition to the second moment, the spin-2 fourth moments, $e^{(4)}$, are defined as:
\begin{align*}
e^{(4)}_1 &= M_{\text{40}} - M_{\text{04}} \\
e^{(4)}_2 &= 2\left(M_{\text{31}} - M_{\text{13}}\right),
\end{align*}
where $M_{\text{pq}}$ are the standardized higher moments as defined in \cite{2023MNRAS.520.2328Z} measured on stars and PSF models.
\figref{fig:psf_residuals_whisker_ECDFS} shows
the whisker plots of $e$, $e^{(4)}$ (top rows), and $\delta e$, $\delta e^{(4)}$
in the \gls{ECDFS} field. 
The direction of a whisker represents the orientation of the \gls{shape}, while the length represents the amplitude $|e|$ or $|e^{(4)}|$.
We observe coherent patterns in both the \gls{PSF} moments and the residuals, the latter of which warrants further investigation if it persists in future data releases.

\begin{figure}[htb!]
\includegraphics[width=\linewidth]{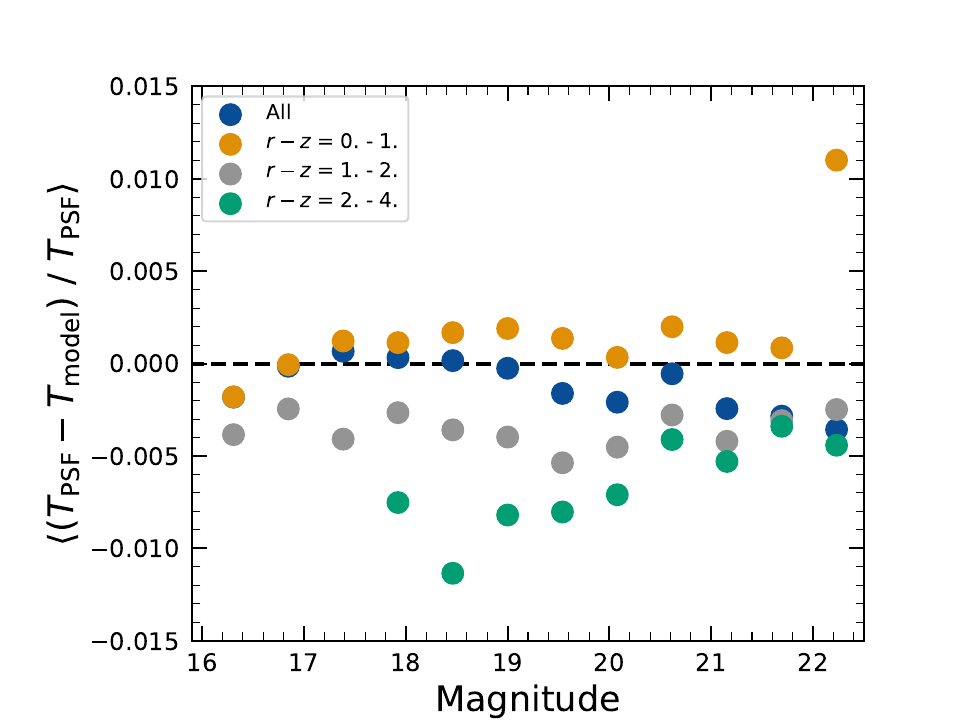}
\caption{Binned $\delta T/T$ as a function of magnitude across all visits and filters and in bins of stellar colors.}
\label{fig:psf_residuals_mag_color}
\end{figure}

\figref{fig:psf_residuals_mag_color} shows a plot of  $\delta T/T$ versus stellar magnitude, which can reveal any dependencies between \gls{PSF} size and flux.
We also repeat this analysis in color bins to probe chromatic effects.
Binning by color uncovers a clear color dependence, as was also seen in \gls{DES} \citep{DES:2020vau}.
The residual is consistent with \tabref{tab:psf-1d_stats} and its cause is unknown.
\gls{DP1} does not include the color correction implemented in the DES Year 6 analysis, \citet{2025OJAp....8E..26S}. 
This will be included in processing of future data releases.

As noted in  \cite{PSTN-019}, two key Piff features were not used in the \gls{DP1} processing.
PSF color dependence was not implemented, and, while Rubin software allows Piff to work with sky coordinates (including WCS transformations), 
it does not yet correct for sensor-induced astrometric distortions such as tree rings \citep{2017JInst..12C5015P}.
Both features are planned for upcoming releases.

\subsection{Astrometry}
\begin{figure}[htb!]
\includegraphics[width=\linewidth]{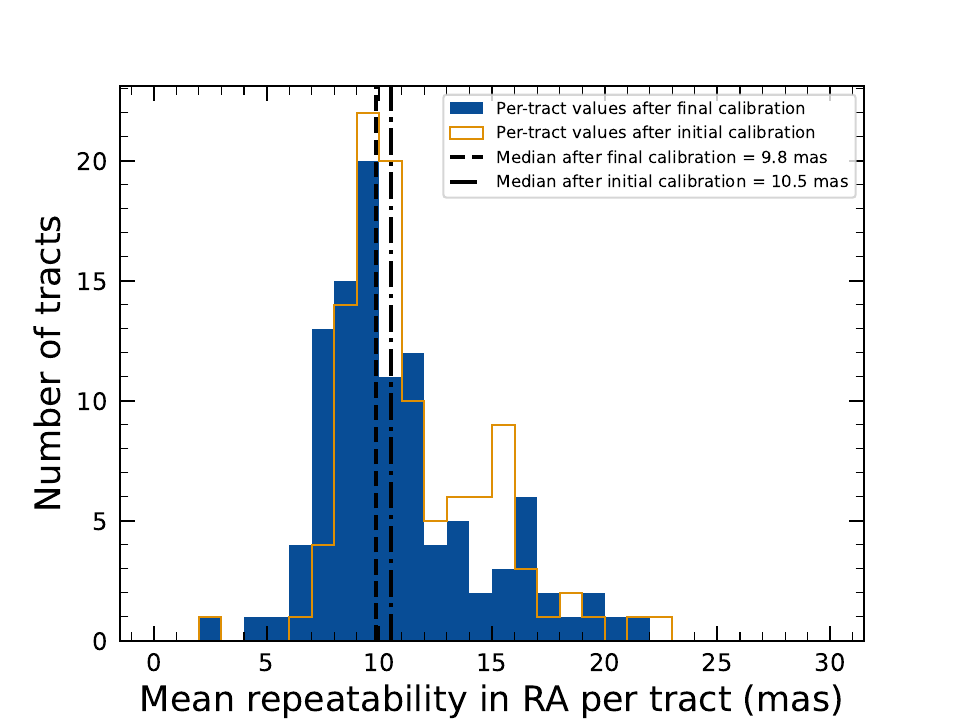}
\caption{Mean per-tract astrometric repeatability of measurements of isolated point sources in RA in visits across all bands.}
\label{fig:dmAstroErr}
\end{figure}

\begin{figure}[htb]
\includegraphics[width=\linewidth]{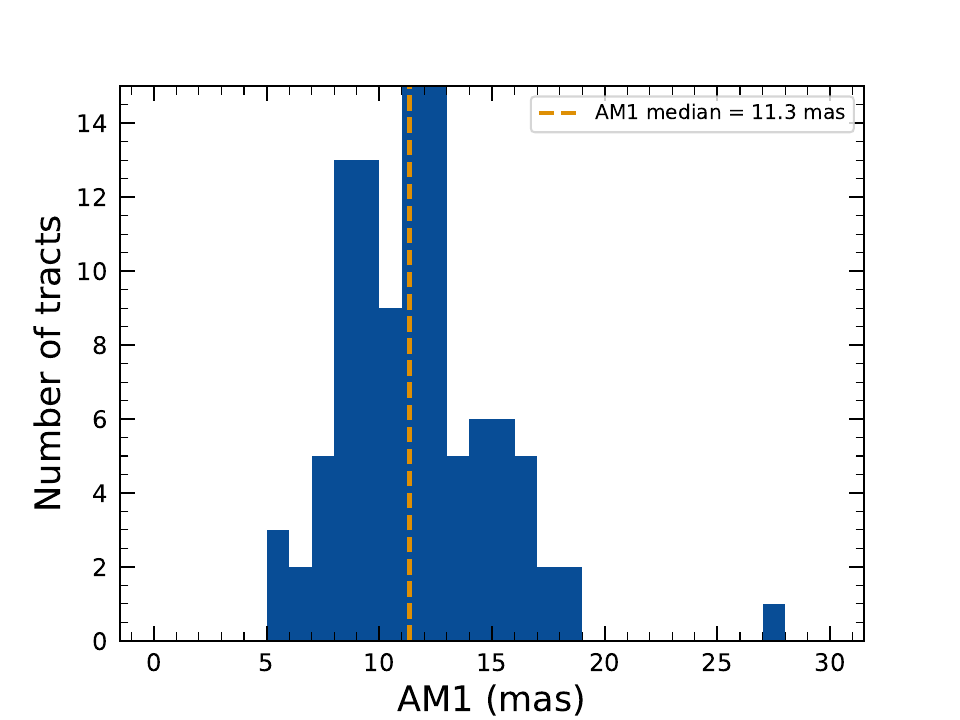}
 \caption{Median per-tract repeatability in separations between isolated point sources 5~arcmin apart (AM1) in visits across all bands.}
 \label{fig:AM1}
 \end{figure}

To characterize astrometric performance, we evaluate both internal consistency and agreement with an external reference.
The primary measure of internal consistency is the repeatability of position measurements for the same object, defined as the RMS of the astrometric distance distribution for 
stellar pairs having a specified separation in arcminutes. 
We associate isolated point sources across visits and compute the rms of their fitted positions, rejecting any stars with another star within  2\arcsec.
\figref{fig:dmAstroErr} shows the mean per-\gls{tract} rms astrometric error in RA  for all isolated point sources, both after the initial calibration and after the final calibration, which includes proper motion corrections.
The results indicate that the astrometric solution is already very good after the initial \gls{calibration}.
Global calibration yields only modest improvement, likely due to the short time span of \gls{DP1} and the minimal distortions in the LSSTComCam.
In the main survey, the longer time baseline and greater distortions near the \gls{LSSTCam} field edges will make global calibration more impactful.
An additional measure of internal consistency is the repeatability of separations between objects at a given distance.
To compute this, we identify pairs of objects that are separated by a specified distance and measure their precise separation during each visit in which both objects are observed.
The scatter in these separation measurements provides an indication of the internal consistency of the astrometric model.
\figref{fig:AM1} shows the median separation for pairs of objects separated by approximately 5 arcminutes (referred to as ``AM1''), computed per tract after the final calibration.
These values are already approaching the design requirement of $10$ mas.

\begin{figure*}[htb]
\includegraphics[width=\linewidth]{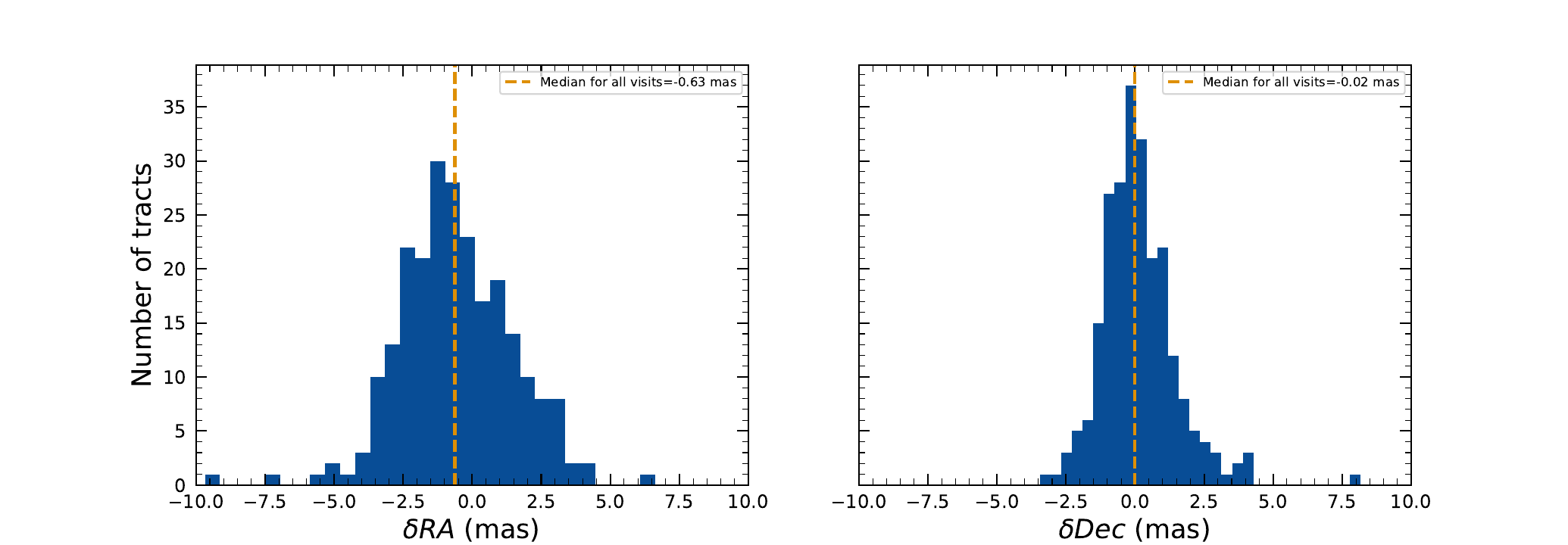}
\caption{Median absolute offset for all visits in $r$-band in \gls{tract} 4849 in the ECDFS field. 
The offset is the difference between the positions of isolated point sources that were reserved from the astrometric fit and matched objects from the Gaia DR3 catalog.}
\label{fig:AA1}
\end{figure*}

\begin{figure*}[htb]
\includegraphics[width=\linewidth]{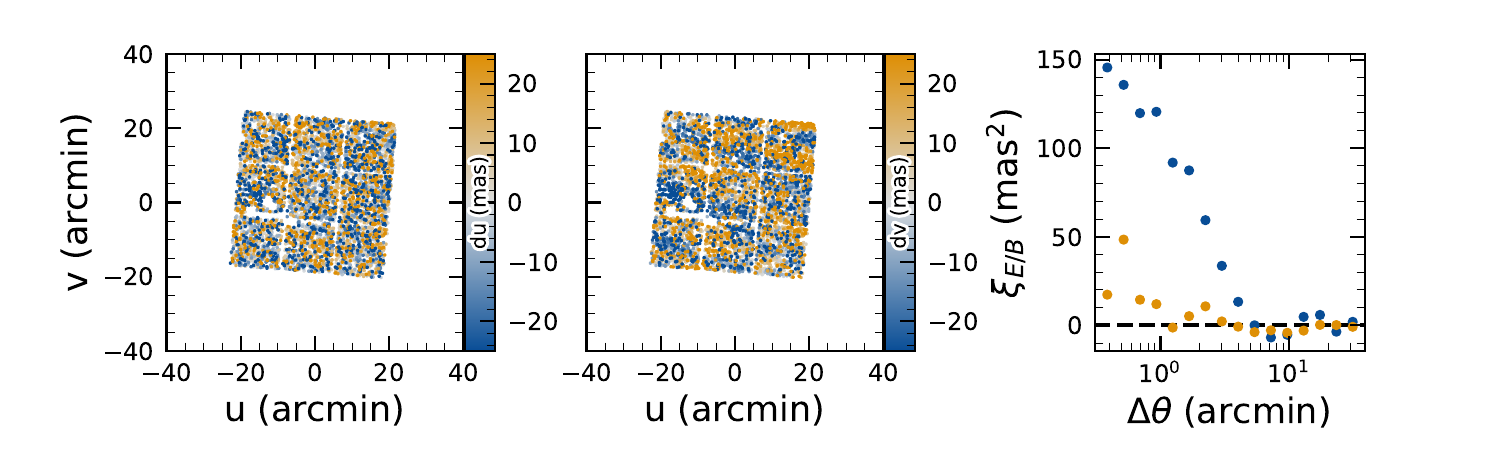}
\caption{Astrometric residuals in $u$ (left panel) and $v$ (center panel) directions with the E (blue) and B (orange) modes of the two-point correlation function (right panel) 
seen in visit 2024120200359 in tract 2393 in $u$ band.
The residuals show a wave-like pattern characteristic of atmospheric turbulence, and there is significant E-mode and negligible B-mode in the correlation function.}
\label{fig:Astrometry_Emode}
\end{figure*}

\begin{figure*}[htb!]
\includegraphics[width=\linewidth]{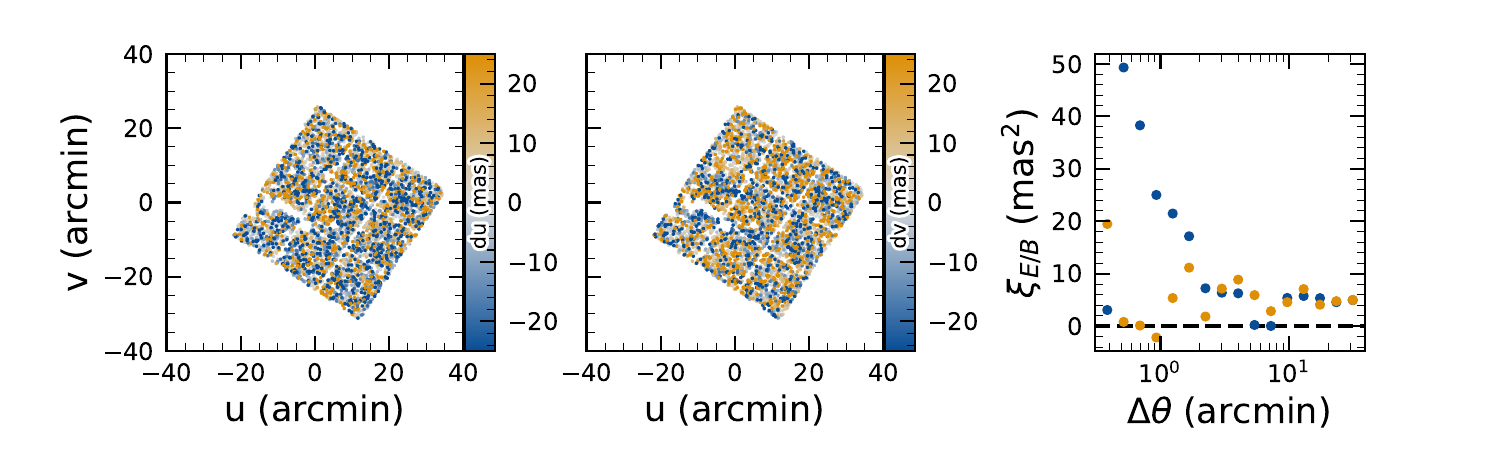}
\caption{Astrometric residuals in $u$ (left panel) and $v$ (center panel) directions, with the E (blue)  and B (orange) 
modes of the two-point correlation function (right panel) seen in visit 2024120700527 in tract 2393 in $u$ band.
There are coherent residuals, but without the wave-like pattern seen in \figref{fig:Astrometry_Emode}, and the correlation function has significant values for both E and B-modes.}
\label{fig:Astrometry_EBmode}
\end{figure*}

To assess external consistency, we consider the median separation between sources not included in the astrometric fit and associated objects from a reference catalog (\secref{sssec:monster}).
For this, we use the Gaia \gls{DR3} catalog, with the object positions shifted to the observation epoch using the Gaia proper motion parameters.
\figref{fig:AA1} shows the median separation for each visit in the $r$-band in \gls{tract} 4849 in the ECDFS fields.
The calculated values are almost all within $5$\xspace mas, well below the design requirement of $50$\xspace mas for the main survey.
By examining the astrometric residuals, we can assess whether there are distortions not accounted for by the astrometric model. 
In some cases, residuals from a single visit exhibit behavior consistent with atmospheric turbulence, as shown in \figref{fig:Astrometry_Emode}, which is characterized by a curl-free gradient field in the two-point correlation function of the residuals (E-mode),  \citet{2021A&A...650A..81L} and \citet{2021AJ....162..106F}. 
However, as seen in \figref{fig:Astrometry_EBmode}, the residuals in many visits also have correlation functions with a non-negligible divergence-free B-mode,
indicating that some of the remaining residuals are due to unmodeled instrumental effects, such as rotations between visits.

\begin{figure}[htb!]
\includegraphics[width=\linewidth]{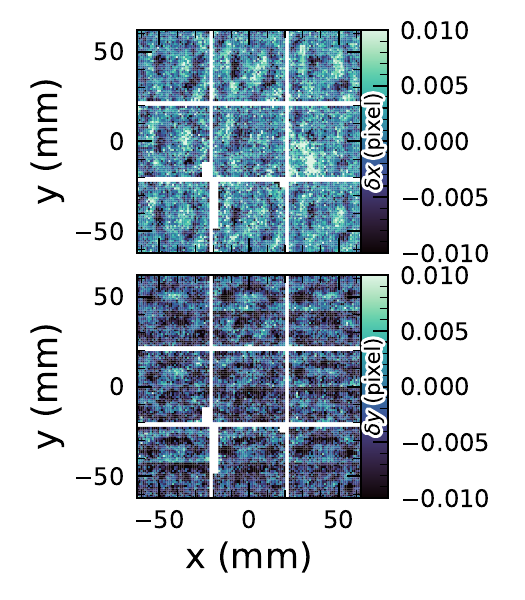}
\caption{Median astrometric residuals  as a function of focal plane position, shown in the left panel for the $x$ direction and in the right panel for the $y$ direction, for all nine \gls{LSSTComCam} CCDs independently.
The range of the color scale is $\pm$ 0.01 pixels, corresponding to 2 mas, showing that the effect is small. }
\label{fig:Astrometry_FoV}
\end{figure}

\begin{figure}[htb!]
\includegraphics[width=\linewidth]{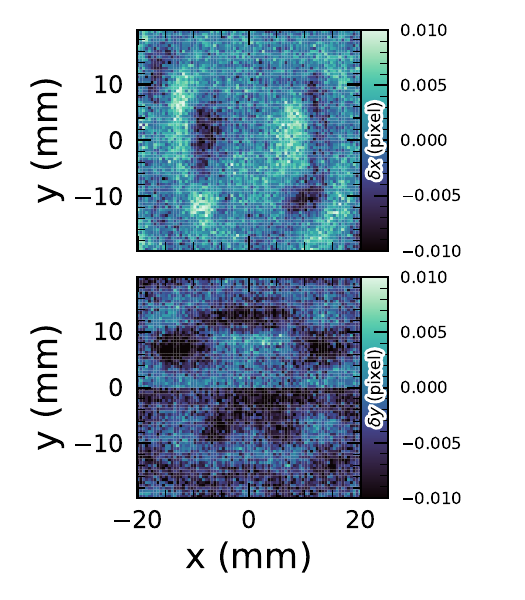}
\caption{Median residuals as a function of pixel position, shown in the left panel for the $x$ direction and in the right panel for the $y$ direction. 
These residuals are aggregated across all nine CCDs that comprise the central \gls{LSSTComCam} raft.
The range of the color scale is $\pm$ 0.01 pixels, corresponding to 2 mas, showing that the effect is small.}
\label{fig:Astrometry_CCD}
\end{figure}

We can see unmodeled camera distortions by stacking the astrometric residuals over many visits as a function of the focal plane position.
\figref{fig:Astrometry_FoV} shows the median residuals in $x$ and $y$ directions for \nvisits visits.
Spatial structures are evident at the \gls{CCD} level, as well as at the mid-line break,  the discontinuity between the two rows of amplifiers,  in the y-direction residuals.
Further stacking all the detectors makes certain effects particularly clear.
\figref{fig:Astrometry_CCD} shows distortions very similar to those measured for an \gls{LSSTCam} \gls{ITL} sensor in a laboratory setting in \citet{2023PASP..135k5003E}.

\subsection{Differential Chromatic Refraction}
\label{sec:differential_chromatic_refraction}
\begin{figure}[htb!]
\includegraphics[width=\linewidth]{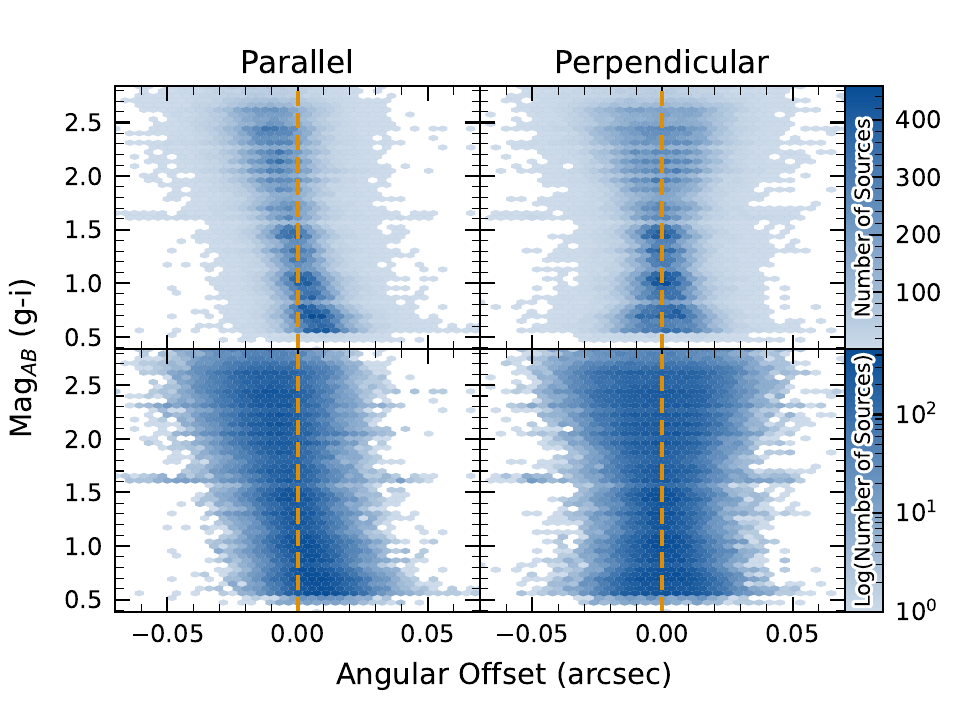}
\caption{Visualization of \gls{Differential Chromatic Refraction} (DCR) observed in the \gls{LSSTComCam} commissioning campaign. 
The $g-i$ color is computed for every source in the reference catalog (\secref{sssec:monster}) that is matched to a direct source in the science image, and the binned density for the full survey is plotted against the angular offset between the reference and detected positions. The angular offset is projected along coordinates parallel and perpendicular to the parallactic angle of the observation, and shows a characteristic correlation along the parallel axis with no correlation along the perpendicular axis. The orange vertical dashed line indicates the expected $g-i$ magnitude distribution at zero angular offset.}
\label{fig:dcr}
\end{figure}

\gls{DCR} occurs when light passes through Earth’s atmosphere, refracting more for shorter wavelengths, which causes blue light to appear shifted closer to the zenith.
This wavelength-dependent effect results in the smearing of point sources along the zenith direction, specifically parallel to the parallactic angle.
The DCR effect is observable in \gls{LSSTComCam} data, particularly in the angular offset versus $g-i$ band magnitude difference plots,  as shown in \figref{fig:dcr}. 
These plots include 228 visits selected to maximize the range of observed airmass, which spans 1.01–1.30 with a mean value of 1.13.
When looking at data perpendicular to the parallactic angle, sources exhibit no discernible DCR effect, which is expected, and form a clear vertical distribution on the two-dimensional density plots in \figref{fig:dcr}.

In contrast, sources aligned with the parallactic angle exhibit a tilted, linear distribution, clearly demonstrating that the relationship between angular offset and the $g-i$ band magnitude difference, thereby providing a visual indication of the \gls{DCR} effect.
The DCR effect will be addressed in future releases. 

\subsection{Stellar Photometry}
The photometric repeatability for isolated bright unresolved sources following the \gls{FGCM} fits was excellent. 
For the 10\% of unresolved sources withheld from the fit and having signal-to-noise ratios greater than 100, the photometric repeatability 
after applying chromatic correction was 7.1, 5.4, 5.4, 5.1, 5.9, and 6.5  mmag in the $ugrizy$ bands respectively, across all fields.
After accounting for photometric noise, the intrinsic photometric repeatability was approximately 4.8, 2.7, 1.7, 1.0, 2.0, and 1.1 mmag in $ugrizy$.
The \gls{DP1} processing does not yet include chromatic corrections in the final photometry. 
In this case  the delivered photometric repeatability was 3-8 mmag  for grizy.  

\begin{figure*}[htb!]
\includegraphics[width=\linewidth]{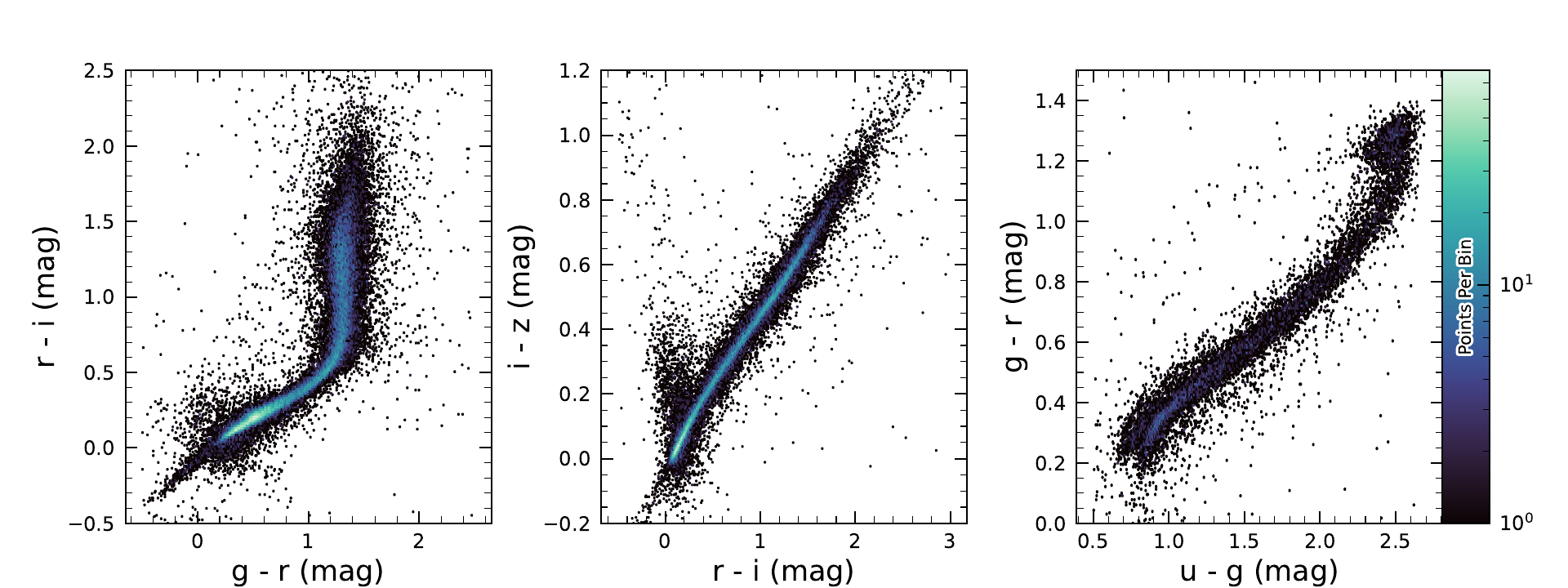}
\caption{Examples of stellar loci for unresolved sources from the \gls{DP1} dataset. From left to right: 
$gri$ stellar locus containing 63,236 stars with signal-to-noise ratio $>$ 200 in the $i$ band;
$riz$ stellar locus containing 46,760 stars with signal-to-noise ratio $>$ 200 in the $i$ band
$ugr$ stellar locus containing 12,779 stars with signal-to-noise ratio $>$ 50 in the $u$ band.}
\label{fig:stellarloci}
\end{figure*}

In \figref{fig:stellarloci}, we show the stellar loci for $ugriz$ for unresolved sources in the \gls{DP1} \texttt{Object} table (\secref{ssec:catalogs}).
These unresolved sources  were selected using the extendedness parameter (\secref{ssec:catalogs}) in the \texttt{Object} catalog. 
This parameter is assigned a value of 0 (unresolved) or 1 (resolved) in each band based on the difference between the PSF and CModel magnitudes. 
The extendedness is set to 1 when this magnitude difference exceeds 0.016 mag, as the PSF flux for extended sources is biased low relative to the CModel flux. 
This method has been previously employed by the SDSS pipelines, and its statistical properties, including the optimal combination of information from different bands and repeated measurements, are discussed in \cite{2020AJ....159...65S}.

\begin{figure*}[t]
\includegraphics[width=\linewidth]{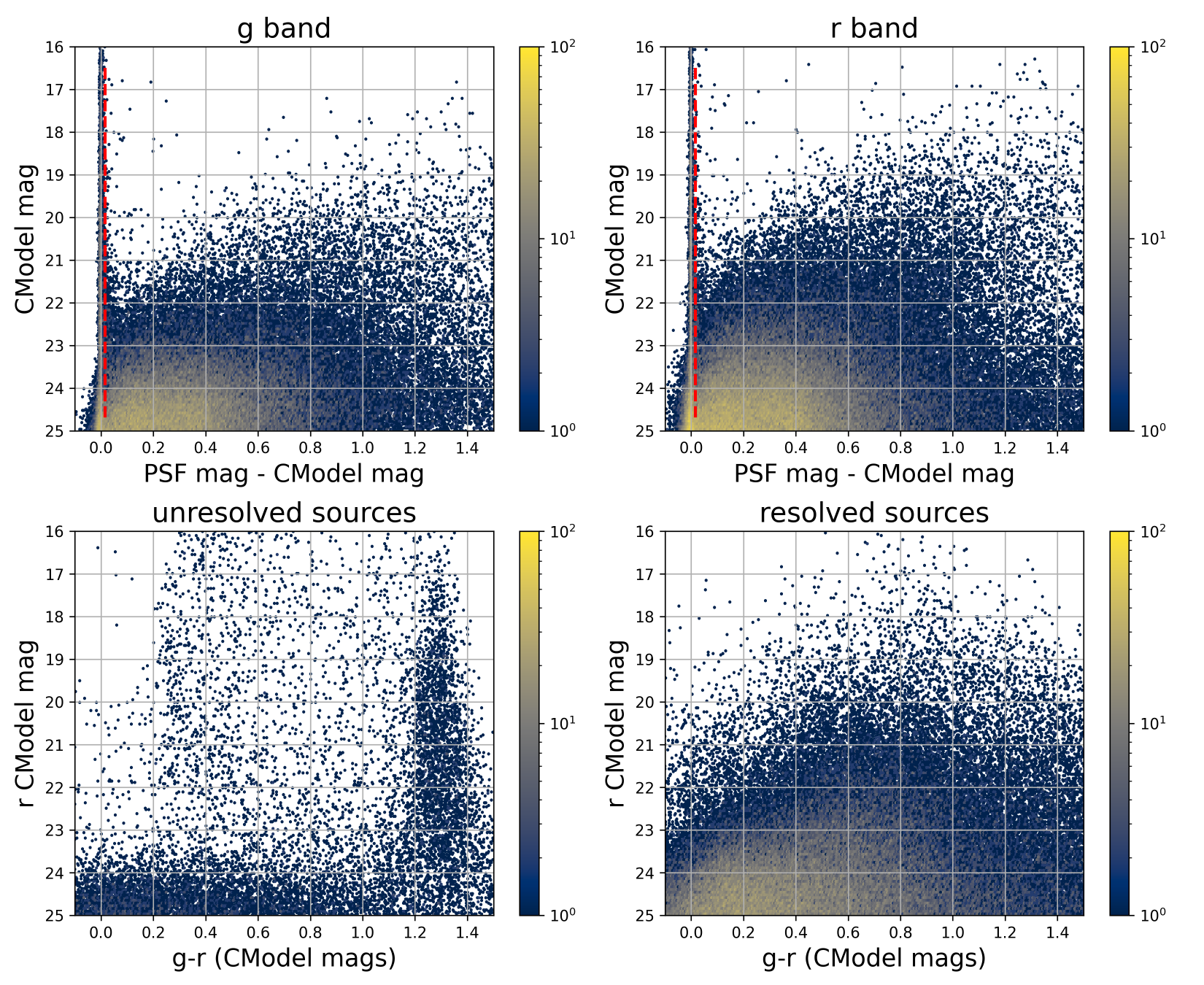}
\caption{The top two panels shows the difference between the PSF and CModel magnitudes as a function of CModel magnitude in the
$g$ and $r$ bands for 178,547 sources with $CModel_r < 25$ from the ECDFS field. The vertical dashed line in each panel
marks the minimum value (0.016 mag) for setting the extendedness parameter to 1. The bottom two panels show the $r$ vs.
$g-r$ color-magnitude diagrams for 14,701 unresolved (left) and 163,666 resolved (right) sources. Note the unresolved sample
suffers from increasing contamination by galaxies for $r>24$.}
\label{fig:extendedness}
\end{figure*}

\figref{fig:extendedness} illustrates the behavior of the extendness parameter. 
Its behavior in the $g$ and $r$ bands is similar, with unresolved sources scattered around the vertical line centered on zero. 
The width of the distribution increases towards fainter magnitudes.
Resolved sources are found to the right and the dashed lines in the top panels show the adopted ``star-galaxy" separation boundary.
The morphology of the two color-magnitude diagrams in the bottom panels suggest that the unresolved sample suffers from increasing contamination by galaxies for $r>24$. 
This behavior is consistent with simulation-based predictions from \cite{2020AJ....159...65S}.

\subsection{Detection Completeness on Coadds}
\label{ssec:detection_completeness}
We characterize completeness by injecting synthetic sources into coadded images, and by comparing source detections to external catalogs.
In both cases, we use a greedy, probabilistic matching \gls{algorithm} that matches reference objects, in order of descending brightness, to the most likely target within a $0\farcs5$ radius.

We inject sources in 12 of the patches of the \gls{ECDFS} region with the deepest coverage.
The input catalog contains stars and galaxies from part of the \gls{DC2} simulations \citep{2021ApJS..253...31L}, where the galaxies consist of an exponential disk and de Vaucouleurs \citep{1948AnAp...11..247D,1953MNRAS.113..134D} bulge.
To avoid deblender failures from excessive increases in object density, stars with a total \gls{flux} (i.e., summed across all six bands) brighter than 17.5~${\rm mag}$ are excluded, as are galaxies whose total \gls{flux} is brighter than 15~${\rm mag}$ or fainter than 26.5~${\rm mag}$.
Half of the remaining objects are selected for injection.
Afterwards, individual bulge and disk components fainter than 29~${\rm mag}$ are also excluded, both for computational expediency and because their structural properties are less likely to be representative of real galaxies.
\begin{figure}[htb]
\includegraphics[width=\linewidth]{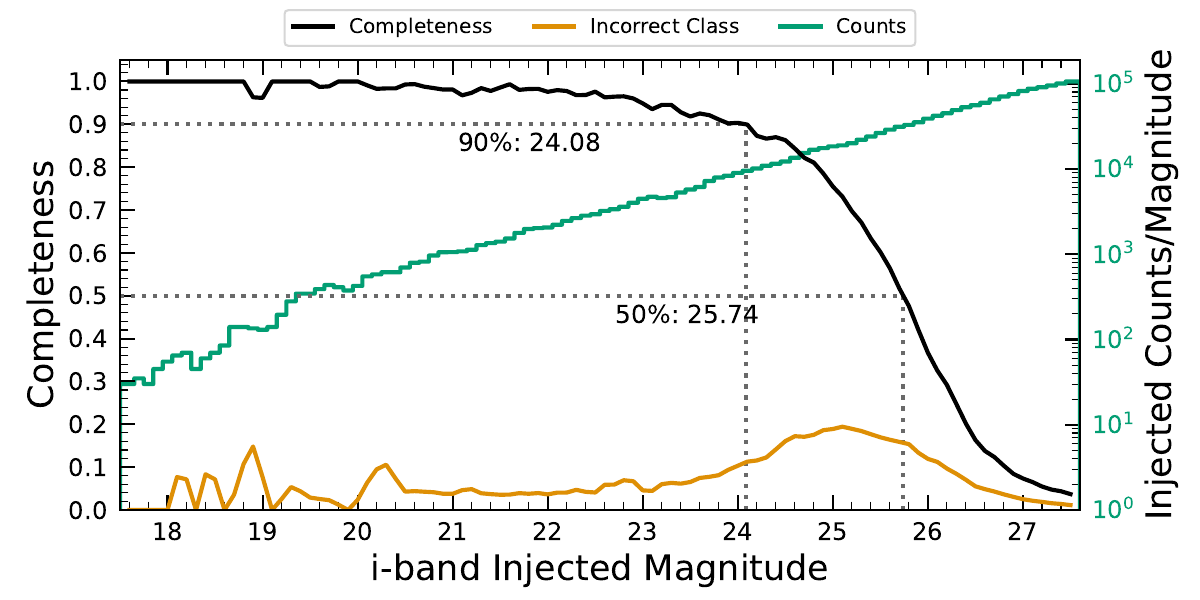}
\caption{Completeness and incorrect classification fraction as a function of $i$-band CModel magnitude (Reference Magnitude) for DC2-based injected objects into a portion of the ECDFS field. 
The ``Incorrect Class'' line shows the proportion of objects that are matched but classified incorrectly by their reference-band extendedness, i.e. stars with extendedness of 1 or galaxies with extendedness of 0 in the reference band.}
\label{fig:injected_lsst_cells_v1_5063_i_completeness_any}
\end{figure}

\figref{fig:injected_lsst_cells_v1_5063_i_completeness_any} shows completeness as a function of magnitude for these injected objects in the \gls{ECDFS} field.
These completeness estimates are comparable to results from matching external catalogs. 
Matching to the Hubble Legacy Field catalog \citep{2016arXiv160600841I, 2019ApJS..244...16W} reaches 50\% completeness at $F775W=26.13$, or about $i=25.83$ from differences in matched object magnitudes.
Similarly, completeness drops below 90\% at $VIS=23.80$ from matching to Euclid Q1 \citep{2025arXiv250315305E} objects, equivalent to roughly $i=23.5$. 
The Euclid imaging is of comparable or shallower depth, so magnitude limits at lower completeness percentages than 90\% are unreliable, whereas the HST images cover too small and irregular of an area to accurately characterize 80-90\% completeness limits.

At the 80\% completeness limit, nearly 20\% of objects, primarily injected galaxies, are incorrectly classified as stars based on their reference band extendedness.
Similarly, the fraction of correctly classified injected stars drops to about 50\% at $i=23.8$ (corresponding to 90\% completeness).

This analysis has several caveats.
The selection of objects for matching in any catalog is not trivial.
Some fraction of the detections are spurious, particularly close to bright stars and their diffraction spikes.
Additionally, some objects lie in masked regions of one survey but not another, which has not been accounted for. 
For injected source matching, the reference catalog (\secref{sssec:monster}) does not include real on-sky objects.
Based on prior analyses of the \gls{DC2} simulations, purity is generally greater than completeness at any given magnitude.
Similarly, for bright ($i<23$) objects classified as stars by reference band extendedness, $<5\%$ are either unmatched to a Euclid or HST object, or misclassified - that is, selecting on extendedness alone yields a fairly pure but incomplete sample of stars.
We expect to remedy some of these shortcomings in future releases.

\subsection{Model Flux and Shape Measurement}
\label{ssec:fluxes}

\begin{figure*}[hbt!]
  \centering
  \begin{subfigure}[t]{0.45\textwidth}
  \includegraphics[width=\linewidth]{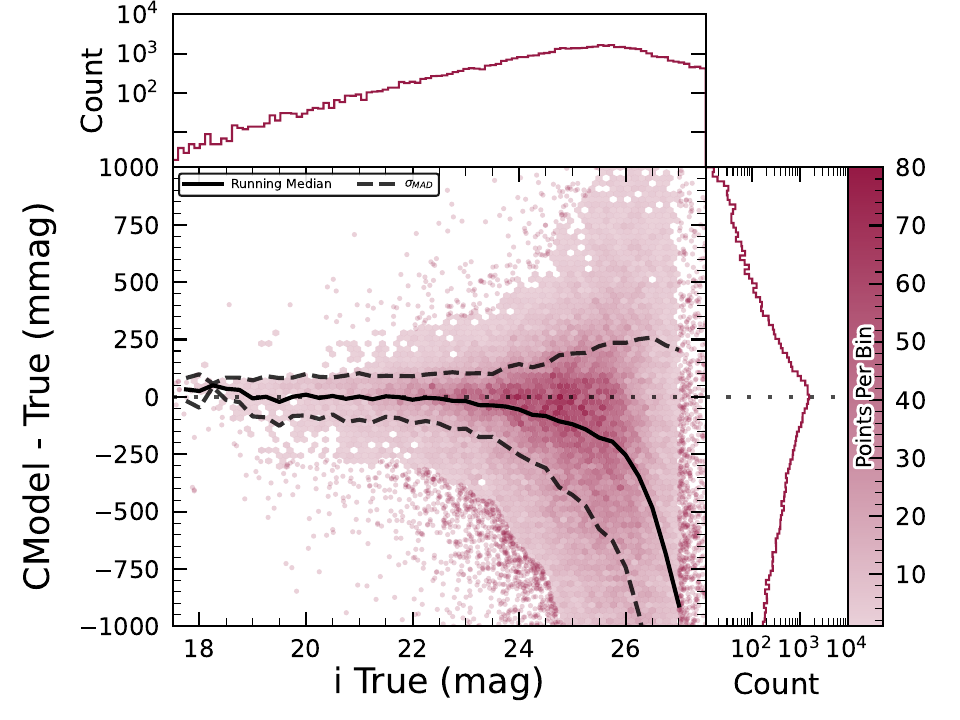}
  \caption{$i$-band magnitude residuals for CModel measurements of injected galaxies.}
  \end{subfigure}\hfill
  \begin{subfigure}[t]{0.45\textwidth}
  \includegraphics[width=\linewidth]{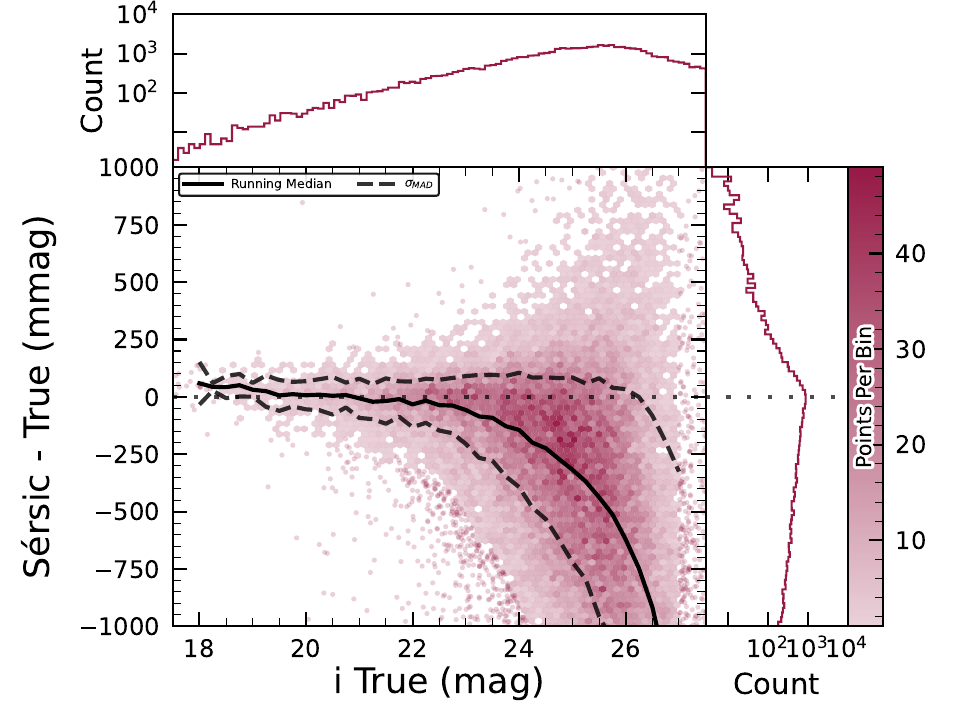}
  \caption{$i$-band magnitude residuals for S\'ersic model measurements of injected galaxies.}
  \end{subfigure}\hfill
\caption{$i$-band magnitude residuals for matched injected DC2 galaxies with the CModel and S\'ersic algorithms in a portion of the \gls{ECDFS} region, including the median and scatter thereof.
The black line is the median.}
\label{fig:injected_lsst_cells_v1_5063_i_mag}
\end{figure*}

\figref{fig:injected_lsst_cells_v1_5063_i_mag} shows $i$-band magnitude residuals for CModel and S\'ersic measurements using the matched injected galaxies described in \secref{ssec:detection_completeness}.
Similar behavior is seen in other bands.
S\'ersic fluxes show reduced scatter for galaxies with $i<22.5$, though CModel fluxes are less biased, with median residuals closer to zero and less magnitude-dependent.
For fainter objects, S\'ersic fluxes are more biased and less accurate.
The magnitude of this bias is considerably larger than previously seen in simulated data.
Subsequent testing indicates that this bias can be (roughly) halved by fitting an exponential model first, and then using those parameters to initialize a free S\'ersic fit.
This approach will be adopted in future releases.
Aperture fluxes - including Kron and \gls{GAaP} - are not shown as they are not corrected to yield total fluxes.
The correction for Kron fluxes can be derived from the S\'ersic index \citep{2005PASA...22..118G}, but this correction is not provided in object tables.

\begin{figure*}[hbt!]
  \centering
  \begin{subfigure}[t]{0.31\textwidth}
\includegraphics[width=\linewidth]{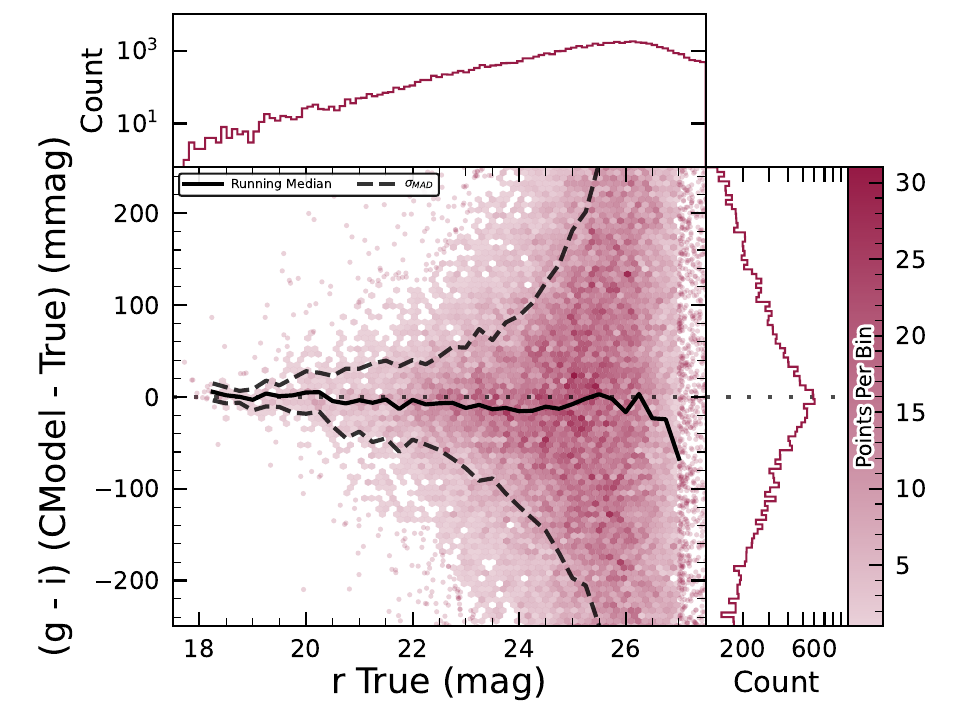}
  \caption{$g-i$ color residuals for CModel measurements of injected galaxies.}
  \end{subfigure}\hfill
  \begin{subfigure}[t]{0.31\textwidth}
\includegraphics[width=\linewidth]{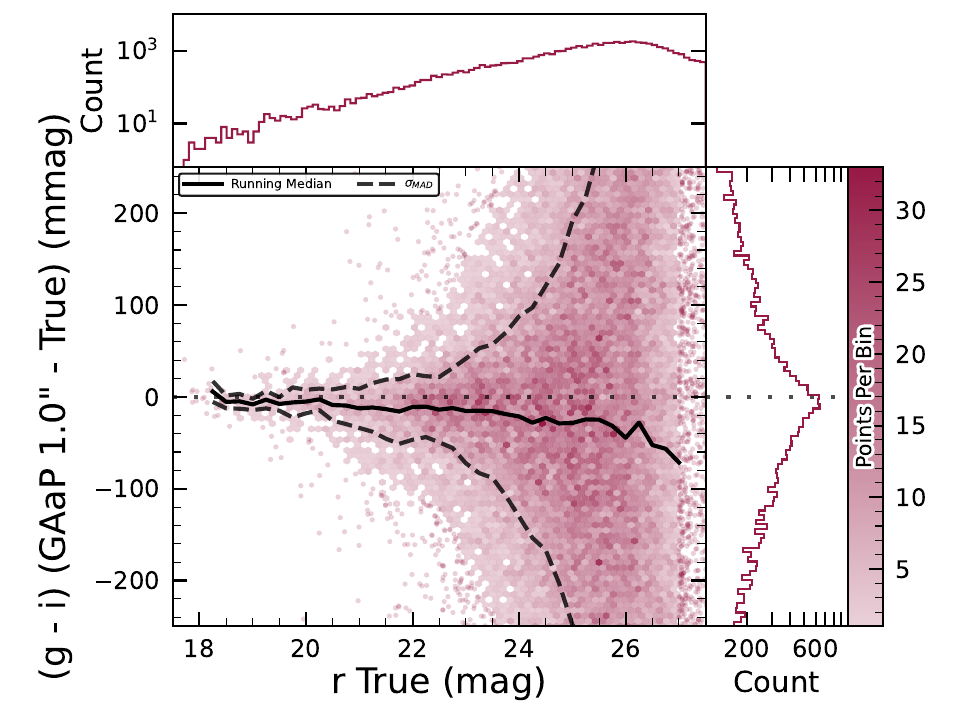}
  \caption{$g-i$ color residuals for \gls{GAaP} measurements of injected galaxies.}
  \end{subfigure}\hfill
    \begin{subfigure}[t]{0.31\textwidth}
\includegraphics[width=\linewidth]{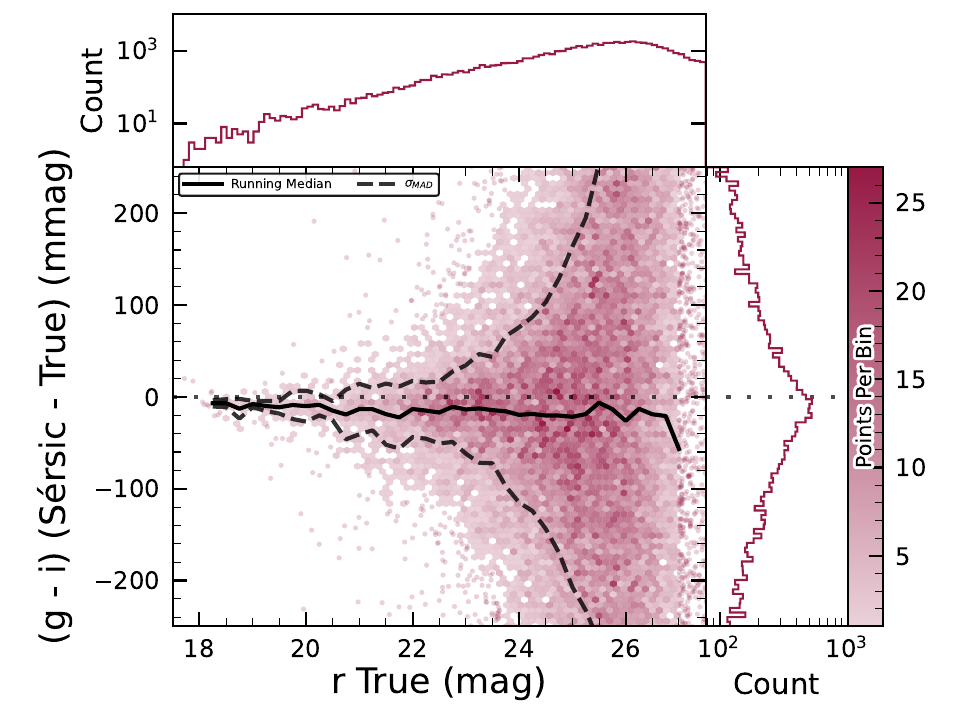}
  \caption{$g-i$ color residuals for S\'ersic model measurements of injected galaxies.}
  \end{subfigure}\hfill
\caption{$g-i$ color residuals versus true $r$-band magnitude for matched injected DC2 galaxies with the CModel, \gls{GAaP} and S\'ersic algorithms in a portion of the \gls{ECDFS} region.}
\label{fig:injected_lsst_cells_v1_5063_r_color_g_minus_i}
\end{figure*}

\figref{fig:injected_lsst_cells_v1_5063_r_color_g_minus_i} shows $g-i$ color residuals versus $r$-band magnitude for the same sample of galaxies as \figref{fig:injected_lsst_cells_v1_5063_i_mag}.
For this and most other colors, \gls{GAaP} (with a $1''$ aperture) and S\'ersic colors both yield lower scatter; however, the CModel colors have the smallest bias.
Curiously, the \gls{GAaP} bias appears to be magnitude-dependent, whereas the S\'ersic bias remains stable from $19<r<26$.
Any of these color measurements are suitable for use for deriving quantities like photometric redshifts, stellar population parameters, etc.

In addition to photometry, some algorithms include measurements of structural parameters like size, ellipticity, and S\'ersic index.
One particular known issue is that many (truly) faint objects have significantly overestimated sizes and fluxes.
This was also seen in the Dark Energy Survey \citep{2025arXiv250105739B}, who dubbed such objects ``super-spreaders''.
These super-spreaders contribute significantly to overestimated fluxes at the faint end (see e.g. \figref{fig:injected_lsst_cells_v1_5063_i_mag}), and are particularly problematic for the Kron algorithm \citep{1980ApJS...43..305K}, which should only be used with caution.

As mentioned in \secref{ssec:coadd_processing}, the S\'ersic fits include a free centroid, which is initialized from the fiducial centroid of the object.
Preliminary analyses of matched injected objects suggest that the S\'ersic model galaxy \gls{astrometry} residuals are somewhat smaller than for the standard centroids used in other measurements, and so users of the S\'ersic photometry should also use these centroid values.
One caveat is that for faint objects and/or in crowded regions with unreliable deblending, free centroids can drift significantly and potentially towards other objects, so objects with large differences between the fiducial and S\'ersic \gls{astrometry} should be discarded or used with caution.

S\'ersic model parameter uncertainties are estimated by computing and inverting the Hessian matrix with the best-fit parameter values, after replacing the pixel data (but not uncertainties) by the best-fit model values.
Currently, only the on-diagonal dispersion term (square root of the variance) is provided as an error estimate for each parameter.
Future releases may provide more off-diagonal terms of the covariance matrix - particularly for the structural parameters, which are known to be correlated.

A major outstanding issue is that many parameter uncertainties - including but not limited to those for fluxes - are underestimated.
This is at least partly (but not wholly) due to the fact that coaddition introduces covariance between pixels, which is not captured in per-pixel variances.

\begin{figure*}[hbt!]
  \centering
  \begin{subfigure}[t]{0.45\textwidth}
  \includegraphics[width=\linewidth]{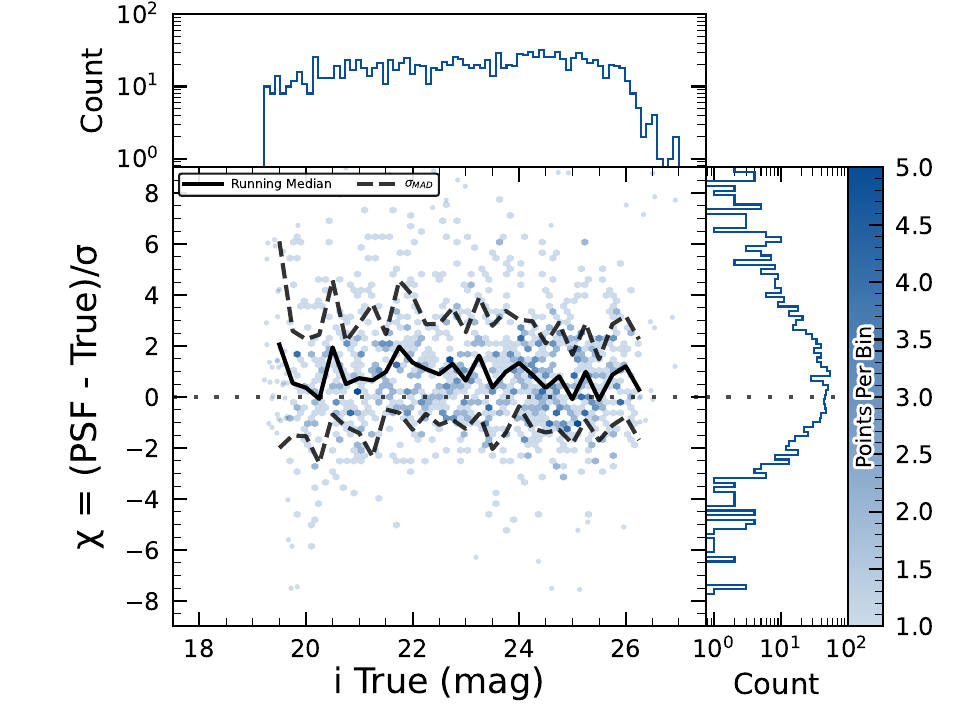}
  \caption{$i$-band flux uncertainty-scaled residuals for PSF model measurements of injected stars.}
  \end{subfigure}\hfill
  \begin{subfigure}[t]{0.45\textwidth}
  \includegraphics[width=\linewidth]{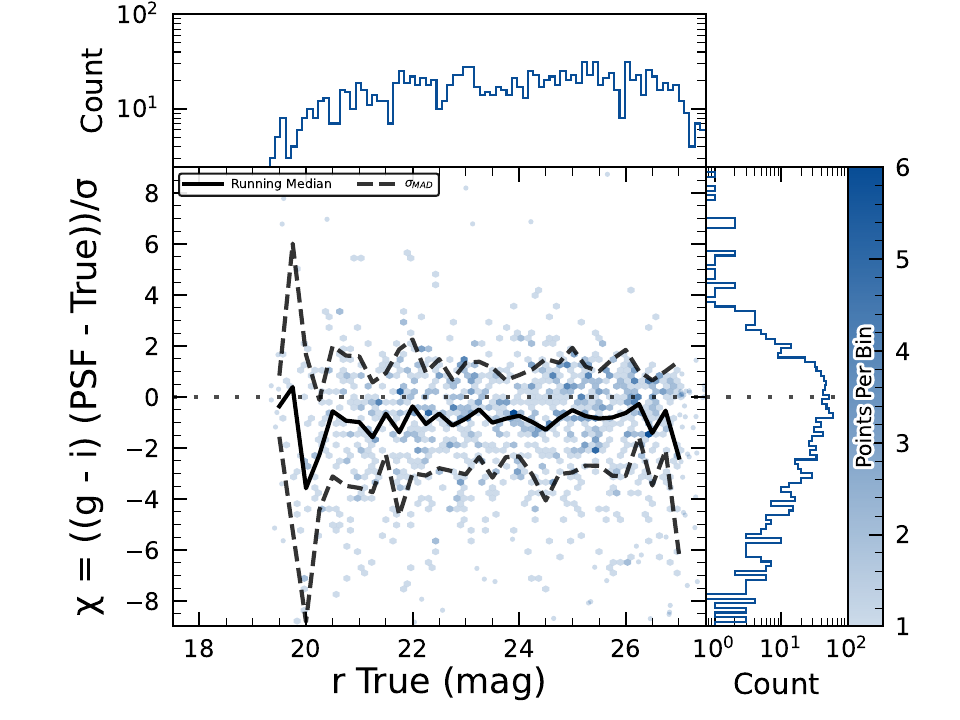}
  \caption{$g-i$ color uncertainty-scaled residuals for PSF model measurements of injected stars.}
  \end{subfigure}\hfill
\caption{Color and flux uncertainty-scaled residuals for matched injected DC2 stars' PSF model measurements in a portion of the \gls{ECDFS} region.}
\label{fig:injected_lsst_cells_v1_5063_star_psf_chi}
\end{figure*}

The degree to which uncertainties are underestimated can depend on the parameter in question and on the brightness of the object.
In plots of uncertainty-scaled residuals, the ideal behavior is for the median (i.e. the bias) to lie close to zero, and for the $\pm1\sigma$ lines to lie at $\pm1$, without any dependence on magnitude.
\figref{fig:injected_lsst_cells_v1_5063_star_psf_chi} shows that flux and color uncertainties for PSF model magnitudes of injected stars are both underestimated, but by a factor of approximately $1.7-2$ that is not very sensitive to \gls{SNR}.
This holds for astrometric/centroid parameters as well.

\begin{figure*}[hbt!]
  \centering
  \begin{subfigure}[t]{0.45\textwidth}
  \includegraphics[width=\linewidth]{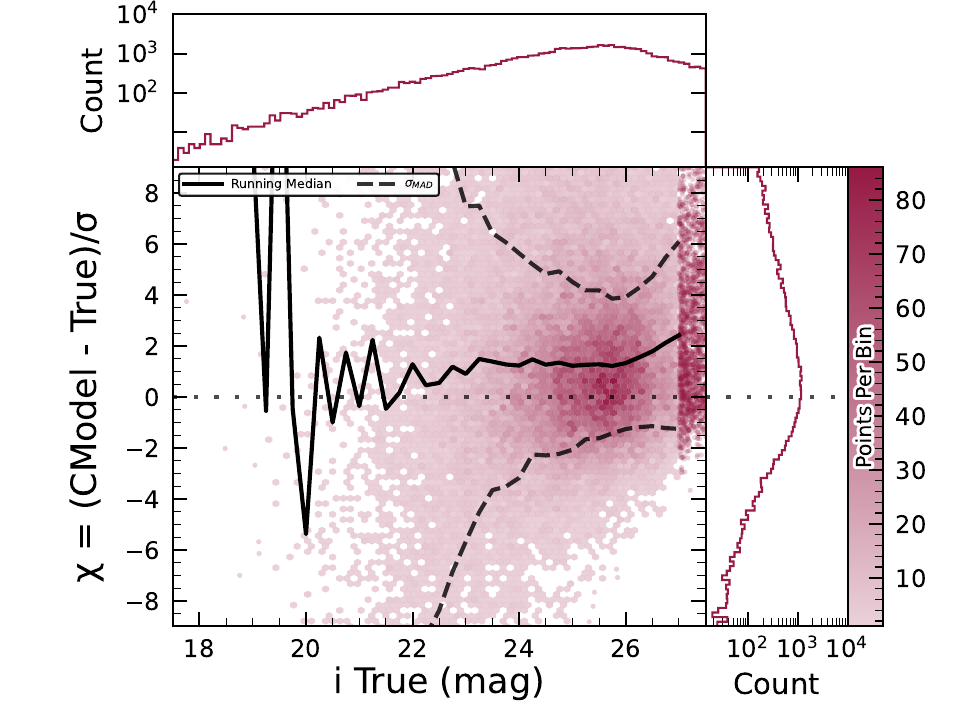}
  \caption{$i$-band flux uncertainty-scaled residuals for CModel measurements of injected galaxies.}
  \end{subfigure}\hfill
  \begin{subfigure}[t]{0.45\textwidth}
  \includegraphics[width=\linewidth]{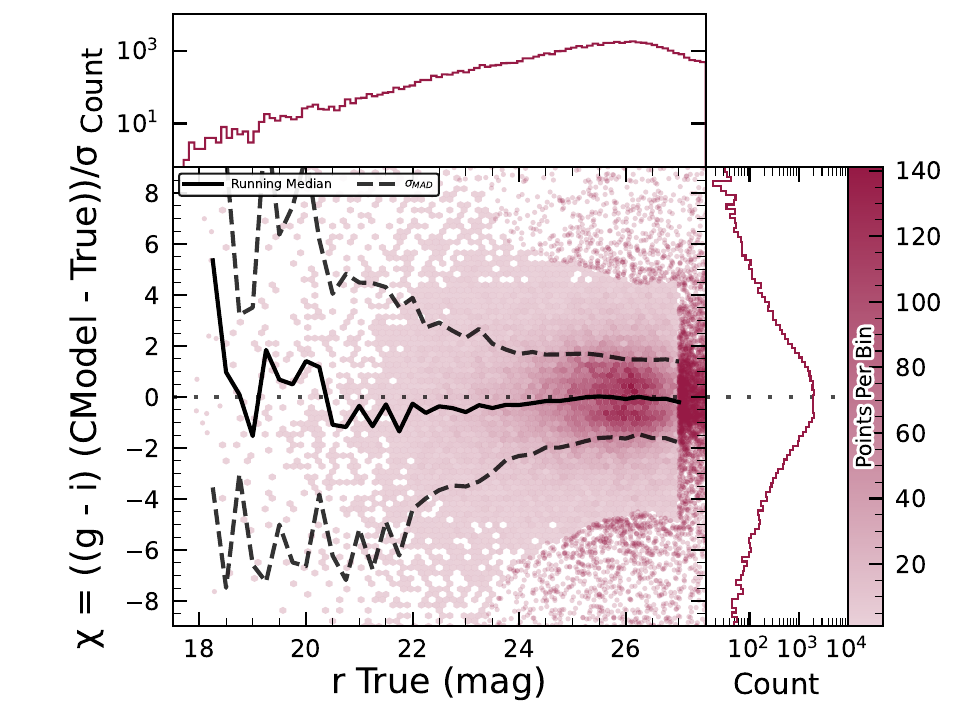}
  \caption{$g-i$ color uncertainty-scaled residuals for CModel measurements of injected galaxies.}
  \end{subfigure}\hfill
\caption{Color and flux uncertainty-scaled residuals for matched injected DC2 galaxies' CModel measurements in a portion of the \gls{ECDFS} region.}
\label{fig:injected_lsst_cells_v1_5063_galaxy_cmodel_chi}
\end{figure*}

\begin{figure*}[hbt!]
  \centering
  \begin{subfigure}[t]{0.45\textwidth}
  \includegraphics[width=\linewidth]{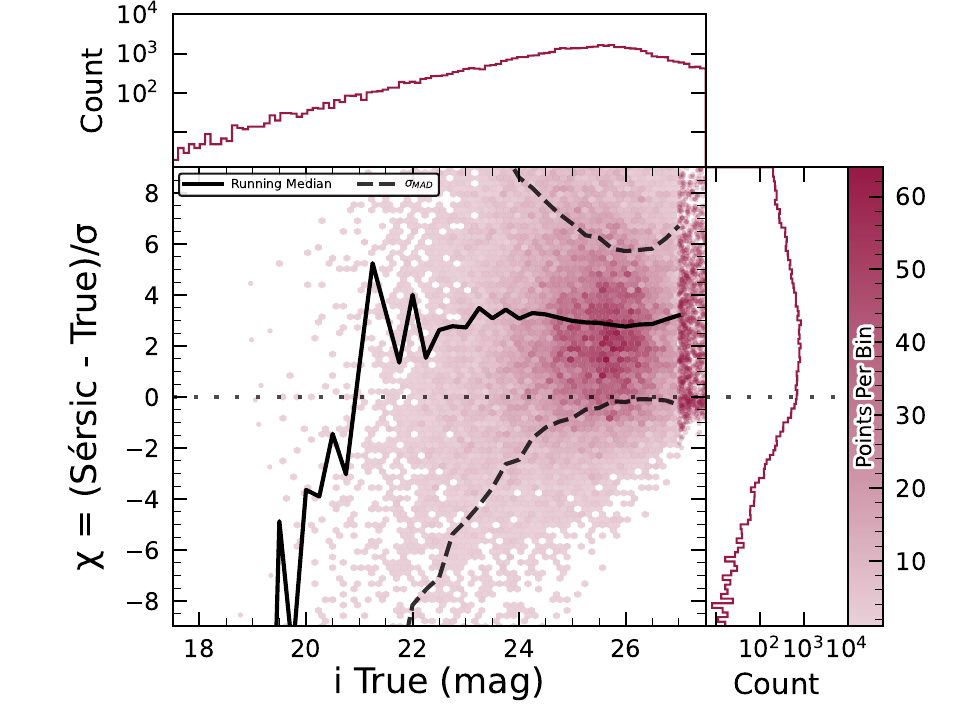}
  \caption{$i$-band flux uncertainty-scaled residuals for S\'ersic model measurements of injected galaxies.}
  \end{subfigure}\hfill
  \begin{subfigure}[t]{0.45\textwidth}
  \includegraphics[width=\linewidth]{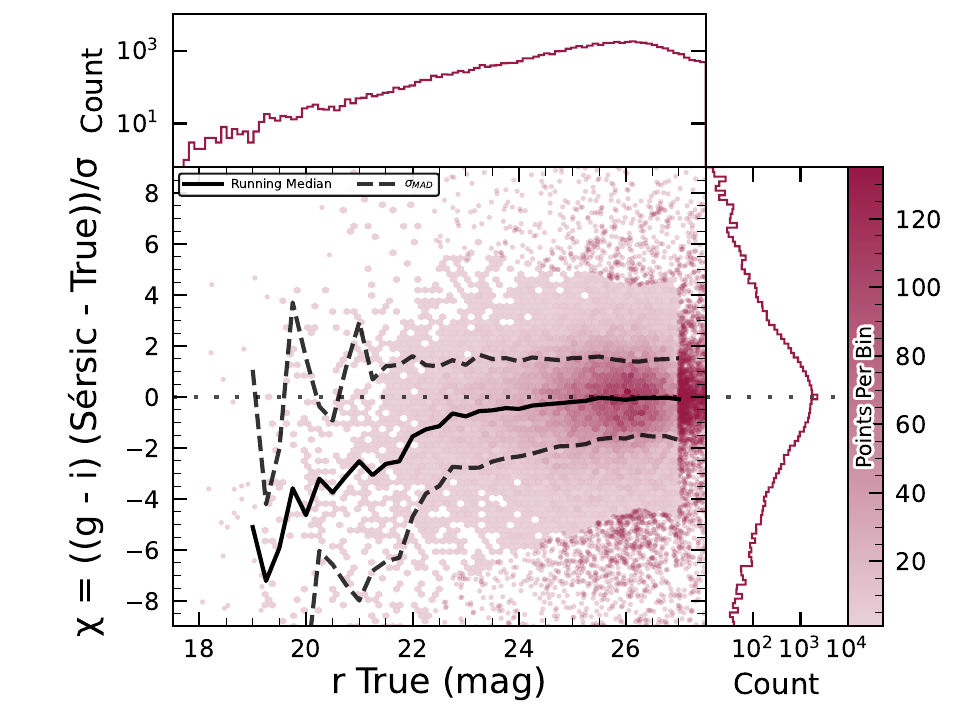}
  \caption{$g-i$ color uncertainty-scaled residuals for S\'ersic model measurements of injected galaxies.}
  \end{subfigure}\hfill
\caption{Color and flux uncertainty-scaled residuals for matched injected DC2 galaxies' S\'ersic measurements in a portion of the \gls{ECDFS} region.}
\label{fig:injected_lsst_cells_v1_5063_galaxy_sersic_chi}
\end{figure*}

In turn, \figref{fig:injected_lsst_cells_v1_5063_galaxy_cmodel_chi} shows that CModel color uncertainties of galaxies are underestimated by a similar factor at the faint end, but with appreciable scaling with magnitude (and thereby \gls{SNR}).
Flux error underestimation is both larger than for colors and scales more strongly with \gls{SNR}.
This indicates that systematic effects dominate the errors in fluxes, particularly for bright galaxies.
This is also at least partly but not wholly due to so-called model inadequacy - that is, the fact that galaxy models, parameteric or otherwise, are insufficiently complex to capture the structure of real galaxies.

\figref{fig:injected_lsst_cells_v1_5063_galaxy_sersic_chi} shows that S\'ersic model fluxes and colors have similar behavior as CModel, but with a greater degree of overestimation.
This may be partly due to the fact that S\'ersic parameter uncertainties are estimated along with the free centroid and structural (shape and S\'ersic index) parameters, whereas the forced CModel fluxes and errors are derived from linear flux fits with a fixed shape and centroid.

Efforts are underway to investigate and quantify the origin of uncertainty underestimates and future releases will, at the least, provide recommendations for mitigations.

\subsection{Difference Imaging} 
\label{ssec:performance_dia}
We assessed the performance of image differencing using both human vetting (\secref{sssec:perf_dia_purity}) and source injection (\secref{sssec:perf_dia_completeness}).

\subsubsection{Difference Imaging Purity} 
\label{sssec:perf_dia_purity}
\begin{figure}[htb!]
\includegraphics[width=\linewidth]{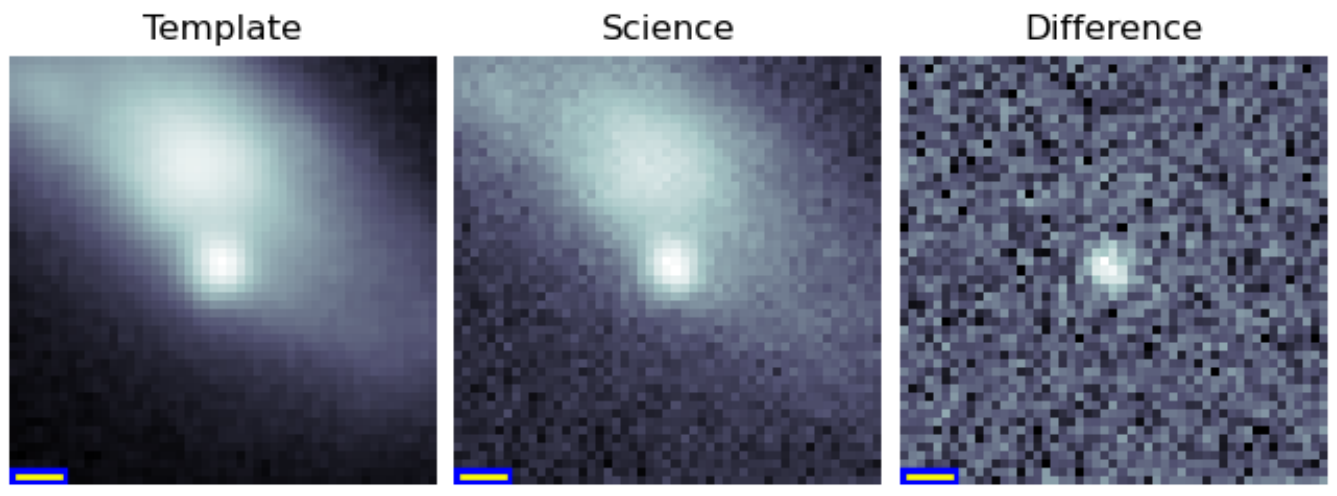}
\caption{An example an image triplet consisting of cutouts showing, from left to right,  the template, science, and difference images that volunteers were asked to label.}
\label{fig:sample_dia_labelling}
\end{figure}

Members of the \gls{DP1} team labeled more than 11,000 DIASource image triplets, each consisting of cutouts from the science, template, and difference images. 
An internal labeling service (\texttt{tasso}) was deployed within the USDF environment. 
A random subset of approximately 16,000 DIASources was selected and uploaded to the service, which remained active for roughly three months and labeled by members of the \gls{DP1} team. 
Users labeled  DIASource PNG images triplets, each consisting of cutouts from the science, template, and difference images. Each stamp had dimensions of 51$\times$51 pixels, matching the input size required by the machine-learning model.
Access to the labeling service was granted to all individuals with commissioning data access. 
Each DIASource was classified exactly once, with a total of 35 volunteers contributing labels. 
\figref{fig:sample_dia_labelling} show an example of one of the image triplets consisting of cutouts from the science, template, and difference images that volunteers were asked to label.

The labeled sources were classified into multiple categories representing real astrophysical events and artifacts.
Prior to any filtering, the raw artifact-to-real ratio was approximately 9:1. 
Bright stars were identified as the dominant source of artifacts, while correlated noise, particularly in the $u$ and $g$ bands, also produced spurious detections near the flux threshold. 
We expect to be able to mitigate these effects in future \gls{LSSTCam} data.

Applying a reliability threshold based on the Machine Learning reliability model described in \secref{ssec:diffim_analysis} improved the purity of transient detections but had limited impact on variable stars. 
This limitation arises from technical constraints at the time of model training, which prevented the injection of variable stars into the synthetic training set.
Future reliability models for \gls{LSSTCam} data, described in \secref{ssec:diffim_analysis}, will be trained using a broader and more representative range of input data.

\begin{figure}[htb]
\includegraphics[width=\linewidth]{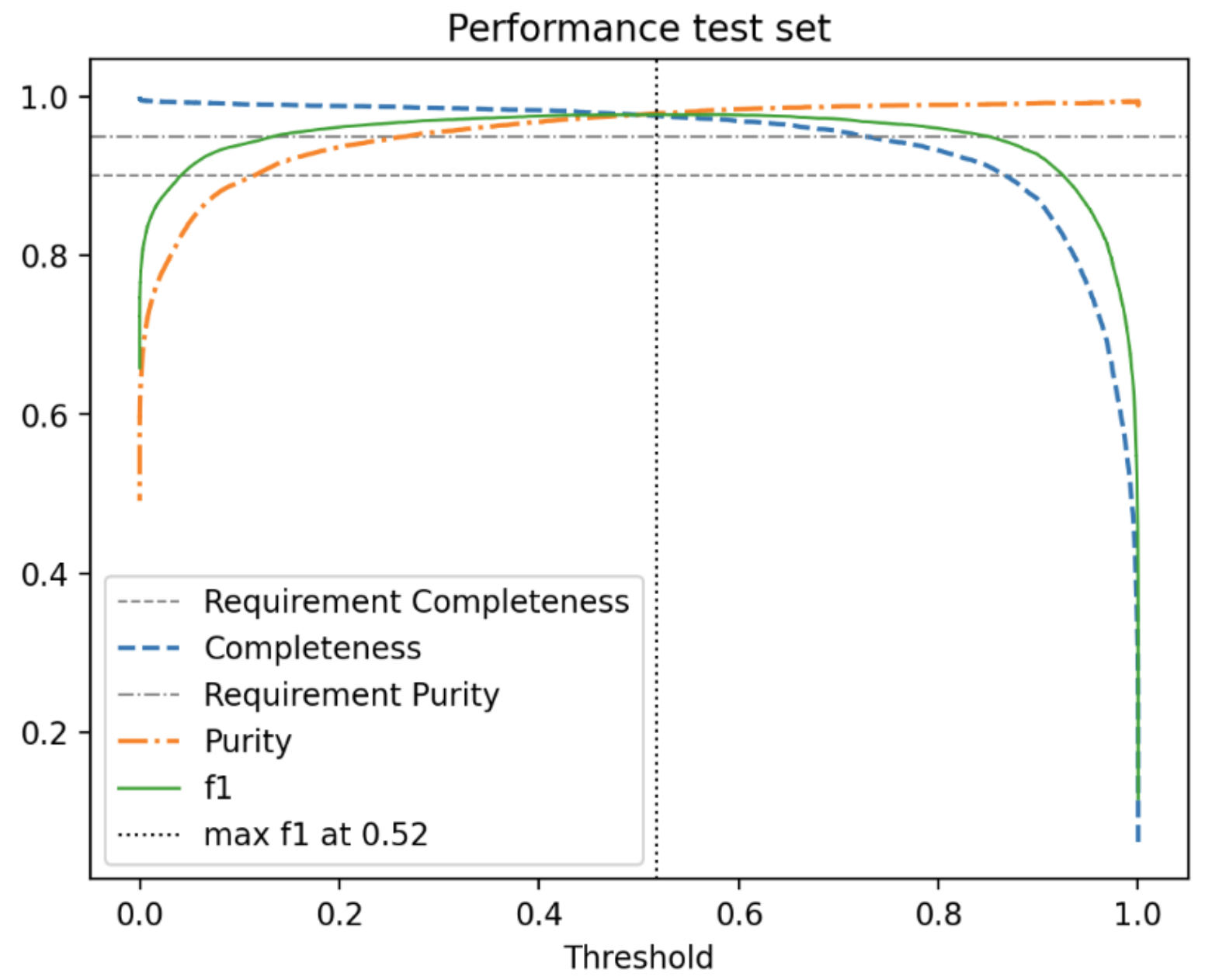}
\caption{The purity and completness of the reliability score is shown as a function of reliability threshold based on the testing data. A vertical line marks the threshold where the highest F1-score is obtained. The F1 score is the harmonic mean of completeness and purity. See \secref{ssec:diffim_analysis} for details on the model and model training.}
\label{fig:completenessvpurity}
\end{figure}

The performance of the reliability model on the test data (\secref{ssec:diffim_analysis}) is shown in \figref{fig:completenessvpurity}.
The rate of true positives and false negatives obtained by thresholding the reliability score at 0.5 is reported for transients (99 stamps), and variable stars (316 stamps) vetted in \texttt{tasso} in \tabref{tab:mlreliability}. 
\setlength{\tabcolsep}{8pt} 
\begin{deluxetable}{lccc}
\caption{The rate of true positives (TP) and false negatives (FN) obtained by thresholding the reliability score at 0.5  for Solar system objects, transients and  variable stars.}
\label{tab:mlreliability}
\tablehead{
  \colhead{\textbf{Object Type}} & 
  \colhead{\textbf{Number}} & 
  \colhead{\textbf{TP Rate}} & 
  \colhead{\textbf{FN Rate} }
}
\startdata
Solar System  & 5,988 & 93.5\% & 6.5\%\\
Transients & 99 & 73.7\% & 26.3\% \\
Variables & 316 & 3.5\% &. 96.5\% \\
\enddata 
\end{deluxetable} Additionally we crossmatched stamps with Solar System Objects with known orbits retrieving 5,988 Solar System Objects stamps.

\subsubsection{Difference Imaging Detection Completeness} 
\label{sssec:perf_dia_completeness}
We assess the performance of our difference imaging \gls{pipeline} using synthetic source injection on the science images prior to differencing.
We construct a catalog of injected sources by joining two different samples of point sources, a set of hosted sources to emulate transients in galaxies and second set of hostless sources.
The hosts are selected from the pipeline source catalog that is produced upstream by imposing a cut on their extendedness measurement and selecting $N_{\rm src}={\rm min}(100, N\times0.05)$ of the $N$ available sources per detector.
For each host we pick a random position angle and radius using its light profile shape to decide where to place the source, and also a random value of brightness for the injected source, with magnitudes higher than the host source.

The hostless sources instead have random positions in the \gls{CCD} focal plane, and magnitudes chosen 
from a random uniform distribution with $20 \geq m \geq m_{lim} + 1$,  where $m_{lim}$ is the limiting magnitude of the image.
We used the \gls{LSST}  \texttt{source\_injection} package\footnote{\nolinkurl{https://pipelines.lsst.io/modules/lsst.source.injection/index.html}} to include these sources in our test images.
We performed a coordinate cross-match task, with a threshold of $0\farcs5$ to find which of these sources were detected and which were lost, enabling the calculation of a set of performance metrics.

\begin{figure}[htb!]
\includegraphics[width=\linewidth]{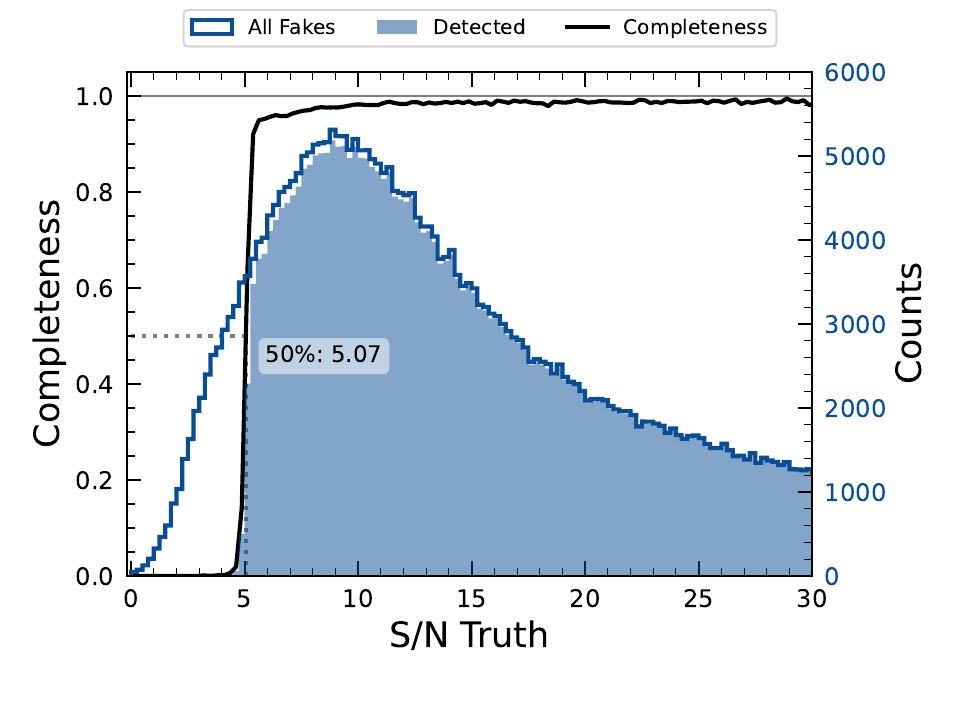}
\caption{The difference image detection completeness for injected sources in the \gls{ECDFS} field, for filters \textit{griz}, as a function of the estimated signal to noise ratio SNR. 
This completeness is the ratio between the found fake sources (shaded histogram) and all the sources (solid line). 
The horizontal dashed line represents where the $50\%$ completeness level is reached, at approximately SNR $\simeq 5.07$.}
\label{fig:eff_snr_griz}
\end{figure}

\begin{figure}[htb!]
\includegraphics[width=\linewidth]{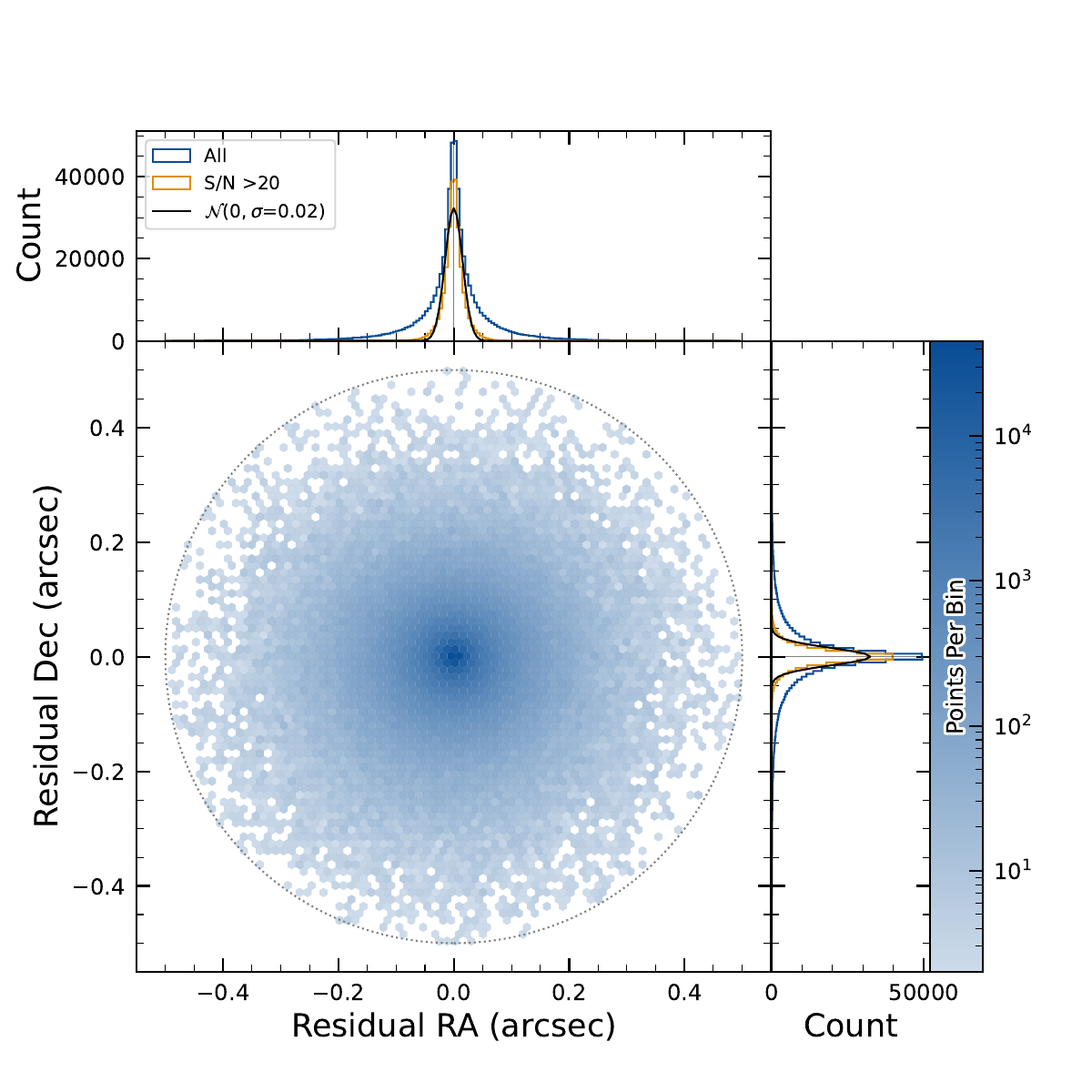}
\caption{Coordinate residuals for detected synthetic sources in difference images, between recovered and true position of the sources in the \gls{ECDFS} field. 
In the top and right panels we include the distribution of these offsets, for all sources as well as for sources with SNR$>20$. 
These high SNR sources show gaussian coordinate residual distributions with $\sigma=0\farcs02$ (black solid lines). 
The circle reflects the matching radius of $0\farcs5$.}
\label{fig:coordinate_offset_diffim_fakes}
\end{figure}

\begin{figure}
\includegraphics[width=\linewidth]{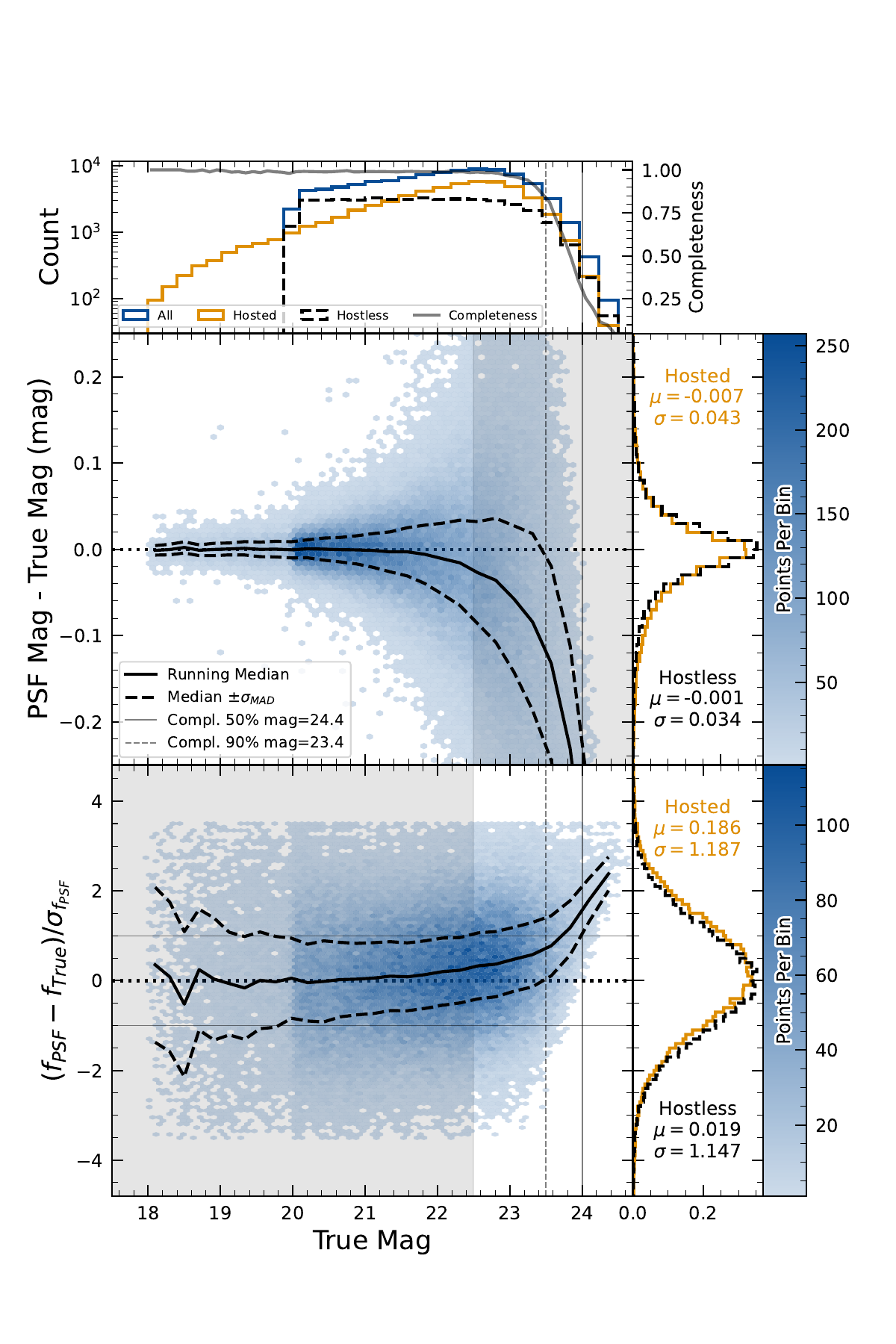}
\caption{Magnitude residuals and flux pulls for $i$-band \gls{PSF} photometry on difference images for ECDFS field in $i$ for detected injected sources.
Top panel: Distribution of true magnitudes for injected sources (blue), and split  into hostless (black dash) and hosted (orange) sources, with detection completeness as a function of true magnitude (gray line). 
Vertical dashed lines indicate the 90\% and 50\% completeness magnitude limits.
Center left panel: 2D hexbin plot of PSF magnitude residuals (measured minus true) versus true magnitude for detected sources, with running median (solid black) and $\sigma_{MAD}$ (dashed black) overlaid.
Center right panel: Marginalized distributions of PSF magnitude residuals for hostless (blue) and hosted (orange) sources with true magnitude $m_i < 22.5$, annotated with robust mean and standard deviation. 
Bottom left panel: 2D hexbin plot of PSF flux pulls versus true magnitude for detected sources, with running median (solid black) and $\sigma_{MAD}$ (dashed black) overlaid.
Bottom right panel: Marginalized distributions of PSF flux pulls for hostless (blue) and hosted (orange) sources with true magnitude $m_i < 22.5$, annotated with robust mean and standard deviation.}
\label{fig:phot_residual_diffim_fakes}
\end{figure}

In \figref{fig:eff_snr_griz} we show the detection completeness as a function of the \gls{SNR}, for sources in the \gls{ECDFS} field, for filters $griz$. 
We observe a completeness $>95\%$ for sources with \gls{SNR}$> 6$, with mean completeness $\simeq 99\%$ and standard deviation of $\simeq 0.7\%$.
In \figref{fig:coordinate_offset_diffim_fakes} we show the distribution of the residuals of the recovered sky coordinates for the detected synthetic sources. The marginal distributions are both centered at zero, and for sources of SNR $>20$ the residuals are compatible with normal distributions $\mathcal{N}(\mu=0, \sigma^2=(0''.02)^2)$.
In \figref{fig:phot_residual_diffim_fakes} we show photometry results for our detected synthetic sources in the \textit{i} filter, using \gls{PSF} photometry on the difference images. 
We include both the magnitude residuals as well as the flux pulls, defined as $(f_{PSF}-f_{\rm{True}})/\sigma_{f_{PSF}}$, where $f_{\rm True}$ is the true flux, $f_{PSF}$ is the PSF flux and $\sigma_{f_{PSF}}$ is its uncertainty, as a function of the true magnitude of the synthetic sources, including the running median and median absolute deviation (MAD) for the whole brightness range. 
We also include the true magnitude distribution as well as the detection completeness on the top panel, and for reference the $90\%$ and $50\%$ completeness magnitude values in vertical lines. 
On the right panels we include the marginal distribution for sources brighter than $22.5$~mag, splitting the data into hosted and hostless, as well as the robust mean and standard deviation.
From this figure we can see that our \gls{flux} measurements are accurate within a wide range of magnitudes, for both hosted and hostless synthetic sources. 
We find that the median offset is below $0.002$ mag for true magnitudes below $21$, and with a maximum $\sigma_{MAD}$ scatter of about $0.02$ mag in this range.
For true $m_i < 22.5$, the robust running median PSF magnitudes residuals are $<0.02$ mag, and when splitting into hosted and hostless both robust median are well below $0.01$, and robust $\sigma$, i.e. $\sigma_{MAD}$ are also well below $0.05$.
For all sources with $m_i<21.5$ the running median is always $|\left<\delta\right>| <0.1$, and MAD $\sigma_\delta < 1$. 
Extending to sources with $m_i<22.5$ then  hostless sources have  a robust mean pull below $0.02$, with a robust standard deviation $<1.15$, while these parameters increase to $0.2$ and $1.2$ for hosted sources, suggesting that we might have contamination from host background sources potentially biasing our fluxes.

\subsection{Solar System}
\label{sec:performance:solsys}

\subsubsection{Asteroid Linking Performance}

The evaluation of asteroid linking performance in \gls{DP1} focused on demonstrating discovery capability.
The solar system discovery pipeline produced 269,581 tracklets, 5,691 linkages, and 281 post-processed candidates.

As described in \secref{sec:drp:solsys}, post-processing of the \texttt{heliolinc} output with \texttt{link\_purify} produced a final set of 281 candidate linkages, ranked with the most promising first. 
We then used \texttt{find\_orb} \citep{findorb} to derive orbit fits for each candidate, sorting the resulting list by $\chi\_{\rm dof}^2$, a measure of fit quality. 
A conservative manual investigation of these candidates yielded a curated list of \nnewasteroiddiscoveries probable new asteroid discoveries.
Manual inspection of the linkages indicated that those ranked 0--137 corresponded to unique real asteroids; ranks 138--200 contained additional real objects intermixed with some spurious linkages; and ranks higher than 200 were essentially all spurious.
This analysis indicates that it will be possible to identify cuts on quality metrics such as $\chi^2$ to define discovery candidate samples with high purity; determining the exact quantitative cut values requires more data with \gls{LSSTCam}.
We next removed all observations matched to known asteroids (using \gls{MPC}'s MPChecker service), reducing the number of candidates to 97.
Of these, four had strong astrometric and/or photometric outliers, likely due to self-subtraction in difference images due to the unavoidable limitations of template generation from the limited quantity of data available from  \gls{LSSTComCam}.
We suspect these four linkages do correspond to real objects, but have chosen to discard them out of an abundance of caution.
The remaining \nnewasteroiddiscoveries were submitted to the Minor Planet Center and accepted as  discoveries, demonstrating the \gls{LSST} pipelines are able to successfully discover new solar system objects.

\subsubsection{Asteroid Association Performance}
\label{ssec:asteroid_association}

During the Solar System association step,  \nsolarsystemsources DiaSources were linked to \nsolarsystemobjects  unique Solar System objects, 
These include 3,934 DiaSources with 338 previously known objects cataloged by the \gls{MPC}, and 2,054 DiaSources with the \nnewasteroiddiscoveries  newly-discovered objects, all of which are main belt asteroids. 
An additional 143 detections of these newly discovered objects were also recovered. 
These detections were not initially identified by the discovery pipelines, as they did not meet the required criteria for tracklet formation, specifically the minimum number of detections and/or the maximum allowed time span between observations.

\begin{figure}[htb!]
\includegraphics[width=\linewidth]{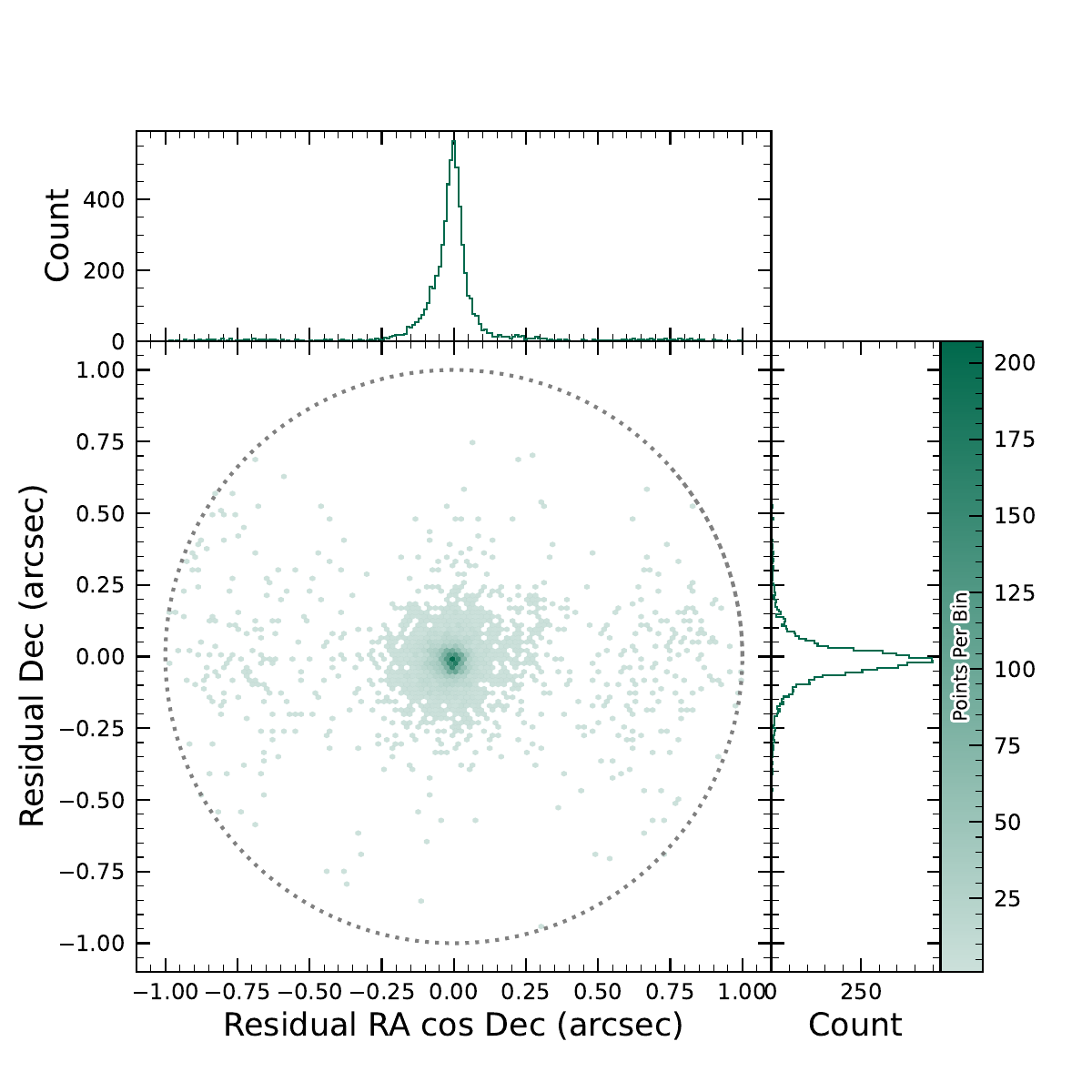}
\caption{Astrometric residuals between expected and observed positions of Solar System Objects in \gls{DP1}. 
The median residuals are $0\farcs001$ and $-0\farcs016$ in R.A./Dec direction, with  standard deviations of $0.''19$ and $0.''10$, respectively. 
No detectable systematic offset from zero indicates there are no major errors in either timing or astrometry delivered by the Rubin system. 
The wider scatter in the RA direction is due to objects whose measured orbital elements are less well constrained, translating to larger along-track positional errors in the predicted positions.}
\label{fig:sso_residuals}
\end{figure}

The astrometric residuals of known asteroid associations are shown in \figref{fig:sso_residuals}.
The astrometric precision for solar system sources is excellent, with the majority of objects detected within $0\farcs1$ of their expected positions.

By analyzing the signed median residuals to search for biases, we find that previously-known objects have mean residuals of $0\farcs001$ and $-0\farcs016$ in the RA and Dec directions respectively, whereas newly-discovered objects have mean residuals of $-0\farcs035$ and $-0\farcs010$ in the RA and Dec directions, respectively.
These mean residuals are small enough to eliminate the possibility of a timing offset greater than the second-scale shutter motion, which is consistent with the timing studies presented in \secref{ssec:comcam_timing}.

The wider scatter in the RA residuals is due to objects whose measured orbital elements are less well constrained, translating to larger along-track positional errors in the predicted positions
Observations of objects with large residuals are the most valuable ones from the point of view of improving the orbit, which is why we kept a generous matching radius.
However, in future releases we are likely to couple this with either orbit fitting to verify the ``singleton" match, or require two near-in-time observations (a tracklet) that match the expected motion vector as well.

Optimal moving source attribution is an area of  active work that we expect to fully converge in time of \gls{DR1}.
In the meantime, for \gls{DP1} we've opted to start with simple, more easily understandable, criteria.

\subsection{Crowded Fields}
\label{ssec:crowded_fields}
Among the seven Rubin \gls{DP1} target fields, two stand out for their severe stellar crowding: the globular cluster 47\,Tucanae (47\_Tuc) and the Fornax dwarf spheroidal galaxy (Fornax dSph).
These fields were selected in part to stress-test the LSST Science Pipelines under high-density conditions. 
While both exhibit high stellar densities, the nature and spatial extent of the crowding differ significantly.

47\,Tuc presents extreme crowding across much of the field, encompassing its dense core and the eastern regions influenced by the Small Magellanic Cloud (SMC). 
This pervasive crowding leads to persistent challenges for deblending and reliable source detection, exposing field-wide limitations in the current pipeline performance \citep{2025arXiv250701343C}.
In contrast, Fornax\,dSph shows significant crowding only in its central region, with outer areas remaining well resolved and easier to process. 

In both 47\,Tuc and Fornax, extreme crowding led to the deblending step being skipped frequently when memory or runtime limits were exceeded, typically due to an excessive number of peaks, or large parent footprints.
However, the impact of these limitations differed: in 47\,Tuc, deblending was often skipped across the entire field, resulting in large gaps and substantially reduced completeness. 
In Fornax, these issues were largely confined to the central region, with much better recovery in the outskirts. 
This contrast highlights how the pipeline’s limitations depend on the spatial extent of high-density regions: 47\,Tuc exposed systematic, field-wide challenges, whereas Fornax revealed more localized, density-driven limits.

\citet{2025RNAAS...9..171W} explored the Rubin \gls{DP1} \texttt{DiaObject} catalog (\secref{ssec:catalogs}) in the 47\,Tuc field, which contains sources detected in difference images. 
Because forced photometry is performed at these positions across all single-epoch images, this dataset bypasses the coadd-based detection and deblending stages that often fail in crowded regions. 
By computing the median of the forced photometry for each  \texttt{DiaObject} across available visits, they recovered approximately three times more candidate cluster members than found in the standard \texttt{Object table} \citep{2025arXiv250701343C}. 
This result underscores the value of difference-imaging–based catalogs for probing dense stellar regions inaccessible to standard coadd processing in \gls{DP1}.

Although the \gls{DP1} pipeline was not optimized for crowded-field photometry, these early studies of 47\,Tuc and Fornax provide critical benchmarks. 
They highlight both the limitations and opportunities for science with Rubin data in crowded environments, and they inform future pipeline development aimed at robust source recovery in complex stellar fields.
 \section{Rubin Science Platform}
\label{sec:data_services}

The \gls{RSP} \citep{LSE-319} is a powerful, cloud-based environment for scientific research and analysis of petascale-scale astronomical survey data.
It serves as the primary interface for scientists to access, visualize, and conduct next-to-the-data analysis of Rubin and \gls{LSST} data.
The  \gls{RSP} is designed around a  ``bring the compute to the data'' principle, eliminating the need for users to download massive datasets.
Although \gls{DP1} is much smaller in size (\sizeinbytes) than many current survey datasets, future \gls{LSST} datasets will be far larger and more complex, making it crucial to co-locate data and analysis for effective scientific discovery.

\begin{figure}[htb!]
\centering
\includegraphics[width=0.98\linewidth]{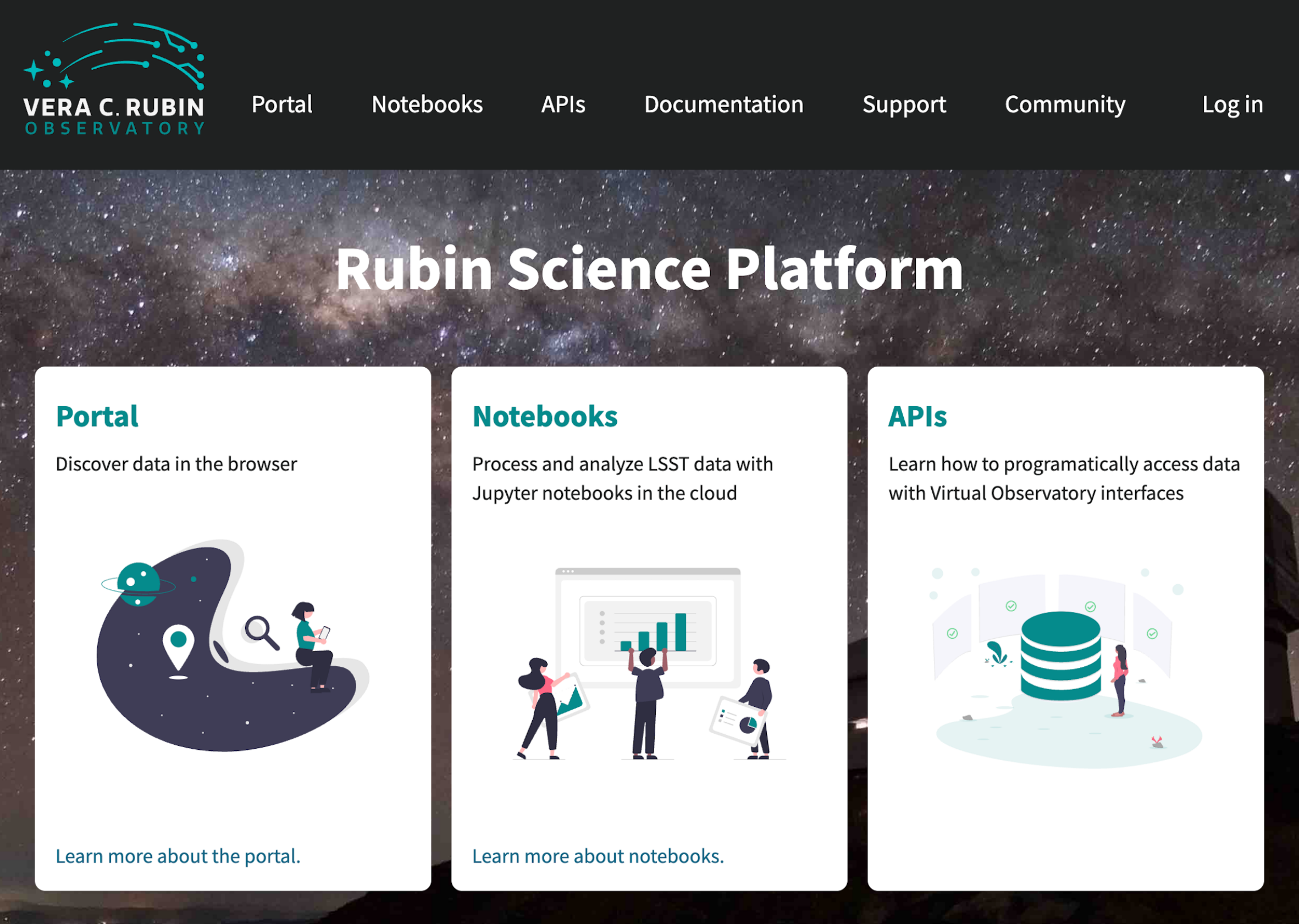}
\caption{The Rubin Science Platform landing page at \url{https://data.lsst.cloud/} showing the three user-facing Aspects as well as links to documentation and support information.}
\label{fig:rsp_landing_page}
\vspace{0.1cm}
\end{figure}

The \gls{RSP} provides users with access to data and services through three distinct user-facing Aspects: a \emph{Portal}, which facilitates interactive exploration of the data; a JupyterLab-based \emph{Notebook} environment for data analysis using Python; and an extensive set of \emph{\glspl{API}} that enable programmatic access to both data and services.
The three Aspects are designed to be fully integrated, enabling seamless workflows across the \gls{RSP}.
The data products described in \secref{sec:data_products} are accessible via all three Aspects, and the system facilitates operations such as starting a query in one Aspect and retrieving its results in another.
\figref{fig:rsp_landing_page} shows the Rubin \gls{Science Platform} landing page in the Google cloud.

The \gls{RSP} is supported by a number of back-end services, including databases, files, and batch computing.
Support for collaborative work through shared workspaces is also included in the \gls{RSP}.

A preview of the \gls{RSP} was launched on Google Cloud in 2022, operating under a shared-risk model to support \href{https://dp0.lsst.io/}{Data Preview 0} \citep{2024ASPC..535..227O}. This allowed the community to test the platform, begin preparations  for science, and provide valuable feedback to inform ongoing development.
It was the first time an astronomical research environment was hosted in a \gls{cloud} environment.
The \gls{DP1} release brings major updates to \gls{RSP} services, enhancing scientific analysis capabilities.
The \gls{RSP} remains under active development, with incremental improvements being rolled out as they mature.
During the Rubin Early Science Phase, the \gls{RSP} will continue to operate under a shared-risk model.
This section outlines the RSP functionality available at the time of the \gls{DP1} release and provides an overview of planned future capabilities.

\subsection{Rubin Data Access Center}
\label{ssec:usdac}
The Rubin US Data Access Center (US DAC) utilizes a novel hybrid on-premises-\gls{cloud} architecture, which combines on-premises infrastructure at the \gls{USDF} at SLAC with flexible and scalable resources in the Google \gls{cloud}.
This architecture has been deployed and tested using the larger simulated data set of DP0.2 \citep{2024SPIE13101E..2BO}.

In this hybrid model, user-facing services are deployed in the \gls{cloud} to support dynamic scaling in response to user demand and to simplify the provisioning and management of large numbers of science user accounts.
The majority of the static data products described in \secref{sec:data_products} are stored on-premises at the \gls{USDF} to benefit from cost-effective mass storage and close integration with Rubin data processing infrastructure, also located at the \gls{USDF}.
For imaging data, the Data Butler (\secref{sssec:data_butler}) provides the interface between the \gls{cloud}-based users and data services, and the on-premises data.
For catalog data, a \gls{cloud}-based \gls{TAP} client (\secref{sssec:ivoa_services}) submits queries to the on-premises \gls{Qserv} database cluster (\secref{ssec:databases}) and retrieves the results.
In the initial \gls{DP1} deployment, catalog data is hosted at the \gls{USDF} while image data is stored in the cloud.
The full hybrid model will be rolled out and further tested following the release of \gls{DP1}.
The RSP features a single-sign-on authentication and authorization system to provide secure access for Rubin data rights holders \citep{rdo-013}.

\subsection{API Aspect}
\label{ssec:rsp_api}
The \gls{API} Aspect provides a comprehensive set of user-facing interfaces for programmatic access to the \gls{DP1} data products, through both \gls{IVOA}-compliant services and the Rubin Data Butler. IVOA services enable standard queries and integration with existing tools, while the Butler facilitates advanced data processing within the LSST Science Pipelines.

At the time of the \gls{DP1} release, some \gls{IVOA} services are unavailable, and certain data products are only accessible via the Butler.
This section provides an overview of the available \gls{IVOA} services and Butler access.

\subsubsection{IVOA Services}
\label{sssec:ivoa_services}

Rubin has adopted a \gls{VO}-first design philosophy, prioritizing compliance with \gls{IVOA} standard interfaces to foster interoperability, standardization, and collaboration.
In cases where standardized protocols have yet to be established, additional services have been introduced to complement these efforts.
This approach ensures that the RSP can be seamlessly integrated with community-standard tools such as \gls{TOPCAT} \citep{2011ascl.soft01010T} and Aladin \citep{2000A&AS..143...33B, 2014ASPC..485..277B, 2022ASPC..532....7B}, as well as libraries such as  PyVO \citep{2014ascl.soft02004G}.

The user-facing \glspl{API} are also used internally within the \gls{RSP}, creating a unified design that ensures consistent and reproducible workflows across all three Aspects.
This reduces code duplication, simplifies maintenance, and ensures all users, both internal and external, access data in the same way.
For example, an \gls{ADQL} query on the \gls{Object} catalog via TAP yields identical results whether run from the Portal, Notebook, or an external client.

The following \gls{IVOA} services are available at the time of the \gls{DP1} release:
\begin{itemize}
\vspace{0.1cm}
\item \textbf{Table Access Protocol (TAP) Service}: A TAP service \citep{2019ivoa.spec.0927D} enables queries of catalog data via the IVOA-standard \gls{ADQL}, a dialect of SQL92 with spherical geometry extensions.
The main \gls{TAP} service for \gls{DP1} runs on the Rubin-developed \gls{Qserv} database (\S~\ref{ssec:databases}), which hosts the core science tables described in \secref{ssec:catalogs}, as well as the Visit database.
It also provides image metadata in the IVOA ObsCore format via the standard \texttt{ivoa.ObsCore} table, making it an ``ObsTAP'' service \citep[ObsTAP;][]{2017ivoa.spec.0509L}.
The TAP service is based on the \gls{CADC}'s open-source Java TAP implementation\footnote{https://github.com/opencadc/tap}, modified for the exact query language accepted by Qserv.
It currently supports a large subset of ADQL, with limitations documented in the data release materials (see \secref{ssec:documentation}) and exposed via the TAP \textbf{capabilities} endpoint where possible. \

The TAP service provides metadata annotations consistent with the standard, including table and column descriptions, indications of foreign-key relationships between tables, and column metadata such as units and \gls{IVOA} Unified Content Descriptors (UCDs).

\vspace{0.1cm}
\item \textbf{Image Access Services}: Rubin image access services are compliant with \gls{IVOA} SIAv2 \citep[Simple Image Access Protocol, version 2;][]{2025arXiv250100544J,2015ivoa.spec.1223D}
for discovering and accessing astronomical images based on \gls{metadata}.
SIAv2 is a \gls{REST}-based protocol designed for the discovery and retrieval of image data. It allows, for instance, querying all images in a given band over a defined sky region and time period.

Users identify an image or observation of interest and query the service.
The result set includes \gls{metadata} about the image, such as the sky position, time, or band, and a data access URL, which includes an IVOA Identifier uniquely identifying the dataset \citep{DMTN-302}, allowing the dataset to be retrieved or a cutout requested via \gls{SODA}.

\vspace{0.1cm}
\item \textbf{Image Cutout Service}: The Rubin Cutout Service \citep{SQR-063, SQR-093} is based on the  IVOA SODA standard \citep{2017ivoa.spec.0517B}.
Users submit requests specifying sky coordinates and the cutout size as the radius from the coordinates, and the  service performs the operation on the full image and returns a result set.
For \gls{DP1}, the  cutout service is a single cutout service only where N cutout requests will require $N$ independent synchronous calls.
We expect some form of bulk cutout service by mid 2026.

\vspace{0.1cm}
\item \textbf{HiPS Data Service}: An authenticated \gls{HiPS}
\citep{2017ivoa.spec.0519F}
data service for seamless pan-and-zoom access to large-scale co-adds.
It supports fast interactive progressive image exploration at a range of resolutions.

\vspace{0.1cm}
\item \textbf{WebDAV}: A \gls{WebDav} service is provided to enable users to remotely manage, edit, and organize files and directories on the \gls{RSP} as if they were local files on their own computer. This is especially useful for local development.
\end{itemize}

\subsubsection{Data Butler}
\label{sssec:data_butler}

The Rubin Data Butler \citep{2022SPIE12189E..11J,2023arXiv230303313L},  is a high-level interface designed to facilitate seamless access to data for both users and software systems.
This includes managing storage formats, physical locations, data staging, and database mappings.
A \gls{Butler} repository contains two components:
\begin{itemize}
    \item the \emph{Data Store}: A physical storage system for datasets, e.g., a \gls{POSIX} file system or S3 object store; and
    \item the \emph{Registry}: An \gls{SQL}-compatible database that stores metadata about the datasets in the data store.
\end{itemize}
For \gls{DP1}, the Butler repository is hosted in the Google Cloud, using an \gls{S3}-compatible store for datasets and AlloyDB, a PostgreSQL-compatible database, for the registry.

In the context of the \gls{Butler}, a \emph{dataset} refers to a unique data product, such as an image, catalog or map, generated by the observatory or processing pipelines
Datasets belong to one of the various types of data products, described in \secref{sec:data_products}.
The \gls{Butler} ensures that each dataset is uniquely identifiable by a combination of three pieces of information: a data coordinate, a dataset type, and a run collection.
For example, a dataset that represents a single raw image  in the $i$ band taken on the night starting 2024-11-11 with exposure ID 2024111100074 would be represented as
\texttt{dataId={'exposure':2024111100074, 'band':'i', 'instrument':'LSSTComCam'}} and is associated with the \texttt{raw} DatasetType.
For a deep coadd on a \gls{patch} of sky in the Seagull field, there would be no exposure dimensions and instead the tract, \gls{patch} and band would be specified as  \texttt{dataId={'tract':7850, 'patch': 6, 'band':'g', 'instrument':'LSSTComCam', skymap='lsst\_cells\_v1'}} and is associated with the \texttt{deep\_coadd} DatasetType.
The tract identification numbers and corresponding target names for these tracts are listed in \tabref{tab:dp1_tracts}.

\setlength{\tabcolsep}{6pt}  
\begin{deluxetable}{ll}
\caption{Tract coverage of each DP1 field. The size of a tract is larger than the LSSTComCam field of view; however, since each observed field extends across more than one tract, each field covers multiple tracts.}
\label{tab:dp1_tracts}
\tablehead{
  \colhead{\textbf{Field Code}} & \colhead{\textbf{Tract ID}} 
}
\startdata
47\_Tuc & \parbox[t]{5cm}{453, 454} \\
ECDFS & \parbox[t]{5cm}{4848, 4849, 5062, 5063, 5064} \\
EDFS\_comcam & \parbox[t]{5cm}{2234, 2235, 2393, 2394} \\
Fornax\_dSph & \parbox[t]{5cm}{4016, 4017, 4217, 4218} \\
Rubin\_SV\_095\_-25 & \parbox[t]{5cm}{5305, 5306, 5525, 5526} \\
Rubin\_SV\_38\_7 & \parbox[t]{5cm}{10221, 10222, 10463, 10464, 10704, 10705} \\
Seagull & \parbox[t]{5cm}{7610, 7611, 7849, 7850} \\
\enddata
\end{deluxetable} 
\setlength{\tabcolsep}{6pt}  
\begin{deluxetable*}{lll}
\caption{Descriptions of and valid values for the key data dimensions in DP1. YYYYMMDD signifies date and \# signifies a single 0--9 digit.}
\label{tab:dp1_dimensions}
\tablehead{
  \colhead{\textbf{Dimension}} &
  \colhead{\textbf{Format/Valid values}} &
  \colhead{\textbf{Description}}
}
\startdata
\texttt{day\_obs} & YYYYMMDD & \parbox[t]{10cm}{A day and night of observations that rolls over during daylight hours.} \\
\texttt{visit} & YYYYMMDD\#\#\#\#\# & \parbox[t]{10cm}{A sequence of observations processed together; synonymous with ``exposure'' in DP1.} \\
\texttt{exposure} & YYYYMMDD\#\#\#\#\# & \parbox[t]{10cm}{A single exposure of all nine ComCam detectors.} \\
\texttt{instrument} & LSSTComCam & \parbox[t]{10cm}{The instrument name.} \\
\texttt{detector} & 0--8 & \parbox[t]{10cm}{A ComCam detector.} \\
\texttt{skymap} & \texttt{lsst\_cells\_v1} & \parbox[t]{10cm}{A set of tracts and patches that subdivide the sky into rectangular regions with simple projections and intentional overlaps.} \\
\texttt{tract} & See \tabref{tab:dp1_tracts} & \parbox[t]{10cm}{A large rectangular region of the sky.} \\
\texttt{patch} & 0--99 & \parbox[t]{10cm}{A rectangular region within a tract.} \\
\texttt{physical\_filter} & \parbox[t]{4cm}{u\_02, g\_01, i\_06, r\_03, z\_03, y\_04} & \parbox[t]{10cm}{A physical filter.} \\
\texttt{band} & u, g, r, i, z, y & \parbox[t]{10cm}{An conceptual astronomical passband.} \\
\enddata
\end{deluxetable*} 
The data coordinate is used to locate a dataset in multi-dimensional space, where dimensions are defined in terms of scientifically meaningful concepts, such as instrument, visit, detector or band.
For example, a calibrated single-visit image (\secref{ssec:science_images}) has dimensions including band, instrument, and detector.
In contrast, the visit table (\secref{ssec:catalogs}), a catalog of all calibrated single-epoch visits in \gls{DP1}, has only the instrument dimension.
The main dimensions used in \gls{DP1} are listed, together with a brief description, in \tabref{tab:dp1_dimensions}.
To determine which dimensions are relevant for a specific dataset, the \gls{Butler} defines dataset types, which associate each dataset with its specific set of relevant dimensions, as well as the associated Python type representing the dataset.
The dataset type defines the kind of data a dataset represents, such as  a raw image  (\texttt{raw}), a processed catalog (\texttt{object\_forced\_source}), or a \gls{sky map} (\texttt{skyMap}).
\tabref{tab:butlerdatasets} lists all the dataset types available via the Butler in \gls{DP1}, together with the dimensions needed to uniquely identify a specific dataset and the number of unique datasets of each type.

It is important to highlight a key difference between accessing catalog data via the \gls{TAP} service versus the Butler.
While the \gls{TAP} service contains entire catalogs, many of the same catalogs in the Butler are split into multiple separate catalogs.
This is partly due to how these catalogs are generated, but also because of the way data is stored within and retrieved from the Butler repository -- it is inefficient to retrieve the entire \texttt{Source} catalog, for example, from the file system.
Instead, because the \texttt{Source} catalog contains data for sources detected in the \texttt{visit\_image}s, there is one \texttt{Source} catalog in the Butler for each \texttt{visit\_image}. Similarly, there is one \texttt{Object} catalog for each \texttt{deep\_coadd}.
All the catalogs described in \secref{ssec:catalogs}, aside from the \texttt{CcdVisit}, \texttt{SSObject}, \texttt{SSSource}, and \texttt{Calibration} catalogs, are split within the Butler.
\setlength{\tabcolsep}{8pt}  
\begin{deluxetable}{llcc}
\caption{The name and number of each type of data product in the Butler and the dimensions required to identify a specific dataset.}
\label{tab:butlerdatasets}
\tablehead{
  \textbf{Data Product} &
  \textbf{Name in Butler} &
  \textbf{Required Dimensions} &
  \textbf{Number in DP1}\
}
\startdata
\multicolumn{4}{l}{\textbf{Image Data Products}} \\
\texttt{raw} & \texttt{raw} & instrument, detector, exposure & 16125 \\
\texttt{visit\_image} & \texttt{visit\_image} & instrument, detector, visit & 15972 \\
\texttt{deep\_coadd} & \texttt{deep\_coadd} & band, skymap, tract, patch & 2644 \\
\texttt{template\_coadd} & \texttt{template\_coadd} & band, skymap, tract, patch & 2730 \\
\texttt{difference\_image} & \texttt{difference\_image} & instrument, detector, visit & 15972 \\
\multicolumn{4}{l}{\textbf{Catalog Data Products}} \\
\texttt{Source} & \texttt{source} & instrument, visit & 1786 \\
\texttt{Object} & \texttt{object} & skymap, tract & 29 \\
\texttt{ForcedSource} & \texttt{object\_forced\_source} & skymap, tract, patch & 636 \\
\texttt{DiaSource} & \texttt{dia\_source} & skymap, tract & 25 \\
\texttt{DiaObject} & \texttt{dia\_object} & skymap, tract & 25 \\
\texttt{ForcedSourceOnDiaObject} & \texttt{dia\_object\_forced\_source} & skymap, tract, patch & 597 \\
\texttt{SSSource} & \texttt{ss\_source} & -- & 1 \\
\texttt{SSObject} & \texttt{ss\_object} & -- & 1 \\
\texttt{Visit} & \texttt{visit\_table} & instrument & 1 \\
\texttt{CcdVisit} & \texttt{visit\_detector\_table} & instrument & 1 \\
\enddata
\end{deluxetable} 
A dataset is associated with one or more \emph{Collections}; logical groupings of  datasets within the \gls{Butler} system that were created or processed together by the same batch operation.
Collections allow multiple datasets with the same data coordinate to coexist without conflict.
Collections support flexible, parallel processing by enabling repeated analyses of the same input data using different configurations.
The \gls{DP1} Butler is read-only; a writeable Butler is expected by mid-2026.

\subsubsection{Remote Programmatic Access}
\label{sssec:remote_api}
The Rubin \gls{RSP} \gls{API} can be accessed from a local system by data rights holders outside of the \gls{RSP}, by creating a user security token. This token can then be used as a bearer token for \gls{API} calls to the \gls{RSP} TAP service.
This capability is especially useful for remote data analysis using tools such as \gls{TOPCAT}, as well as enabling third-party systems, e.g., Community Alert Brokers, to access Rubin data.
Additionally, it supports remote development, allowing for more flexible workflows and integration with external systems.

\subsection{Portal Aspect}
\label{ssec:rsp_portal}
The Portal Aspect provides an interactive web-based environment for exploratory data discovery, filtering, querying ,and visualization of both image and catalog data, without requiring programming expertise.
It enables users to access and analyze large datasets via tools for catalog queries, image browsing, time-series inspection, and cross-matching.

The Portal is built on \gls{Firefly} \citep{2019ASPC..521...32W}, a  web application framework developed by the Infrared Processing and Analysis Center (IPAC).
\gls{Firefly} provides interactive capabilities such as customizable table views, image overlays, multi-panel visualizations, and synchronized displays linking catalog and image data.

Designed to support both exploratory data access and detailed scientific investigation, the Portal delivers an intuitive user experience, allowing users to visually analyze data while retaining access to underlying metadata and query controls.

\subsection{Notebook Aspect}
\label{subsec:notebook}
The Notebook Aspect provides an interactive, web-based environment built on Jupyter Notebooks, enabling users to write and execute Python code directly on Rubin and \gls{LSST} data without downloading it locally.
It offers programmatic access to Rubin and LSST data products, allowing users to query and retrieve datasets, manipulate and display images, compute derived properties, plot results, and reprocess data using the LSST Science Pipelines (\secref{ssec:pipelines}).
The environment comes pre-installed with the pipelines and a broad set of widely used astronomical \gls{software} tools, supporting immediate and flexible data analysis.

\subsection{Databases}
\label{ssec:databases}
The user-facing Aspects of the \gls{RSP} are supported by several backend databases that store catalog data products, image metadata, and other derived datasets.
The \gls{schema} for \gls{DP1} and other Rubin databases are available online at \url{https://sdm-schemas.lsst.io}.

\subsubsection{Qserv}
\label{sssec:qserv}
The final 10-year \gls{LSST} catalog is expected to reach \tenyearcatalogsize and contain measurements for billions of stars and galaxies across trillions of detections.
To support efficient storage, querying, and analysis of this dataset,  Rubin Observatory developed Qserv \citep{Wang:2011:QDS:2063348.2063364, C15_adassxxxii} -- a scalable, parallel, distributed SQL database system.
\gls{Qserv} partitions data over approximately equal-area regions of the celestial sphere, replicates data to ensure resilience and high availability, and uses shared scanning to reduce overall I/O load.
It also supports a package of scientific user-defined functions (SciSQL: \url{https://smonkewitz.github.io/scisql/}) simplifying complex queries involving spherical geometry, statistics, and photometry.
\gls{Qserv} is built on robust production-quality components, including MariaDB (\url{https://www.mariadb.org/}) and XRootD (\url{https://xrootd.org/}).
Qserv runs at the \gls{USDF} and user access to catalog data is via the TAP service (\secref{sssec:ivoa_services}).
This enables catalog-based analysis through both the \gls{RSP} Portal and Notebook Aspects.

Although the small \gls{DP1} dataset does not require Qserv’s full capabilities, we nevertheless chose to use it for \gls{DP1} to accurately reflect the future data access environment and to gain experience with scientifically-motivated queries ahead of full-scale deployment.
\gls{Qserv} is open-source and available on GitHub: \url{https://github.com/lsst/qserv}.

 \section{Support for Community Science}
\label{sec:community_science}

Rubin Observatory has a science community that encompasses thousands of individuals worldwide, with a broad range of experience and expertise in astronomy in general, and in the analysis of optical imaging data specifically.

Rubin's model to support this diverse community to access and analyze \gls{DP1} emphasizes self-help via documentation and tutorials, and employs an open platform for asynchronous issue reporting that enables crowd-sourced solutions.
These two aspects of community support are augmented by virtual engagement activities.
In addition, Rubin supports its Users Committee to advocate on behalf of the science community, and supports the eight \gls{LSST} Science Collaborations (\secref{ssec:science_collaborations}).

All of the resources for scientists that are discussed in this section are discoverable by browsing the \textit{For Scientists} pages of the Rubin Observatory website\footnote{\nolinkurl{https://rubinobservatory.org/for-scientists}}.

\subsection{Documentation}
\label{ssec:documentation}

The data release documentation for \gls{DP1}\footnote{\nolinkurl{https://dp1.lsst.io}} provides an overview of the \gls{LSSTComCam} observations, detailed descriptions of the data products, and a high-level summary of the processing pipelines. 
Although much of its content overlaps significantly with this paper, the documentation is presented as a searchable, web-based resource built using Sphinx\footnote{\nolinkurl{https://www.sphinx-doc.org/}}, with a focus on enabling scientific use of the data products.

\subsection{Tutorials}
\label{ssec:tutorials}

A suite of tutorials \citep{10.11578/rubin/dc.20250909.20} that demonstrate how to access and analyze \gls{DP1} using the RSP accompanies the \gls{DP1} release\footnote{\nolinkurl{https://dp1.lsst.io/tutorials}}.
Jupyter Notebook tutorials are available via the ``Tutorials'' drop-down menu within the Notebook aspect of the \gls{RSP}.
Tutorials for the Portal and API aspects of the \gls{RSP} can be found in the data release documentation.

These tutorials are designed to be inclusive, accessible, clear, focused, and consistent.
Their format and contents follow a set of guidelines \citep{RTN-045} that are informed by modern standards in technical writing.

\subsection{Community Forum}
\label{ssec:forum}

The venue for all user support is the Rubin Community Forum\footnote{\nolinkurl{https://community.lsst.org/}}. Questions about any and all aspects of the Rubin data products, pipelines, and services, including \gls{DP1}, should be posted as new topics in the Support category.
This includes beginner-level and ``how-to'' questions, advanced scientific analysis questions, technical bug reports, account and data access issues, and everything in between.
The Support category of the Forum is monitored by Rubin staff, who follow an established internal workflow for following-up and resolving all reported issues.

The Rubin Community Forum is built on the open-source Discourse platform.
It was chosen because, for a worldwide community of ten thousand Rubin users, a traditional (i.e., closed) help desk represents a risk to Rubin science (e.g., many users with the same question having to wait for responses).
The open nature of the Forum enables self-help by letting users search for similar issues, and enables crowd-sourced problem solving (and avoids knowledge bottlenecks) by letting users help users.

The Rubin Community Forum, and the internal staff workflows for user support, were set up, tested, and refined with \gls{DP0} so that it was ready for use with \gls{DP1}.

\subsection{Engagement Activities}
\label{ssec:engagement}

A variety of live virtual and in-person workshops and seminars offer learning opportunities to scientists and students working with the Rubin data products, services, and tools.

\begin{itemize}
\item Rubin Science Assemblies (weekly, virtual, 1 hour): alternates between hands-on tutorials based on the most recent data release and open drop-in ``office hours'' with Rubin staff.
\item Rubin Data Academy (annual, virtual, 3-4 days): an intense set of hands-on tutorials based on the most recent data release, along with co-working and networking sessions.
\item Rubin Community Workshop (annual, virtual, 5 days), a science-focused conference of contributed posters, talks, and sessions led by members of the Rubin science community and Rubin staff.
\end{itemize}

Following the release of \gls{DP1}, all of these engagement activities focused on use of \gls{DP1} by the science community.
In particular, the 2025 Rubin Data Academy was run the week of the \gls{DP1} release, in order to immediately facilitate community access.
The 2025 Rubin Community Workshop had several sessions to introduce people to the \gls{DP1} dataset and demonstrate how to access and analyze it with the \gls{RSP}.

For schedules, connection information, zoom recordings, and associated materials, visit the \textit{For Scientists} pages of the Rubin Observatory website\footnote{\nolinkurl{https://rubinobservatory.org/for-scientists/events-deadlines}}.
Requests for custom tutorials and presentations for research groups are also accommodated.

\subsection{Users Committee}
\label{ssec:users_committee}

This committee is charged with soliciting feedback from the science community, advocating on their behalf, and recommending science-driven improvements to the \gls{LSST} data products and the Rubin Science Platform tools and services.
Community members are encouraged to attend their virtual meetings and raise issues to their attention, so they can be included in the committee's twice-yearly reports to the Rubin Observatory \gls{Director}.

Like the Forum, the Users Committee was established and began its work with \gls{DP0}, and that feedback was implemented for \gls{DP1}. 
The community's response to \gls{DP1} will be especially valuable input to \gls{DP2} and \gls{DR1}, and the Users Committee encourages all users to interact with them.
For a list of members and contact information, visit the \textit{For Scientists} pages of the Rubin Observatory website.

\subsection{Science Collaborations}
\label{ssec:science_collaborations}

The eight \gls{LSST} Science Collaborations are independent, worldwide communities of scientists, self-organized into collaborations based on their research interests and expertise.
Members work together to apply for funding, build software infrastructure and analysis algorithms, and incorporate external data sets into their \gls{LSST}-based research.

The Science Collaborations also provide valuable advice to Rubin Observatory on the operational strategies and data products to accomplish specific science goals, and Rubin Observatory supports the collaborations via staff liaisons and regular virtual meetings with Rubin operations leadership.

The Science Collaborations have been functioning for many years, and their engagement and feedback on \gls{DP0} was implemented into the community science model for \gls{DP1}, as it will for future data releases.
 \section{Summary and Future Releases}
\label{sec:summary}

Rubin Data Preview 1 offers an initial look at the first on-sky data products and access services from the Vera C. Rubin Observatory. \gls{DP1} forms part of Rubin's Early Science Program, and provides the scientific community with an early opportunity to familiarize themselves with the data formats and access infrastructure for the forthcoming Legacy Survey of Space and Time.
This early release has a proprietary period of two years, during which time it is  available to Rubin data rights holders only via the cloud-based \gls{RSP}.

In this paper we have described the completion status of the observatory at the time of data acquisition, the commissioning campaign that forms the basis of \gls{DP1}, and the processing pipelines used to produce early versions of data products.
We provide details on the data products, their characteristics and known issues, and describe the Rubin Science Platform for access to and analysis of \gls{DP1}.

The data products described in this paper derive from observations obtained by \gls{LSSTComCam}. \gls{LSSTComCam} contains only around 5\% the number of CCDs as the full LSST Science Camera (\gls{LSSTCam}), yet the \gls{DP1} dataset that it has produced will already enable a very broad range of science.
At \sizeinbytes in size, \gls{DP1} covers a total area of \totalarea and contains \nexposures single-epoch images, \ndeepcoadds deep coadded images and \nobjects distinct astrophysical objects, including  \nnewasteroiddiscoveries  new asteroid discoveries.

While some data products anticipated from the \gls{LSST} are not yet available, e.g., cell-based coadds, \gls{DP1} includes several products that will not be provided in future releases.
Notably, difference images are included in \gls{DP1} as pre-generated products; in future releases, these will instead be generated on demand via dedicated services.
The inclusion of pre-generated difference images in \gls{DP1} is feasible due to the relatively small size of the dataset, an approach that will not scale to the significantly larger data volumes expected in subsequent releases.

The \gls{RSP} is continually under development, and new functionality will continue to be deployed incrementally as it becomes available, and independent of the future data release schedule.
User query history capabilities, context-aware documentation and a bulk cutout services are just a few of the services currently under development.

Coincident with the release of \gls{DP1}, Rubin Observatory begins its Science Validation Surveys with the LSST Science Camera (i.e., \gls{LSSTCam}).
This final commissioning phase will produce a dataset that will form the foundation for the second Rubin Data Preview, \gls{DP2}.
Full operations, marking the start of the \gls{LSST}, are expected to commence in 2026. 
\begin{acknowledgments}.
This material is based upon work supported in part by the National Science Foundation through Cooperative Agreements AST-1258333 and AST-2241526 and Cooperative Support Agreements AST-1202910 and AST-2211468 managed by the Association of Universities for Research in Astronomy (AURA), and the Department of Energy under Contract No.\ DE-AC02-76SF00515 with the SLAC National Accelerator Laboratory managed by Stanford University.
Additional Rubin Observatory funding comes from private donations, grants to universities, and in-kind support from LSST-DA Institutional Members.

This work has been supported by the French National Institute of Nuclear and Particle Physics (IN2P3) through dedicated funding provided by the National Center for Scientific Research (CNRS).

This work has been supported by STFC funding for UK participation in LSST, through grant ST/Y00292X/1.
\end{acknowledgments}
\vspace{5mm}

\facilities{Rubin:Simonyi (LSSTComCam), Rubin:USDAC}

\software{Rubin Data Butler \citep{2022SPIE12189E..11J},
          LSST Science Pipelines \citep{PSTN-019},
          LSST Feature Based Scheduler v3.0  \citep{peter_yoachim_2024_13985198, Naghib_2019}
          Astropy \citep{2013A&A...558A..33A, 2018AJ....156..123A, 2022ApJ...935..167A}
          PIFF \citep{DES:2020vau},
          GBDES \citep{2022ascl.soft10011B},
          Qserv \citep{Wang:2011:QDS:2063348.2063364, C15_adassxxxii}, 
          Slurm, HTCondor, CVMFS, FTS3, ESNet
          }

\appendix
\printglossaries

\bibliographystyle{aasjournal}
\bibliography{local,lsst,ivoa,lsst-dm,refs_ads,refs,books}
\end{document}